\documentclass{pasa}

\usepackage{graphicx}
\usepackage{newtxtext,newtxmath}

\usepackage{amsmath}	
\usepackage{amssymb}	
\usepackage{longtable}
\usepackage{comment}
\usepackage{xspace}
\usepackage{ccaption}
\usepackage{pdflscape}
\usepackage[letterpaper]{geometry}
\usepackage{acronym}

\newcommand{\snr}{\ensuremath{S/N}\xspace}

\newcommand{\res}[1]{#1\xspace}

\jid{PASA}
\doi{10.1017/pas.\the\year.xxx}
\jyear{\the\year}

\usepackage{aas_macros}
\usepackage{hyperref} 
\hypersetup{colorlinks,citecolor=blue,linkcolor=blue,urlcolor=blue}

\title[MeerTime MSP Census]{The MeerTime Pulsar Timing Array -- \\
A Census of Emission Properties and Timing Potential}

\newcommand{\jbca}{1}
\newcommand{\cas}{2}
\newcommand{\ozgrav}{3}
\newcommand{\mpi}{4}
\newcommand{\icrar}{5}
\newcommand{\sarao}{6}
\newcommand{\atnf}{7}
\newcommand{\ox}{8}
\newcommand{\skao}{9}
\newcommand{\uwc}{10}
\newcommand{\aut}{11}
\newcommand{\cnrs}{12}
\newcommand{\usn}{13}
\newcommand{\luth}{14}

\author[R.~Spiewak et al.]{
R.~Spiewak,$^{\jbca,\cas,\ozgrav}$\thanks{E-mail: \href{mailto:renee.spiewak@manchester.ac.uk}{renee.spiewak@manchester.ac.uk (RS)}}\hspace{4pt} 
M.~Bailes,$^{\cas,\ozgrav}$ 
M.~T.~Miles,$^{\cas,\ozgrav}$ 
A.~Parthasarathy,$^{\mpi,\cas,\ozgrav}$ 
D.~J.~Reardon,$^{\cas,\ozgrav}$ 
M.~Shamohammadi,$^{\cas}$ 
R.~M.~Shannon,$^{\cas,\ozgrav}$ 
N.~D.~R.~Bhat,$^{\icrar}$
S.~Buchner,$^{\sarao}$
A.~D.~Cameron,$^{\cas,\ozgrav}$
F.~Camilo,$^{\sarao}$
M.~Geyer,$^{\sarao}$
S.~Johnston,$^{\atnf}$
A.~Karastergiou,$^{\ox}$
M.~Keith,$^{\jbca}$
M.~Kramer,$^{\mpi,\jbca}$
M.~Serylak,$^{\skao,\uwc}$
W.~van Straten,$^{\aut}$
G.~Theureau,$^{\cnrs,\usn,\luth}$
V.~Venkatraman Krishnan$^{\mpi}$

\affil{$^{\jbca}$Jodrell Bank Centre for Astrophysics, Department of Physics and Astronomy, University of Manchester, Manchester M13 9PL, UK}
\affil{$^{\cas}$Centre for Astrophysics and Supercomputing, Swinburne University of Technology, PO Box 218, Hawthorn, VIC 3122, Australia}
\affil{$^{\ozgrav}$ARC Centre of Excellence for Gravitational Wave Discovery (OzGrav), Mail H29, Swinburne University of Technology, PO Box 218, Hawthorn, VIC 3122, Australia }
\affil{$^{\mpi}$Max-Planck-Institut f\"{u}r Radioastronomie, Auf dem H\"{u}gel 69, D-53121 Bonn, Germany }
\affil{$^{\icrar}$International Centre for Radio Astronomy Research, Curtin University, Bentley, WA 6102, Australia}
\affil{$^{\sarao}$South African Radio Astronomy Observatory, 2 Fir Street, Observatory 7925, South Africa }
\affil{$^{\atnf}$Australia Telescope National Facility, CSIRO Space and Astronomy, PO~Box~76, Epping NSW~1710, Australia }
\affil{$^{\ox}$Oxford Astrophysics, Denys Wilkinson Building, Keble Road, OX1 3RH, UK}
\affil{$^{\skao}$Square Kilometre Array Observatory, Jodrell Bank Observatory, Macclesfield, Cheshire SK11 9DL, United Kingdom}
\affil{$^{\uwc}$Department of Physics and Astronomy, University of the Western Cape, Bellville, Cape Town, 7535, South Africa}
\affil{$^{\aut}$Institute for Radio Astronomy \& Space Research,
Auckland University of Technology, Private Bag 92006, Auckland 1142, New Zealand}
\affil{$^{\cnrs}$Laboratoire de Physique et Chimie de l'Environnement et de l'Espace 
LPC2E UMR7328, Universit\'e d'Orl\'eans, CNRS, F-45071 Orl\'eans, France}
\affil{$^{\usn}$Station de Radioastronomie de Nan\c cay, Observatoire de Paris, PSL University, 
CNRS, Universit\'e d'Orl\'eans, 18330 Nan\c cay, France}
\affil{$^{\luth}$Laboratoire Univers et Th\'eories, Observatoire de Paris, Universit\'e PSL, CNRS, Universit\'e de Paris, 92190 Meudon, France}
}

\begin{document}
\begin{frontmatter}

\maketitle

\acrodef{gw}[GW]{gravitational wave}
\acrodef{msp}[MSP]{millisecond pulsar}
\acrodef{ipta}[IPTA]{International Pulsar Timing Array}
\acrodef{nstar}[NS]{neutron star}
\acrodef{rm}[RM]{rotation measure}
\acrodef{lofar}[LOFAR]{Low-Frequency Array}
\acrodef{fast}[FAST]{Five-hundred-meter Aperture Synthesis Telescope}
\acrodef{gbt}[GBT]{Green Bank Telescope}
\acrodef{ska}[SKA]{Square Kilometre Array}
\acrodef{gmrt}[GMRT]{Giant Metrewave Radio Telescope}
\acrodef{snr}[\snr]{signal-to-noise ratio}
\acrodef{pta}[PTA]{Pulsar Timing Array}
\acrodef{mpta}[MPTA]{MeerTime Pulsar Timing Array}
\acrodef{dm}[DM]{dispersion measure}
\acrodef{ppta}[PPTA]{Parkes Pulsar Timing Array}
\acrodef{toa}[ToA]{time of arrival}
\acrodefplural{toa}[ToAs]{times of arrival}
\acrodef{sarao}[SARAO]{South African Radio Astronomy Observatory}
\acrodef{nanograv}[NANOGrav]{North American Nanohertz Observatory for Gravitational Waves}
\acrodef{epta}[EPTA]{European Pulsar Timing Array}
\acrodef{inpta}[InPTA]{Indian Pulsar Timing Array}
\acrodef{gc}[GC]{globular cluster}
\acrodef{rfi}[RFI]{radio frequency interference}
\acrodef{ptuse}[PTUSE]{pulsar timing user supplied equipment}
\acrodef{atnf}[ATNF]{Australia Telescope National Facility}


\begin{abstract}
MeerTime is a five-year Large Survey Project to time pulsars with MeerKAT, the 64-dish South African precursor to the \acl{ska}. 
The science goals for the programme include timing \acp{msp} to high precision ($<1\,\upmu$s) to study the Galactic \ac{msp} population and to contribute to global efforts to detect nanohertz \aclp{gw} with the \ac{ipta}. 
In order to plan for the remainder of the programme \res{and} to use the allocated time most efficiently, we have conducted an initial census \res{with the MeerKAT ``L-band'' receiver} of 189 \acp{msp} visible to MeerKAT and here present their \res{\aclp{dm}, polarization profiles, polarization fractions, \aclp{rm}, flux density measurements, spectral indices, and timing potential}. 
As all of these observations are taken with the same instrument (which uses coherent dedispersion, interferometric polarization calibration techniques, and a uniform flux scale), they present an excellent resource for population studies. 
We used wideband pulse portraits as timing standards for each \ac{msp} and demonstrated that the \ac{mpta} can already contribute significantly to the \ac{ipta} as it currently achieves better than 1\,$\upmu$s timing accuracy on 89 \acp{msp} (observed with fortnightly cadence). 
By the conclusion of the initial five-year MeerTime programme in July 2024, the \ac{mpta} will be extremely significant in global efforts to detect the \acl{gw} background with a contribution to the detection statistic comparable to other long-standing timing programmes.

\end{abstract}

\acresetall

\begin{keywords}
pulsars: general; \res{methods: observational}
\end{keywords}
\end{frontmatter}



\section{Introduction}\label{sec:mtc_int}
Millisecond pulsars (MSPs)\acused{msp} are the sub-population of pulsars that are characterised by short spin periods ($P<50$\,ms) and low magnetic field strengths ($B_\textrm{surf} <10^{10}$\,G)\footnote{This definition of an MSP is relatively simple \citep[c.f.][]{lgy+12} but useful for this work and in relation to the other MeerTime themes described in section~\ref{sec:mtc_int}.}. 
The first binary pulsar, PSR~B1913$+$16 (J1915$+$1606; \citealt{ht75}), possessed an unusually \res{low} spin period ($P=59$\,ms) and magnetic field strength ($B_\textrm{surf} = 2 \times 10^{10}\,$G) compared to the bulk of the population known at the time, which, apart from the young Crab and Vela pulsars, consisted mainly of solitary pulsars with spin periods of typically 0.1-2\,s and magnetic fields of order 10$^{12}$\,G. 
Prior to the discovery of PSR~B1913+16, \citet{bkk74} had predicted that mass accretion onto \acp{nstar} may decrease their magnetic fields, and PSR~B1913+16 appeared to agree with their prediction as at least one of the \acp{nstar} in this system must have accreted matter during a previous stage of mass transfer. 
When the first \ac{msp}, PSR~B1937+21 (J1939+2134), was discovered with $P=1.56$\,ms and $B_\textrm{surf}=4\times10^{8}$\,G \citep{bkh+82}, it was soon proposed that a (now missing) companion had been responsible for its spin-up and the low field made spin-up to its millisecond period possible \citep{acrs82,rs82}. 
As more recycled pulsars were discovered, a self-consistent model began to emerge \citep[e.g.,][]{bailes89} in which most pulsars lost their binary companions in the initial explosion due to impulsive kicks and mass loss at the time of the first explosion, but those that remained bound had a chance for a second life upon accreting material from their evolved companions that not only reduced their magnetic field strengths but made it possible for them to be spun-up to millisecond periods. 
NSs with lower-mass companions accrete material for longer than those with high-mass companions because their evolutionary timescales are longer, and \res{they} are able to be spun up to very short periods ($P\sim1.5$\,ms) with an associated reduction of their magnetic fields to $\sim10^8$\,G. 
\acp{msp} with heavier white dwarf companions tend to spin more slowly ($P\sim 10$\,-\,$16$ ms) with some notable exceptions, such as PSR~J1614$-$2230 \citep{crh+06a,dprrh10}, and those with \ac{nstar} companions slower still ($P\sim 16$\,-\,$60$ ms). 
For an early paper describing these models, see \citet{bvdh91}, and, for a more recent discussion of their binary properties, \citet{tkf+17}. 
Support for these models \citep[see, e.g.,][]{phinney92} comes in part from the large fraction of \acp{msp} observed to have binary companions compared to the fraction of ``normal'' pulsars in binaries ($\sim74$\% of \acp{msp} are in binaries but only $\sim2$\% of \res{longer-period} pulsars are in binaries; from the \ac{atnf} Pulsar Catalogue, v.1.64, \citealt{mhth05}).

Studies of the properties of \acp{msp} and the Galactic \ac{msp} population overall have often been limited or biased by small sample sizes \citep[e.g.,][]{jb91} or by instrumentation that prohibits studies of the polarization or novel features of the pulse profile, which are probably determined by the pulsar's magnetosphere and spin period. 
\res{See, for example, \citet{kxl+98}, \citet{xkj+98}, and \citet{kll+99} for a comprehensive study of 27 \acp{msp}, or the review by \citet{kx00}, and references therein.} 
The majority of \acp{msp} have been discovered by quasi-uniform surveys by large-aperture telescopes such as the Parkes 64-m radio telescope (also known as \textit{Murriyang}; \res{e.g., \citealt{mlc+01,kjvs+10}}), the Arecibo 305-m dish \res{\citep[e.g.,][]{cfl+06}}, and the Green Bank 100-m Telescope (GBT; \res{e.g., \citealt{blr+13,slr+14}}). \acused{gbt}
More recently, other instruments like the \ac{lofar} and \ac{gmrt} have discovered pulsars in a combination of quasi-uniform surveys and \res{targeted observations of} gamma-ray point sources with pulsar characteristics \res{\citep[see, e.g.,][]{rap+12,bcm+16,scb+19}}. 
\res{In many cases, the only published pulse profiles are those formed using survey data and thus limited in resolution due to survey data rates as well as the use of incoherent dispersion correction.} 
It is not uncommon for pulsars with large \acp{dm} to have narrow features smeared beyond recognition using the discovery hardware which cannot \textit{a priori} know the \ac{dm} of the pulsars.

There is also some inconsistency in reported \ac{msp} flux densities in the \ac{atnf} Pulsar Catalogue and the literature. 
This is for a variety of reasons that include uncertainties in the pulsar position within the primary beam during initial timing, pulsar scintillation that makes pulsars more likely to have been brighter in their discovery observations, and design limitations that make absolute flux calibration a secondary consideration in the construction of some pulsar survey hardware. 
\res{For example, the flux density of PSR~J1843$-$1113 was first determined by \citet{hfs+04} to be $S_{1400} = 0.1$\,mJy, but was revised to 0.3\,mJy by \citet{lbb+13}, and then to 0.6\,mJy by \citet{dcl+16}.}

Analyses of \ac{msp} polarization profiles and their evolution with radio frequency, which could potentially lead to a more consistent theory of the pulsar emission mechanism, have only been performed on samples of a few to $\sim20$ \acp{msp} \citep[e.g.,][and references therein]{rah+17}, and the field could significantly benefit from a large comprehensive analysis derived from high \ac{snr} observations. 
Additional areas of pulsar astronomy that would benefit from such a dataset include probing the Galactic magnetic field with \ac{rm} measurements \citep[e.g.,][]{hmvd18,sbg+19}, improving Galactic electron density models with new independent distance measurements \citep[see, e.g.,][]{mnf+16,ymw17}, and constraining models of the \ac{nstar} equation of state through \ac{msp} mass measurements \citep{of16}.

Estimates of the total \res{\ac{msp}} \citep{lbb+13} and binary \ac{nstar}-\ac{nstar} populations \citep[e.g.,][]{bdap+03} rely upon \res{our understanding of} the \res{\ac{msp}} luminosity function, which in turn depends upon \res{measured} \ac{msp} distances, flux densities and their pulse shapes. 
Hence an authoritative survey of \acp{msp} with the same flux density scale, coherent dedispersion to ascertain the true pulse shape, and ultimately pulsar parallaxes from timing is very valuable when we want to estimate their underlying populations and ascertain potential yields for future pulsar surveys and timing programmes such as those planned for the Square Kilometre Array \citep[SKA; e.g.,][]{lab+18}. \acused{ska}

The new precursor to the \ac{ska} in South Africa, MeerKAT \res{(\textit{Meer} Karoo Array Telescope)}, \res{presents} an opportunity to revisit the fluxes, spectral indices, pulse shapes, polarimetry, and timing potential of the \acp{msp} visible to it. 
The MeerTime \res{project} \citep{bja+20} is a MeerKAT Large Survey Project and has four major themes. 
These include the Thousand Pulsar Array \citep{jkk+20}, the Relativistic Binary programme \citep{ksvk+21}, the Globular Cluster pulsar timing programme \citep{rgf+21}, and the \ac{mpta}, which studies \acp{msp} with the primary aim of detecting \acp{gw}. 
\citet{pbs+21} \res{analysed the pulse-to-pulse variability (``jitter'') of 29 of these \acp{msp}, placing limits on their timing precision.} 
\res{In this work, we present a census of 189 \acp{msp} visible to MeerKAT at $\sim1400$\,MHz frequencies, providing high-time-resolution, polarization- and flux-calibrated profiles, measured \acp{dm}, polarization fractions, \acp{rm}, sub-banded flux densities, and spectral indices, as well as timing potential for the \ac{mpta}. 
The first results of the \ac{mpta} will be presented by Miles et al. (2022, in prep.), including a full release of timing residuals and refined ephemerides, and closely followed by an analysis of red noise and \ac{gw} search (Middleton et al., 2022, in prep..}

The 64 13.5-m offset Gregorian MeerKAT dishes have a high aperture efficiency and low system temperature that achieves a system equivalent flux density of just \res{$\approx7.0$}\,Jy at $\sim1400$\,MHz frequencies with the full array \citep{jonas+16}. 
With an antenna gain roughly 4 times that of Parkes, a lower system temperature ($\sim18$\,K), and a larger bandwidth (856\,MHz of the MeerKAT L-band receiver compared with the 340 MHz of the Parkes multibeam receiver that was used for much of the last decade for the \res{Parkes Pulsar Timing Array \citep[PPTA; ][]{mhb+13}}), MeerKAT has become the best telescope in the southern hemisphere for a large-scale \ac{msp} observing programme and population study \citep{bja+20}. \acused{ppta}
Compared to, for example, the \ac{fast}, which can observe down to Dec\,$>-15$\,degrees, MeerKAT has a lower sensitivity but greater sky coverage (Dec\,$<+40$\,degrees, \res{which includes the MSP-rich Galactic centre}) and significantly higher slew rates (60 and 120\,deg\,min$^{-1}$ in elevation and azimuth respectively). 
The 500-m diameter \ac{fast} has much superior gain but can only observe about one source every ten minutes due to reconfiguration overheads. 
The ability to quickly move between sources with minimal overhead is one of the advantages of large-N (number), small-D (diameter) arrays like MeerKAT.

The timing precision and stability of \acp{msp} can additionally be leveraged to search for low-frequency \acp{gw} through observations of arrays of \acp{msp} like a galactic-scale interferometer, termed Pulsar Timing Arrays \citep[PTAs;][]{hd83}. \acused{pta}
The current global effort, the \ac{ipta}, places stringent constraints on the amplitude of a \ac{gw} background but no detection has been made \citep{pdcd+19}, although there are tantalising indications of similar spectral slopes in the red noise of the timing residuals of many of the \acp{msp} \res{with no spatial correlations indicative of a \ac{gw} background} \citep[e.g.,][]{abb+20ng,ccg+21,gsr+21}. 
While the sensitivity of current \acp{pta} to a \ac{gw} signal scales favourably with time, it can be greatly accelerated by adding more pulsars to the array and increasing the cadence \citep{sejr13}. 
As Parkes is the largest telescope in the southern hemisphere currently contributing to \ac{pta} efforts, MeerTime can contribute significantly to the \ac{ipta} by $i)$ observing the current \ac{ipta} pulsars with higher precision, $ii)$ observing additional pulsars not currently included in the \ac{ipta} due to low precision, and $iii)$ helping offset the recent loss of the Arecibo 305-m telescope.

Throughout this paper, we define \acp{msp} as having rotational periods $<50$\,ms and period derivatives $<2\times10^{-17}$\,s\,s$^{-1}$. 
In section~\ref{sec:mtc_method}, we discuss the sources included in this census and the observational parameters. 
We further discuss the polarization properties in section~\ref{sec:mtc_pol}, the flux density measurements in section~\ref{sec:mtc_flux}, and the preliminary timing results in section~\ref{sec:mtc_time}. 
We conclude with some predictions for the \ac{ska} in section~\ref{sec:mtc_conc}. 
Information on accessing FITS-format files of integrated pulse profiles as well as the full table of results is provided at the end of this manuscript.

\section{Methods}
\label{sec:mtc_method}
All observations included in this paper were performed with the MeerKAT L-band receiver (856-1712\,MHz) with the \res{Pulsar Timing User Supplied Equipement (PTUSE)} \acused{ptuse} back-end that uses coherent dedispersion. 
A full description of the observing system can be found in \citet{bja+20}. 
Typical observations used 54-61 dishes resulting in a system equivalent flux density of 6.3-7.2\,Jy near 1400\,MHz. 
The system temperature varies as a function of frequency and is slightly worse near the band edges but difficult to measure in the \res{areas of the band permanently affected by \ac{rfi}} \citep{bja+20,gsa+21}. 
\res{While the MeerKAT Ultra-High Frequency (UHF) receiver at 544-1088\,MHz is now fully commissioned for pulsar timing, time constraints did not allow for a full \ac{msp} census with the UHF receiver. Using the results of this L-band census, a smaller census of low-\ac{dm}, steep-spectrum sources with the UHF receiver may be feasible, and such sources may be observed with the UHF receiver for the \ac{mpta}. The same applies for high-DM, flat-spectrum sources with the MeerKAT S-band receiver in future.}

As the original network interface card for the first \ac{ptuse} machine was unable to ingest the full bandwidth in ``1K mode'' (using 1024 channels across the 856-MHz band), observations between mid-April 2019 and February 2020 were restricted to 928 frequency channels (dropping 48 channels from the top and bottom of the band where the roll-off adversely affects sensitivity in any case), reducing the recorded bandwidth to 775.75\,MHz. 
In order to maintain consistency throughout our analyses, we therefore chose to reduce the observations with 856\,MHz of recorded bandwidth down to the same 775.75\,MHz. 
The loss in sensitivity is nearly negligible because of the sharp roll-off at the edges of the receiver band; comparisons of \ac{snr} measurements of a sample of 20 pulsars (250 observations) show a median sensitivity loss of only 2.2\% (mean of 4.5\%)\footnote{20\% of the observations in our sample of 250 showed an increase of up to $\approx5$\% in \ac{snr} after trimming the band edges, which is likely an artefact of band-pass calibration over-weighting edge channels.}. 
Almost all of our pulsars have negative spectral indices (typically between $-$1 and $-$3), which makes the loss of the lower part of the band worse than the top of it. 
Broader bands also make determination of \acp{dm} more accurate, which in turn aids in precision timing. 
Since late Feb 2020, all of our observations have used the entire 856 MHz of bandwidth.

\begin{figure}
    \centering
    \includegraphics[width=\columnwidth]{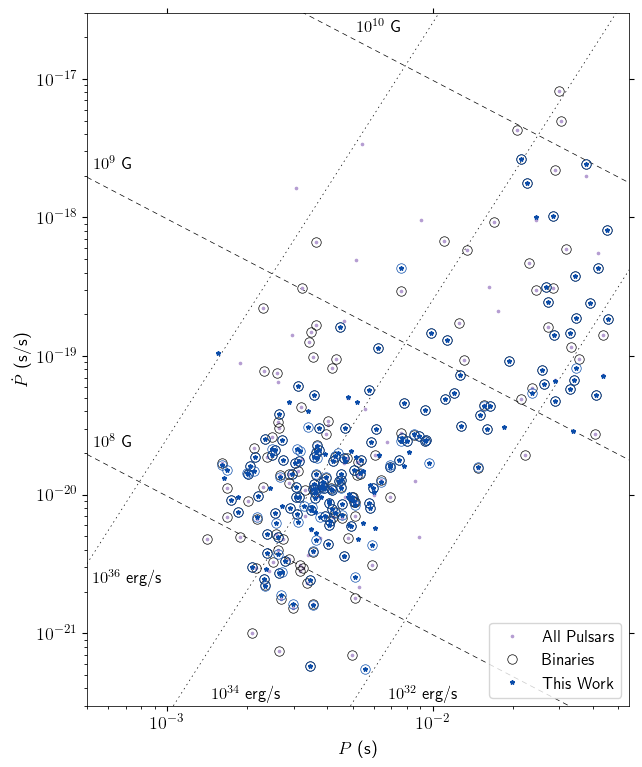}
    \caption[$P$-$\dot P$ diagram of \ac{msp} census sources]{A $P$-$\dot P$ diagram showing the known \acp{msp} (purple dots) and those included in the MeerTime \ac{msp} census project (blue dots). Binary \acp{msp} are outlined with a circle. The data are from the \ac{atnf} pulsar catalogue v1.64 \citep{mhth05}.} 
    \label{fig:mtc_ppdot}
\end{figure}

Our source list was derived from the \ac{atnf} pulsar catalogue \citep[using v.1.63;][]{mhth05}, limited to radio-loud \acp{msp} not associated with \acp{gc} and visible to MeerKAT \res{(Dec $\leq +30$\,degrees to avoid severe time restrictions)}. 
We further restricted the list to exclude sources with positional uncertainties greater than 2\,arcsec (near the resolution of the tied array beam) \res{or without complete timing solutions (ephemerides) available to us}, and we removed 12 Northern sources (Dec\,$>0$\,degrees) with \res{\textit{known}} low flux densities ($S_{1400}<0.1$\,mJy), under the assumption that these could be observed with greater efficiency by Northern telescopes and were unlikely to efficiently contribute to the timing array\footnote{\res{There remain 10 pulsars in this study with Dec~$>0$\,degrees and $S_{1400} < 0.1$\,mJy as none of these had flux densities in v.1.63 of the ATNF pulsar catalogue but the \snr was sufficient in 1024-second observations for inclusion.}}. 
Finally, some sources were removed from our list \res{(after one or more initial observations)} due to insufficient precision in ephemeris parameters (leading to significant ``smearing'' in individual 8-second sub-integrations) or due to low \ac{snr} in 1024-second observations\footnote{\res{Some of the 33 sources that were excluded from the list of census sources were later observed to good precision after, for example, improving the ephemeris, but there were insufficient usable observations during the selected time range for this census}}. 
In total, 33 sources were excluded from our original list. 
To this list, we added 4 unpublished sources (or sources with currently unpublished timing solutions) matching the same sample \res{limits}; these \acp{msp} were discovered in the Einstein@Home reprocessing of the Parkes Multibeam Pulsar Survey \citep{kek+13} and the Green Bank North Celestial Cap survey \citep{slr+14}. 
We list our sources with basic parameters, including number of observations, and references in Table~\ref{tab:mtc_basic1}, where the \acp{dm} indicated are fitted from the brightest epoch per source\footnote{\res{We chose to provide DMs from single observations to demonstrate the precision achievable with the MeerKAT L-band receiver. Long-term time-dependent variations, where significant, are accounted for in our ephemerides using ``DM1'' and ``DM2'' parameters.} We include the epoch for each measured DM, and further information, in a supplementary table.}. 
We also plot the sources in a $P$-$\dot P$ diagram in Figure~\ref{fig:mtc_ppdot}, where $P$ is the rotation period of the pulsar and the dot denotes its time derivative. 
\res{We note that, although initial ephemerides from published sources, the ATNF pulsar catalogue, or private communication were used as a starting point, we regularly refined ephemerides to ensure high-quality observations. 
These updated ephemerides will be presented in future work (Miles et al., 2022, in prep.).}

\begin{landscape}
\newgeometry{left=0.6in,top=1.05in,textwidth=870pt,textheight=526pt,right=0.5in,headsep=30pt}

\begin{table*}
  \centering
  \caption[Auto-generated table]{The 189 pulsars included in the MeerTime census with basic parameters, measured RMs with 1-$\sigma$ uncertainties, percentages of linear polarization ($L/I$), circular polarization ($V/I$) and absolute circular polarization ($|V|/I$), median ToA uncertainties normalised to 256\,seconds (see section 5 for details), flux densities at 1400~MHz, spectral indices from the power law model fit, and references (initial publication). The number of observations included in the analysis for each pulsar is given in the second column. Values less than 1-$\sigma$ for constrained-positive parameters are given as 2-$\sigma$ upper limits. The pulsars included in the regular observing programme (88 out of the 89) are indicated with asterisks. RMs measured from summed observations (not the mean of RMs from multiple observations) are indicated with a dagger ($\dagger$).}
  \label{tab:mtc_basic1}
  \begin{footnotesize}
  \begin{tabular}{ l c c c c c c c c c c c c } \hline
    Source Name 	& $N$ 	& Period 	& DM 	& RM 	& $L/I$ 	& $V/I$ 	& $|V|/I$ 	& Median ToA 	& $S_{1400}$ 	& $\alpha$ 	& Reference \\ 
 	&  	&  	&  	&  	&  	&  	&  	&  Uncertainty 	&  	&  	&  \\ 
     	&  	& (ms) 	& (pc~cm$^{-3}$) 	& (rad~m$^{-2}$) 	& (\%) 	& (\%) 	& (\%) 	& ($\upmu$s) 	& (mJy) 	&  	&  \\ \hline 
    J0023+0923    	& $6$ 	& $3.05$ 	& $14.3216(6)$ 	& $-5.1(9)$ 	& $30(2)$ 	& $3(2)$ 	& $3(2)$ 	& $0.49$ 	& $0.73(12)$ 	& $-1.4(7)$ 	& \citealt{hrm+11} \\ 
    J0030+0451$^*$  	& $33$ 	& $4.87$ 	& $4.33294(11)$ 	& $1.32(13)$ 	& $33(2)$ 	& $0.9(22)$ 	& $2(2)$ 	& $0.91$ 	& $1.09(3)$ 	& $-2.4(2)$ 	& \citealt{lzb+00} \\ 
    J0034$-$0534  	& $6$ 	& $1.88$ 	& $13.748(2)$ 	&   --$^\dagger$  	& $<4.1$ 	& $-3(4)$ 	& $<8.1$ 	& $2.5$ 	& $0.21(5)$ 	& $-1.0(9)$ 	& \citealt{bhl+94} \\ 
    J0101$-$6422$^*$  	& $8$ 	& $2.57$ 	& $11.9210(4)$ 	& $20(10)$ 	& $21(2)$ 	& $2(6)$ 	& $<11.2$ 	& $2.0$ 	& $0.156(11)$ 	& $-1.4(5)$ 	& \citealt{kcj+12} \\ 
    J0125$-$2327$^*$  	& $39$ 	& $3.68$ 	& $9.5974(3)$ 	& $4.78(7)$ 	& $34(2)$ 	& $-3(2)$ 	& $2(2)$ 	& $0.19$ 	& $2.49(12)$ 	& $-1.0(2)$ 	& \citealt{mess+20} \\ 
    J0154+1833    	& $6$ 	& $2.36$ 	& $19.79830(14)$ 	& $-21.9(6)$$^\dagger$  	& $30(2)$ 	& $-0.7(61)$ 	& $<12.1$ 	& $0.89$ 	& $0.11(2)$ 	& $-1.7(8)$ 	& \citealt{mgf+19} \\ 
    J0348+0432    	& $6$ 	& $39.1$ 	& $40.464(7)$ 	& $48.3(3)$ 	& $28(2)$ 	& $-13(3)$ 	& $11(3)$ 	& $1.6$ 	& $0.67(4)$ 	& $-1.8(3)$ 	& \citealt{lbr+13} \\ 
    J0407+1607    	& $6$ 	& $25.7$ 	& $35.593(6)$ 	&   --$^\dagger$  	& $4(2)$ 	& $6(3)$ 	& $5(3)$ 	& $14$ 	& $0.35(2)$ 	& $-2.5(2)$ 	& \citealt{lxf+05} \\ 
    J0437$-$4715$^*$  	& $74$ 	& $5.76$ 	& $2.6411(2)$ 	& $0.15(2)$ 	& $25(2)$ 	& $-5(3)$ 	& $13(3)$ 	& $0.0087$ 	& $126(2)$ 	& $-1.78(11)$ 	& \citealt{jlh+93} \\ 
    J0453+1559    	& $6$ 	& $45.8$ 	& $30.298(2)$ 	& $-35.3(3)$ 	& $20(2)$ 	& $-0.1(31)$ 	& $<6.1$ 	& $1.8$ 	& $0.33(2)$ 	& $-1.3(3)$ 	& \citealt{dsm+13} \\ 
    J0509+0856    	& $6$ 	& $4.06$ 	& $38.3271(8)$ 	& $44.1(3)$ 	& $22(2)$ 	& $2(2)$ 	& $3(2)$ 	& $1.6$ 	& $1.47(3)$ 	& $-2.19(9)$ 	& \citealt{mgf+19} \\ 
    J0557+1550    	& $6$ 	& $2.56$ 	& $102.5700(12)$ 	& $26(1)$ 	& $27(2)$ 	& $0(10)$ 	& $<21$ 	& $3.8$ 	& $0.067(5)$ 	& $-2.0(3)$ 	& \citealt{skl+15} \\ 
    J0557$-$2948  	& $6$ 	& $43.6$ 	& $49.033(2)$ 	& $1(2)$ 	& $17(2)$ 	& $-2(9)$ 	& $<18$ 	& $5.6$ 	& $0.046(2)$ 	& $-2.2(2)$ 	& \citealt{kmlk+18} \\ 
    J0610$-$2100$^*$  	& $38$ 	& $3.86$ 	& $60.6865(3)$ 	& $33.7(2)$ 	& $71(2)$ 	& $-5(2)$ 	& $5(2)$ 	& $0.78$ 	& $0.65(2)$ 	& $-1.24(14)$ 	& \citealt{bjd+06} \\ 
    J0613$-$0200$^*$  	& $36$ 	& $3.06$ 	& $38.7750(5)$ 	& $18.4(2)$ 	& $19(2)$ 	& $0.3(21)$ 	& $3(2)$ 	& $0.18$ 	& $2.15(7)$ 	& $-2.09(13)$ 	& \citealt{lnl+95} \\ 
    J0614$-$3329$^*$  	& $39$ 	& $3.15$ 	& $37.05245(14)$ 	& $38.9(2)$ 	& $31(2)$ 	& $-1(2)$ 	& $2(2)$ 	& $0.52$ 	& $0.68(4)$ 	& $-1.7(2)$ 	& \citealt{rrc+11} \\ 
    J0621+1002    	& $6$ 	& $28.9$ 	& $36.5533(7)$ 	& $52.0(2)$ 	& $21(2)$ 	& $-16(2)$ 	& $14(2)$ 	& $1.1$ 	& $1.72(14)$ 	& $-2.4(3)$ 	& \citealt{cnst96} \\ 
    J0621+2514    	& $6$ 	& $2.72$ 	& $83.633(3)$ 	& $11(2)$ 	& $64(3)$ 	& $10(10)$ 	& $<27$ 	& $11$ 	& $0.079(3)$ 	& $-1.5(2)$ 	& \citealt{san16} \\ 
    J0636$-$3044$^*$  	& $25$ 	& $3.95$ 	& $15.4589(4)$ 	& $36.9(2)$ 	& $41(2)$ 	& $2(2)$ 	& $6(2)$ 	& $1.9$ 	& $1.51(14)$ 	& $-0.9(5)$ 	& \citealt{mbc+19} \\ 
    J0709+0458    	& $6$ 	& $34.4$ 	& $44.273(3)$ 	& $42.0(6)$ 	& $21(2)$ 	& $4(3)$ 	& $<6.9$ 	& $4.3$ 	& $0.256(7)$ 	& $-1.52(11)$ 	& \citealt{mgf+19} \\ 
    J0711$-$6830$^*$  	& $40$ 	& $5.49$ 	& $18.4096(2)$ 	& $23.9(3)$ 	& $15(2)$ 	& $-15(2)$ 	& $14(2)$ 	& $0.76$ 	& $2.6(3)$ 	& $-1.7(4)$ 	& \citealt{bjb+97} \\ 
    J0721$-$2038  	& $6$ 	& $15.5$ 	& $76.093(6)$ 	& $-18.4(4)$ 	& $27(2)$ 	& $-18(4)$ 	& $16(4)$ 	& $5.7$ 	& $0.273(6)$ 	& $-1.72(9)$ 	& \citealt{bkl+13} \\ 
    J0732+2314    	& $6$ 	& $4.09$ 	& $44.670(3)$ 	& $-3(2)$ 	& $22(2)$ 	& $18(3)$ 	& $16(3)$ 	& $5.9$ 	& $0.73(8)$ 	& $-1.3(5)$ 	& \citealt{mgf+19} \\ 
    J0737$-$3039A 	& $10$ 	& $22.5$ 	& $48.916(2)$ 	& $120.46(4)$ 	& $33(2)$ 	& $-0.9(20)$ 	& $4(2)$ 	& $1.1$ 	& $3.06(7)$ 	& $-1.86(13)$ 	& \citealt{bdap+03} \\ 
    J0751+1807    	& $6$ 	& $3.48$ 	& $30.2409(6)$ 	& $41.6(2)$ 	& $27(2)$ 	& $-10(2)$ 	& $10(2)$ 	& $0.50$ 	& $1.35(6)$ 	& $-1.1(2)$ 	& \citealt{lzc95} \\ 
    J0824+0028    	& $6$ 	& $9.86$ 	& $34.5473(12)$ 	& $38.7(3)$ 	& $43(2)$ 	& $6(4)$ 	& $8(4)$ 	& $3.4$ 	& $0.51(2)$ 	& $-2.20(14)$ 	& \citealt{dsm+13} \\ 
    J0900$-$3144$^*$  	& $39$ 	& $11.1$ 	& $75.6862(7)$ 	& $85.43(5)$ 	& $38(2)$ 	& $9(2)$ 	& $10(2)$ 	& $0.58$ 	& $3.84(3)$ 	& $-1.30(4)$ 	& \citealt{bjd+06} \\ 
    J0921$-$5202  	& $8$ 	& $9.68$ 	& $122.710(3)$ 	& $-179(1)$ 	& $32(2)$ 	& $3(7)$ 	& $<14.2$ 	& $8.3$ 	& $0.189(7)$ 	& $-1.6(2)$ 	& \citealt{mlb+12} \\ 
    J0931$-$1902$^*$  	& $22$ 	& $4.64$ 	& $41.4900(5)$ 	& $-100.2(2)$ 	& $34(2)$ 	& $1(2)$ 	& $3(2)$ 	& $0.96$ 	& $0.52(4)$ 	& $-2.4(3)$ 	& \citealt{lsg15} \\ 
    J0955$-$6150$^*$  	& $67$ 	& $2.00$ 	& $160.8997(2)$ 	& $-49(1)$ 	& $4(2)$ 	& $0.5(21)$ 	& $8(2)$ 	& $1.0$ 	& $0.639(12)$ 	& $-3.18(8)$ 	& \citealt{ckr+15} \\ 
    J1012$-$4235$^*$  	& $40$ 	& $3.10$ 	& $71.6515(3)$ 	& $61.8(5)$ 	& $20(2)$ 	& $2(3)$ 	& $4(3)$ 	& $0.82$ 	& $0.258(6)$ 	& $-1.33(11)$ 	& \citealt{ckr+15} \\ 
    J1017$-$7156$^*$  	& $53$ 	& $2.34$ 	& $94.21473(10)$ 	& $-63.36(8)$ 	& $38(2)$ 	& $-21(3)$ 	& $22(3)$ 	& $0.064$ 	& $1.09(3)$ 	& $-2.01(11)$ 	& \citealt{kjb+12} \\ 
    J1022+1001$^*$  	& $36$ 	& $16.5$ 	& $10.2550(8)$ 	& $1.83(5)$ 	& $61(2)$ 	& $-14(2)$ 	& $14(2)$ 	& $0.29$ 	& $3.9(3)$ 	& $-1.9(4)$ 	& \citealt{cnst96} \\ 
    J1024$-$0719$^*$  	& $36$ 	& $5.16$ 	& $6.4864(2)$ 	& $-2.89(14)$ 	& $62(2)$ 	& $3(2)$ 	& $3(2)$ 	& $0.78$ 	& $1.50(8)$ 	& $-1.8(4)$ 	& \citealt{bjb+97} \\ 
    J1035$-$6720  	& $5$ 	& $2.87$ 	& $84.166(11)$ 	& $-53(3)$ 	& $63(3)$ 	& $10(20)$ 	& $<39$ 	& $15$ 	& $0.065(2)$ 	& $-2.9(2)$ 	& \citealt{cpw+18} \\ 
    J1036$-$8317$^*$  	& $39$ 	& $3.41$ 	& $27.0927(2)$ 	& $12.9(2)$ 	& $26(2)$ 	& $5(3)$ 	& $9(3)$ 	& $0.58$ 	& $0.45(2)$ 	& $-1.1(2)$ 	& \citealt{ckr+15} \\ 
    J1038+0032    	& $6$ 	& $28.9$ 	& $26.31(2)$ 	& $20(5)$$^\dagger$  	& $18(2)$ 	& $2(8)$ 	& $<16$ 	& $34$ 	& $0.102(12)$ 	& $-2.1(5)$ 	& \citealt{bjd+06} \\ 
    J1045$-$4509$^*$  	& $40$ 	& $7.47$ 	& $58.1106(6)$ 	& $94.43(9)$ 	& $22(2)$ 	& $12(2)$ 	& $13(2)$ 	& $0.64$ 	& $2.39(6)$ 	& $-2.13(10)$ 	& \citealt{bhl+94} \\ 
    J1056$-$7117  	& $6$ 	& $26.3$ 	& $92.904(5)$ 	& $-31.3(3)$ 	& $17(2)$ 	& $-4(3)$ 	& $8(3)$ 	& $9.8$ 	& $0.50(2)$ 	& $-2.2(2)$ 	& \citealt{nbb+14} \\ 
    J1101$-$6424$^*$  	& $49$ 	& $5.11$ 	& $207.3633(5)$ 	& $-41.2(2)$ 	& $21(2)$ 	& $-14(2)$ 	& $16(2)$ 	& $2.1$ 	& $0.302(3)$ 	& $-1.55(4)$ 	& \citealt{ncb+15} \\ 
    J1103$-$5403$^*$  	& $41$ 	& $3.39$ 	& $103.9114(3)$ 	& $-67.7(4)$ 	& $33(2)$ 	& $-19(4)$ 	& $19(4)$ 	& $0.79$ 	& $0.395(10)$ 	& $-2.43(14)$ 	& \citealt{kjr+11} \\ 
    J1120$-$3618  	& $6$ 	& $5.56$ 	& $45.119(2)$ 	& $-48(8)$$^\dagger$  	& $9(2)$ 	& $-6(5)$ 	& $<10.0$ 	& $16$ 	& $0.30(2)$ 	& $-1.9(3)$ 	& \citealt{rb13} \\ 
    J1125$-$5825$^*$  	& $40$ 	& $3.10$ 	& $124.8083(4)$ 	& $-5.5(2)$ 	& $25(2)$ 	& $-2(2)$ 	& $9(2)$ 	& $0.45$ 	& $1.00(2)$ 	& $-1.40(8)$ 	& \citealt{kjvs+10} \\ 
    J1125$-$6014$^*$  	& $51$ 	& $2.63$ 	& $52.92972(4)$ 	& $8.45(8)$ 	& $31(2)$ 	& $0.2(20)$ 	& $2(2)$ 	& $0.10$ 	& $1.32(2)$ 	& $-1.41(7)$ 	& \citealt{fsk+04} \\ 
    J1142+0119    	& $6$ 	& $5.08$ 	& $19.172(7)$ 	& $0.31(11)$$^\dagger$  	& $16(3)$ 	& $20(20)$ 	& $<34$ 	& $17$ 	& $0.045(5)$ 	& $-1.7(4)$ 	& \citealt{san16} \\ 
    J1157$-$5112  	& $21$ 	& $43.6$ 	& $39.724(3)$ 	& $-4.46(12)$ 	& $47(2)$ 	& $3(2)$ 	& $4(2)$ 	& $6.9$ 	& $0.19(2)$ 	& $-2.8(4)$ 	& \citealt{eb01} \\\hline
  \end{tabular}
  \end{footnotesize}
\end{table*}

\begin{table*}
  \centering
  \contcaption{MeerTime MSP census results}
  \label{tab:mtc_basic21}
  \begin{footnotesize}
  \begin{tabular}{ l c c c c c c c c c c c c } \hline
    Source Name 	& $N$ 	& Period 	& DM 	& RM 	& $L/I$ 	& $V/I$ 	& $|V|/I$ 	& Median ToA 	& $S_{1400}$ 	& $\alpha$ 	& Reference \\ 
 	&  	&  	&  	&  	&  	&  	&  	&  Uncertainty 	&  	&  	&  \\ 
     	&  	& (ms) 	& (pc~cm$^{-3}$) 	& (rad~m$^{-2}$) 	& (\%) 	& (\%) 	& (\%) 	& ($\upmu$s) 	& (mJy) 	&  	&  \\ \hline 
    J1207$-$5050  	& $7$ 	& $4.84$ 	& $50.640(2)$ 	& $-13(1)$ 	& $28(2)$ 	& $-7(4)$ 	& $6(4)$ 	& $3.5$ 	& $0.39(3)$ 	& $-2.1(3)$ 	& \citealt{rap+12} \\ 
    J1216$-$6410$^*$  	& $41$ 	& $3.54$ 	& $47.3933(2)$ 	& $-22.0(3)$ 	& $26(2)$ 	& $-8(3)$ 	& $9(3)$ 	& $0.27$ 	& $1.15(2)$ 	& $-2.00(7)$ 	& \citealt{fsk+04} \\ 
    J1227$-$6208  	& $23$ 	& $34.5$ 	& $362.913(3)$ 	& $47.3(2)$ 	& $14(2)$ 	& $-11(2)$ 	& $10(2)$ 	& $4.1$ 	& $0.272(2)$ 	& $-2.20(3)$ 	& \citealt{btb+15} \\ 
    J1231$-$1411$^*$  	& $12$ 	& $3.68$ 	& $8.0884(4)$ 	& $11.6(2)$ 	& $44(2)$ 	& $-5(3)$ 	& $4(3)$ 	& $3.1$ 	& $0.29(2)$ 	& $-1.9(4)$ 	& \citealt{rrc+11} \\ 
    J1300+1240    	& $6$ 	& $6.22$ 	& $10.1658(9)$ 	& $9.0(8)$ 	& $42(2)$ 	& $1(4)$ 	& $12(4)$ 	& $3.9$ 	& $0.39(7)$ 	& $-1.7(7)$ 	& \citealt{wol90a} \\ 
    J1302$-$3258  	& $6$ 	& $3.77$ 	& $26.1860(7)$ 	& $-30(1)$ 	& $43(2)$ 	& $-8(6)$ 	& $<11.8$ 	& $2.6$ 	& $0.17(2)$ 	& $-2.4(8)$ 	& \citealt{hrm+11} \\ 
    J1312+0051    	& $6$ 	& $4.23$ 	& $15.3417(14)$ 	& $2(1)$ 	& $62(2)$ 	& $-0.8(56)$ 	& $<11.3$ 	& $4.2$ 	& $0.19(2)$ 	& $-1.3(4)$ 	& \citealt{san16} \\ 
    J1327$-$0755$^*$  	& $24$ 	& $2.68$ 	& $27.9068(3)$ 	& $-0.3(5)$$^\dagger$  	& $15(2)$ 	& $4(3)$ 	& $8(3)$ 	& $1.0$ 	& $0.19(2)$ 	& $-2.6(4)$ 	& \citealt{blr+13} \\ 
    J1337$-$6423  	& $6$ 	& $9.42$ 	& $259.891(8)$ 	& $-108.3(9)$ 	& $17(2)$ 	& $3(6)$ 	& $<11.1$ 	& $5.9$ 	& $0.273(5)$ 	& $-1.86(8)$ 	& \citealt{kjb+12} \\ 
    J1400$-$1431  	& $6$ 	& $3.08$ 	& $4.9319(2)$ 	& $5(3)$$^\dagger$  	& $16(2)$ 	& $0.2(42)$ 	& $<8.3$ 	& $2.0$ 	& $0.17(2)$ 	& $-3.5(5)$ 	& \citealt{rsm+13} \\ 
    J1405$-$4656  	& $7$ 	& $7.60$ 	& $13.885(3)$ 	& $-5(1)$ 	& $30(2)$ 	& $4(3)$ 	& $<6.9$ 	& $17$ 	& $0.31(4)$ 	& $-4.0(6)$ 	& \citealt{btb+15} \\ 
    J1420$-$5625  	& $7$ 	& $34.1$ 	& $64.485(5)$ 	& $33.0(7)$ 	& $33(2)$ 	& $11(5)$ 	& $8(5)$ 	& $14$ 	& $0.121(4)$ 	& $-1.2(2)$ 	& \citealt{hfs+04} \\ 
    J1421$-$4409$^*$  	& $38$ 	& $6.39$ 	& $54.6403(5)$ 	& $-24.3(6)$ 	& $17(2)$ 	& $3(2)$ 	& $10(2)$ 	& $1.2$ 	& $1.26(4)$ 	& $-1.49(13)$ 	& \citealt{kbj+18} \\ 
    J1431$-$4715  	& $6$ 	& $2.01$ 	& $59.3473(9)$ 	& $13.3(2)$ 	& $33(2)$ 	& $3(4)$ 	& $<8.5$ 	& $0.68$ 	& $0.67(2)$ 	& $-1.0(2)$ 	& \citealt{btb+15} \\ 
    J1431$-$5740$^*$  	& $40$ 	& $4.11$ 	& $131.3766(6)$ 	& $-42.2(4)$ 	& $19(2)$ 	& $-28(3)$ 	& $27(3)$ 	& $1.1$ 	& $0.359(6)$ 	& $-1.31(8)$ 	& \citealt{bbb+13} \\ 
    J1435$-$6100$^*$  	& $49$ 	& $9.35$ 	& $113.7819(2)$ 	& $-53.73(12)$ 	& $23(2)$ 	& $7(2)$ 	& $6(2)$ 	& $0.74$ 	& $0.313(5)$ 	& $-1.37(7)$ 	& \citealt{clm+01} \\ 
    J1439$-$5501  	& $11$ 	& $28.6$ 	& $14.565(8)$ 	& $-7.9(4)$ 	& $29(2)$ 	& $-6(3)$ 	& $8(3)$ 	& $2.6$ 	& $0.38(4)$ 	& $-2.8(4)$ 	& \citealt{fsk+04} \\ 
    J1446$-$4701$^*$  	& $39$ 	& $2.19$ 	& $55.82787(6)$ 	& $-9.6(2)$ 	& $38(2)$ 	& $-10(3)$ 	& $10(3)$ 	& $0.43$ 	& $0.36(2)$ 	& $-2.0(2)$ 	& \citealt{kjb+12} \\ 
    J1453+1902    	& $6$ 	& $5.79$ 	& $14.057(2)$ 	& $3.2(6)$ 	& $52(2)$ 	& $-0.4(80)$ 	& $<16$ 	& $5.4$ 	& $0.16(2)$ 	& $1.3(5)$ 	& \citealt{lmcs07} \\ 
    J1454$-$5846  	& $15$ 	& $45.2$ 	& $116.091(2)$ 	& $84.3(8)$ 	& $8(2)$ 	& $-0.5(26)$ 	& $<5.2$ 	& $11$ 	& $0.299(4)$ 	& $-1.42(7)$ 	& \citealt{clm+01} \\ 
    J1455$-$3330$^*$  	& $37$ 	& $7.99$ 	& $13.5697(2)$ 	& $14.7(8)$ 	& $18(2)$ 	& $-3(2)$ 	& $5(2)$ 	& $0.99$ 	& $0.73(4)$ 	& $-2.2(3)$ 	& \citealt{lnl+95} \\ 
    J1502$-$6752  	& $6$ 	& $26.7$ 	& $152.242(6)$ 	& $232.8(2)$ 	& $38(2)$ 	& $-16(4)$ 	& $12(4)$ 	& $18$ 	& $0.87(2)$ 	& $-1.3(2)$ 	& \citealt{kjb+12} \\ 
    J1513$-$2550$^*$  	& $13$ 	& $2.12$ 	& $46.8765(3)$ 	&   --$^\dagger$  	& $<4.0$ 	& $8(3)$ 	& $6(3)$ 	& $0.79$ 	& $0.308(12)$ 	& $-3.8(2)$ 	& \citealt{san16} \\ 
    J1514$-$4946$^*$  	& $7$ 	& $3.59$ 	& $31.00857(8)$ 	& $35(2)$ 	& $65(2)$ 	& $-3(5)$ 	& $<10.1$ 	& $0.66$ 	& $0.25(3)$ 	& $0.4(6)$ 	& \citealt{kcj+12} \\ 
    J1525$-$5545$^*$  	& $45$ 	& $11.4$ 	& $126.9640(14)$ 	& $2.9(3)$ 	& $13(2)$ 	& $0.8(22)$ 	& $9(2)$ 	& $1.5$ 	& $0.426(6)$ 	& $-1.40(7)$ 	& \citealt{nbb+14} \\ 
    J1529$-$3828  	& $6$ 	& $8.49$ 	& $73.655(2)$ 	& $-30(8)$ 	& $16(2)$ 	& $9(5)$ 	& $<10.4$ 	& $12$ 	& $0.27(2)$ 	& $-2.0(4)$ 	& \citealt{nbb+14} \\ 
    J1536$-$4948  	& $5$ 	& $3.08$ 	& $37.998(8)$ 	&   --$^\dagger$  	& $4(3)$ 	& $0(20)$ 	& $<39$ 	& $13$ 	& $0.087(10)$ 	& $-3.2(6)$ 	& \citealt{rap+12} \\ 
    J1537+1155    	& $6$ 	& $37.9$ 	& $11.608(3)$ 	& $7.7(5)$ 	& $25(2)$ 	& $1(3)$ 	& $4(3)$ 	& $6.3$ 	& $0.46(7)$ 	& $-1.1(6)$ 	& \citealt{wol91a} \\ 
    J1537$-$5312  	& $9$ 	& $6.93$ 	& $117.687(4)$ 	& $-37.4(8)$ 	& $21(2)$ 	& $-7(6)$ 	& $<11.6$ 	& $7.1$ 	& $0.32(2)$ 	& $-2.3(3)$ 	& \citealt{ccb+20} \\ 
    J1543$-$5149$^*$  	& $38$ 	& $2.06$ 	& $50.9820(4)$ 	& $-27(2)$ 	& $7(2)$ 	& $5(2)$ 	& $6(2)$ 	& $1.0$ 	& $0.82(3)$ 	& $-3.2(2)$ 	& \citealt{kjb+12} \\ 
    J1545$-$4550$^*$  	& $48$ 	& $3.58$ 	& $68.3884(2)$ 	& $2.99(11)$ 	& $50(2)$ 	& $-12(3)$ 	& $14(3)$ 	& $0.31$ 	& $1.07(3)$ 	& $-1.29(12)$ 	& \citealt{bbb+13} \\ 
    J1546$-$5925  	& $6$ 	& $7.80$ 	& $168.282(3)$ 	& $-40(20)$ 	& $22(2)$ 	& $7(8)$ 	& $<16$ 	& $6.6$ 	& $0.269(12)$ 	& $-1.3(2)$ 	& \citealt{mlb+12} \\ 
    J1547$-$5709$^*$  	& $39$ 	& $4.29$ 	& $95.7221(4)$ 	& $119.7(4)$ 	& $21(2)$ 	& $-6(4)$ 	& $<8.0$ 	& $1.5$ 	& $0.343(9)$ 	& $-2.04(12)$ 	& \citealt{ccb+20} \\ 
    J1552$-$4937  	& $6$ 	& $6.28$ 	& $113.803(2)$ 	& $-30(9)$ 	& $11(2)$ 	& $2(6)$ 	& $<11.6$ 	& $4.8$ 	& $0.34(2)$ 	& $-2.6(2)$ 	& \citealt{fsk+04} \\ 
    J1600$-$3053$^*$  	& $40$ 	& $3.60$ 	& $52.3282(2)$ 	& $-11.69(10)$ 	& $30(2)$ 	& $0.8(20)$ 	& $2(2)$ 	& $0.12$ 	& $2.22(5)$ 	& $-0.99(12)$ 	& \citealt{jbo+07} \\ 
    J1603$-$7202$^*$  	& $39$ 	& $14.8$ 	& $38.0497(8)$ 	& $30.01(14)$ 	& $18(2)$ 	& $27(2)$ 	& $30(2)$ 	& $0.41$ 	& $2.46(13)$ 	& $-2.7(3)$ 	& \citealt{llb+96} \\ 
    J1614$-$2230$^*$  	& $38$ 	& $3.15$ 	& $34.4850(3)$ 	& $-29.12(8)$ 	& $63(2)$ 	& $4(2)$ 	& $2(2)$ 	& $0.27$ 	& $1.14(4)$ 	& $-1.8(2)$ 	& \citealt{crh+06a} \\ 
    J1618$-$3921  	& $25$ 	& $12.0$ 	& $117.9312(7)$ 	& $177.5(2)$ 	& $25(2)$ 	& $10(2)$ 	& $9(2)$ 	& $4.2$ 	& $0.631(6)$ 	& $-2.28(4)$ 	& \citealt{eb01b} \\ 
    J1618$-$4624  	& $7$ 	& $5.93$ 	& $125.4068(11)$ 	& $2(2)$$^\dagger$  	& $6(2)$ 	& $23(7)$ 	& $22(7)$ 	& $4.3$ 	& $0.229(8)$ 	& $-1.8(2)$ 	& \citealt{ccb+20} \\ 
    J1622$-$6617  	& $6$ 	& $23.6$ 	& $88.0049(12)$ 	& $9(1)$ 	& $8(2)$ 	& $9(3)$ 	& $6(3)$ 	& $3.8$ 	& $0.44(4)$ 	& $-1.7(4)$ 	& \citealt{kjb+12} \\ 
    J1629$-$6902$^*$  	& $41$ 	& $6.00$ 	& $29.4948(2)$ 	& $36.67(15)$ 	& $27(2)$ 	& $-3(3)$ 	& $7(3)$ 	& $0.39$ 	& $1.01(4)$ 	& $-2.0(2)$ 	& \citealt{eb01b} \\ 
    J1640+2224    	& $6$ 	& $3.16$ 	& $18.42881(12)$ 	& $19(2)$ 	& $12(2)$ 	& $5(3)$ 	& $7(3)$ 	& $0.47$ 	& $0.46(6)$ 	& $-2.1(6)$ 	& \citealt{fcwa95} \\ 
    J1643$-$1224$^*$  	& $37$ 	& $4.62$ 	& $62.3975(4)$ 	& $-304.28(11)$ 	& $17(2)$ 	& $-0.7(21)$ 	& $12(2)$ 	& $0.29$ 	& $3.78(3)$ 	& $-2.12(3)$ 	& \citealt{lnl+95} \\ 
    J1652$-$4838$^*$  	& $39$ 	& $3.79$ 	& $188.1559(8)$ 	& $-29.5(3)$ 	& $21(2)$ 	& $4(2)$ 	& $5(2)$ 	& $1.0$ 	& $0.917(14)$ 	& $-1.11(7)$ 	& \citealt{kek+13} \\ 
    J1653$-$2054$^*$  	& $25$ 	& $4.13$ 	& $56.5213(3)$ 	& $-4.5(3)$ 	& $18(2)$ 	& $3(2)$ 	& $3(2)$ 	& $2.2$ 	& $0.575(14)$ 	& $-2.25(13)$ 	& \citealt{btb+15} \\ 
    J1658$-$5324$^*$  	& $15$ 	& $2.44$ 	& $30.8301(6)$ 	& $7.8(4)$ 	& $64(2)$ 	& $-3(10)$ 	& $<20$ 	& $1.1$ 	& $0.43(4)$ 	& $-2.7(4)$ 	& \citealt{kcj+12} \\ 
    J1705$-$1903$^*$  	& $21$ 	& $2.48$ 	& $57.50473(7)$ 	& $-15.6(2)$ 	& $22(2)$ 	& $2(2)$ 	& $<4.7$ 	& $0.12$ 	& $0.575(14)$ 	& $-1.47(11)$ 	& \citealt{mbc+19} \\
    J1708$-$3506$^*$  	& $25$ 	& $4.51$ 	& $146.7636(14)$ 	& $-10.6(2)$ 	& $15(2)$ 	& $2(2)$ 	& $3(2)$ 	& $1.8$ 	& $1.45(2)$ 	& $-2.23(5)$ 	& \citealt{kjvs+10} \\ \hline
  \end{tabular}
  \end{footnotesize}
\end{table*}

\begin{table*}
  \centering
  \contcaption{MeerTime MSP census results}
  \label{tab:mtc_basic22}
  \begin{footnotesize}
  \begin{tabular}{ l c c c c c c c c c c c c } \hline
    Source Name 	& $N$ 	& Period 	& DM 	& RM 	& $L/I$ 	& $V/I$ 	& $|V|/I$ 	& Median ToA 	& $S_{1400}$ 	& $\alpha$ 	& Reference \\ 
 	&  	&  	&  	&  	&  	&  	&  	&  Uncertainty 	&  	&  	&  \\ 
     	&  	& (ms) 	& (pc~cm$^{-3}$) 	& (rad~m$^{-2}$) 	& (\%) 	& (\%) 	& (\%) 	& ($\upmu$s) 	& (mJy) 	&  	&  \\ \hline 
    J1709+2313    	& $6$ 	& $4.63$ 	& $25.3399(14)$ 	& $37(2)$$^\dagger$  	& $27(2)$ 	& $7(6)$ 	& $<12.8$ 	& $6.1$ 	& $0.19(2)$ 	& $-3.1(6)$ 	& \citealt{fcwa95} \\ 
    J1713+0747$^*$  	& $35$ 	& $4.57$ 	& $15.9904(5)$ 	& $10.88(10)$ 	& $32(2)$ 	& $-2(2)$ 	& $3(2)$ 	& $0.042$ 	& $8.3(5)$ 	& $-1.0(3)$ 	& \citealt{fwc93} \\ 
    J1719$-$1438$^*$  	& $41$ 	& $5.79$ 	& $36.7752(5)$ 	& $17.3(2)$ 	& $25(2)$ 	& $4(2)$ 	& $4(2)$ 	& $1.5$ 	& $0.43(2)$ 	& $-2.5(2)$ 	& \citealt{bbb+11a} \\ 
    J1721$-$2457$^*$  	& $12$ 	& $3.50$ 	& $48.2341(6)$ 	& $-37.7(4)$ 	& $19(2)$ 	& $9(2)$ 	& $7(2)$ 	& $2.6$ 	& $1.03(4)$ 	& $-1.9(2)$ 	& \citealt{eb01b} \\ 
    J1727$-$2946  	& $8$ 	& $27.1$ 	& $60.7336(14)$ 	& $-90.4(9)$ 	& $24(2)$ 	& $3(5)$ 	& $9(5)$ 	& $4.3$ 	& $0.265(11)$ 	& $-2.4(2)$ 	& \citealt{fsk+04} \\ 
    J1730$-$2304$^*$  	& $39$ 	& $8.12$ 	& $9.6257(5)$ 	& $-3.72(9)$ 	& $28(2)$ 	& $-20(3)$ 	& $20(3)$ 	& $0.41$ 	& $3.5(2)$ 	& $-2.0(2)$ 	& \citealt{lnl+95} \\ 
    J1731$-$1847$^*$  	& $6$ 	& $2.34$ 	& $106.4721(7)$ 	& $21.5(4)$ 	& $52(2)$ 	& $-17(4)$ 	& $14(4)$ 	& $0.38$ 	& $0.38(2)$ 	& $-2.4(2)$ 	& \citealt{kjvs+10} \\ 
    J1732$-$5049$^*$  	& $53$ 	& $5.31$ 	& $56.8209(4)$ 	& $-8.2(3)$ 	& $25(2)$ 	& $0.3(22)$ 	& $<4.3$ 	& $0.69$ 	& $2.11(11)$ 	& $-1.9(2)$ 	& \citealt{eb01b} \\ 
    J1737$-$0811$^*$  	& $44$ 	& $4.18$ 	& $55.3060(11)$ 	& $64.86(15)$ 	& $26(2)$ 	& $0.1(21)$ 	& $2(2)$ 	& $1.9$ 	& $1.07(2)$ 	& $-2.19(7)$ 	& \citealt{blr+13} \\ 
    J1738+0333    	& $6$ 	& $5.85$ 	& $33.771(3)$ 	& $34.7(5)$ 	& $22(2)$ 	& $-2(4)$ 	& $<7.1$ 	& $1.3$ 	& $0.34(4)$ 	& $-4.1(5)$ 	& \citealt{jac05} \\ 
    J1741+1351    	& $6$ 	& $3.75$ 	& $24.1962(3)$ 	& $63.5(7)$ 	& $18(2)$ 	& $2(4)$ 	& $<8.6$ 	& $0.80$ 	& $0.29(4)$ 	& $-0.8(5)$ 	& \citealt{jbo+07} \\ 
    J1744$-$1134$^*$  	& $39$ 	& $4.07$ 	& $3.13849(12)$ 	& $1.62(9)$ 	& $90(2)$ 	& $0.2(21)$ 	& $<4.2$ 	& $0.13$ 	& $2.6(2)$ 	& $-1.7(4)$ 	& \citealt{bjb+97} \\ 
    J1745+1017    	& $6$ 	& $2.65$ 	& $23.9711(3)$ 	& $27.2(3)$ 	& $50(2)$ 	& $-10(4)$ 	& $9(4)$ 	& $0.91$ 	& $0.51(4)$ 	& $-2.5(4)$ 	& \citealt{bgc+13} \\ 
    J1745$-$0952  	& $6$ 	& $19.4$ 	& $64.501(4)$ 	& $52(1)$ 	& $19(2)$ 	& $-15(3)$ 	& $12(3)$ 	& $6.3$ 	& $0.92(5)$ 	& $-1.3(2)$ 	& \citealt{eb01b} \\ 
    J1747$-$4036$^*$  	& $40$ 	& $1.65$ 	& $152.9407(8)$ 	& $-43.8(2)$ 	& $16(2)$ 	& $-0.4(22)$ 	& $<4.4$ 	& $0.55$ 	& $1.506(13)$ 	& $-2.84(4)$ 	& \citealt{kcj+12} \\ 
    J1750$-$2536  	& $9$ 	& $34.7$ 	& $179.395(9)$ 	& $173(5)$$^\dagger$  	& $14(2)$ 	& $-14(9)$ 	& $11(9)$ 	& $39$ 	& $0.107(5)$ 	& $-3.7(2)$ 	& \citealt{kek+13} \\ 
    J1751$-$2857$^*$  	& $38$ 	& $3.91$ 	& $42.7890(5)$ 	& $33.20(13)$ 	& $41(2)$ 	& $-2(4)$ 	& $<8.3$ 	& $1.2$ 	& $0.461(9)$ 	& $-1.27(8)$ 	& \citealt{sfl+05} \\ 
    J1754+0032    	& $6$ 	& $4.41$ 	& $70.2804(7)$ 	& $40.0(8)$ 	& $17(2)$ 	& $1(3)$ 	& $<5.6$ 	& $1.4$ 	& $0.61(3)$ 	& $-1.8(2)$ 	& \citealt{mbc+19} \\ 
    J1755$-$3716  	& $5$ 	& $12.8$ 	& $167.566(5)$ 	& $52(4)$ 	& $12(2)$ 	& $-9(5)$ 	& $<10.4$ 	& $12$ 	& $0.524(10)$ 	& $-1.37(9)$ 	& \citealt{nbb+14} \\ 
    J1756$-$2251$^*$  	& $49$ 	& $28.5$ 	& $121.2312(3)$ 	& $-10.78(14)$ 	& $17(2)$ 	& $-2(2)$ 	& $3(2)$ 	& $1.3$ 	& $1.087(13)$ 	& $-1.14(6)$ 	& \citealt{fsk+04} \\ 
    J1757$-$1854  	& $10$ 	& $21.5$ 	& $378.111(13)$ 	& $702.6(8)$ 	& $19(2)$ 	& $9(3)$ 	& $7(3)$ 	& $15$ 	& $0.142(2)$ 	& $-1.40(5)$ 	& \citealt{cck+18} \\ 
    J1757$-$5322$^*$  	& $58$ 	& $8.87$ 	& $30.8047(4)$ 	& $70.15(10)$ 	& $21(2)$ 	& $-17(2)$ 	& $17(2)$ 	& $0.85$ 	& $1.76(5)$ 	& $-1.53(12)$ 	& \citealt{eb01} \\ 
    J1801$-$1417$^*$  	& $37$ 	& $3.63$ 	& $57.2516(2)$ 	& $232.55(12)$ 	& $18(2)$ 	& $3(2)$ 	& $4(2)$ 	& $0.78$ 	& $1.54(4)$ 	& $-1.80(13)$ 	& \citealt{fsk+04} \\ 
    J1801$-$3210  	& $6$ 	& $7.45$ 	& $177.729(2)$ 	& $225(2)$$^\dagger$  	& $19(2)$ 	& $5(5)$ 	& $<9.5$ 	& $6.5$ 	& $0.451(8)$ 	& $-2.47(7)$ 	& \citealt{kjvs+10} \\ 
    J1802$-$2124$^*$  	& $37$ 	& $12.6$ 	& $149.5902(4)$ 	& $290.2(6)$ 	& $17(2)$ 	& $3(2)$ 	& $3(2)$ 	& $0.44$ 	& $0.734(8)$ 	& $-2.38(5)$ 	& \citealt{fsk+04} \\ 
    J1804$-$2717$^*$  	& $6$ 	& $9.34$ 	& $24.6746(5)$ 	& $40.0(2)$ 	& $49(2)$ 	& $15(3)$ 	& $15(3)$ 	& $1.8$ 	& $1.4(2)$ 	& $-1.7(6)$ 	& \citealt{llb+96} \\ 
    J1804$-$2858$^*$  	& $11$ 	& $1.49$ 	& $232.528(2)$ 	& $-421(3)$ 	& $7(2)$ 	& $-8(3)$ 	& $5(3)$ 	& $1.6$ 	& $0.918(11)$ 	& $-1.26(7)$ 	& \citealt{mbc+19} \\ 
    J1806+2819    	& $6$ 	& $15.1$ 	& $18.78(2)$ 	&   --$^\dagger$  	& $<4.8$ 	& $0(10)$ 	& $<21$ 	& $35$ 	& $0.090(11)$ 	& $-2.0(6)$ 	& \citealt{kmlk+18} \\ 
    J1810$-$2005  	& $6$ 	& $32.8$ 	& $240.72(4)$ 	& $-15.7(8)$ 	& $9(2)$ 	& $5(3)$ 	& $<5.7$ 	& $30$ 	& $1.73(4)$ 	& $-2.0(2)$ 	& \citealt{clm+01} \\ 
    J1811$-$2405$^*$  	& $53$ 	& $2.66$ 	& $60.61997(6)$ 	& $30.2(2)$ 	& $27(2)$ 	& $-5(2)$ 	& $5(2)$ 	& $0.28$ 	& $1.33(2)$ 	& $-1.82(6)$ 	& \citealt{kjvs+10} \\ 
    J1813$-$2621  	& $6$ 	& $4.43$ 	& $122.4924(10)$ 	& $142.2(3)$ 	& $32(2)$ 	& $17(3)$ 	& $15(3)$ 	& $3.0$ 	& $0.584(8)$ 	& $-1.79(9)$ 	& \citealt{fsk+04} \\ 
    J1821+0155    	& $7$ 	& $33.8$ 	& $51.7660(4)$ 	& $110.3(3)$ 	& $33(2)$ 	& $0.8(28)$ 	& $4(3)$ 	& $0.77$ 	& $0.29(2)$ 	& $-2.8(4)$ 	& \citealt{rsm+13} \\ 
    J1825$-$0319$^*$  	& $36$ 	& $4.55$ 	& $119.5584(6)$ 	& $179.9(2)$ 	& $19(2)$ 	& $35(4)$ 	& $33(4)$ 	& $1.9$ 	& $0.183(2)$ 	& $-1.47(5)$ 	& \citealt{bbb+13} \\ 
    J1826$-$2415  	& $7$ 	& $4.70$ 	& $81.8625(11)$ 	& $58(1)$$^\dagger$  	& $12(2)$ 	& $1(5)$ 	& $<10.4$ 	& $4.3$ 	& $0.49(2)$ 	& $-1.4(2)$ 	& \citealt{bsb+19} \\ 
    J1828+0625    	& $8$ 	& $3.63$ 	& $22.4331(12)$ 	& $20(4)$$^\dagger$  	& $11(2)$ 	& $-8(5)$ 	& $5(5)$ 	& $1.7$ 	& $0.32(3)$ 	& $-2.6(3)$ 	& \citealt{rap+12} \\ 
    J1829+2456    	& $6$ 	& $41.0$ 	& $13.700(5)$ 	& $0.3(6)$$^\dagger$  	& $68(2)$ 	& $0.8(94)$ 	& $<19$ 	& $17$ 	& $0.037(4)$ 	& $-3.5(6)$ 	& \citealt{clm+04} \\ 
    J1832$-$0836$^*$  	& $16$ 	& $2.72$ 	& $28.19086(11)$ 	& $41.7(8)$ 	& $27(2)$ 	& $-0.9(39)$ 	& $6(4)$ 	& $0.33$ 	& $0.92(4)$ 	& $-1.9(2)$ 	& \citealt{bbb+13} \\ 
    J1835$-$0114  	& $6$ 	& $5.12$ 	& $98.253(3)$ 	& $115.5(6)$ 	& $25(2)$ 	& $-21(7)$ 	& $16(7)$ 	& $5.3$ 	& $0.373(8)$ 	& $-1.27(9)$ 	& \citealt{ekl09} \\ 
    J1843$-$1113$^*$  	& $37$ 	& $1.85$ 	& $59.95797(13)$ 	& $8.91(12)$ 	& $34(2)$ 	& $0.7(30)$ 	& $<6.1$ 	& $0.36$ 	& $0.56(2)$ 	& $-1.7(2)$ 	& \citealt{hfs+04} \\ 
    J1843$-$1448$^*$  	& $6$ 	& $5.47$ 	& $114.5440(9)$ 	& $256(6)$$^\dagger$  	& $6(2)$ 	& $1(4)$ 	& $<8.8$ 	& $4.6$ 	& $0.486(6)$ 	& $-1.83(5)$ 	& \citealt{fsk+04} \\ 
    J1844+0115    	& $8$ 	& $4.19$ 	& $148.250(5)$ 	& $30(2)$$^\dagger$  	& $16(3)$ 	& $0(15)$ 	& $<30$ 	& $10$ 	& $0.094(3)$ 	& $-1.88(14)$ 	& \citealt{csl+12} \\ 
    J1850+0124    	& $7$ 	& $3.56$ 	& $118.88(2)$ 	& $-69(2)$ 	& $35(2)$ 	& $20(20)$ 	& $<30$ 	& $1.8$ 	& $0.188(3)$ 	& $-1.79(6)$ 	& \citealt{csl+12} \\ 
    J1853+1303    	& $7$ 	& $4.09$ 	& $30.5709(4)$ 	& $77.5(2)$ 	& $25(2)$ 	& $4(3)$ 	& $16(3)$ 	& $0.67$ 	& $0.50(4)$ 	& $-3.3(3)$ 	& \citealt{fsk+04} \\ 
    J1855$-$1436  	& $7$ 	& $3.59$ 	& $109.208(9)$ 	& $140(30)$ 	& $27(3)$ 	& $0(20)$ 	& $<35$ 	& $2.4$ 	& $0.046(2)$ 	& $-2.95(14)$ 	& \citealt{san16} \\ 
    J1857+0943    	& $7$ 	& $5.36$ 	& $13.2992(2)$ 	& $23.36(5)$ 	& $14(2)$ 	& $-0.2(21)$ 	& $4(2)$ 	& $0.30$ 	& $5.0(5)$ 	& $-2.4(4)$ 	& \citealt{srs+86} \\ 
    J1858$-$2216  	& $6$ 	& $2.38$ 	& $26.641(5)$ 	& $-12(1)$ 	& $37(3)$ 	& $0(20)$ 	& $<33$ 	& $7.9$ 	& $0.057(7)$ 	& $-2.1(4)$ 	& \citealt{san16} \\
    J1900+0308    	& $7$ 	& $4.91$ 	& $249.86(2)$ 	& $49(6)$$^\dagger$  	& $23(2)$ 	& $-30(10)$ 	& $20(10)$ 	& $18$ 	& $0.140(3)$ 	& $-1.45(8)$ 	& \citealt{csl+12} \\ 
    J1901+0300    	& $7$ 	& $7.80$ 	& $253.747(11)$ 	&   --$^\dagger$  	& $<4.8$ 	& $-10(10)$ 	& $<30$ 	& $7.8$ 	& $0.135(4)$ 	& $-1.66(12)$ 	& \citealt{skl+15} \\ \hline
  \end{tabular}
  \end{footnotesize}
\end{table*}

\begin{table*}
  \centering
  \contcaption{MeerTime MSP census results}
  \label{tab:mtc_basic23}
  \begin{footnotesize}
  \begin{tabular}{ l c c c c c c c c c c c c } \hline
    Source Name 	& $N$ 	& Period 	& DM 	& RM 	& $L/I$ 	& $V/I$ 	& $|V|/I$ 	& Median ToA 	& $S_{1400}$ 	& $\alpha$ 	& Reference \\ 
 	&  	&  	&  	&  	&  	&  	&  	&  Uncertainty 	&  	&  	&  \\ 
     	&  	& (ms) 	& (pc~cm$^{-3}$) 	& (rad~m$^{-2}$) 	& (\%) 	& (\%) 	& (\%) 	& ($\upmu$s) 	& (mJy) 	&  	&  \\ \hline 
    J1902$-$5105$^*$  	& $40$ 	& $1.74$ 	& $36.2505(3)$ 	& $13(2)$ 	& $3(2)$ 	& $2(2)$ 	& $11(2)$ 	& $0.26$ 	& $1.01(2)$ 	& $-3.12(7)$ 	& \citealt{kcj+12} \\ 
    J1903+0327    	& $7$ 	& $2.15$ 	& $297.483(3)$ 	& $227(5)$$^\dagger$  	& $4(2)$ 	& $-13(6)$ 	& $7(6)$ 	& $1.9$ 	& $0.600(11)$ 	& $-0.71(8)$ 	& \citealt{crl+08} \\ 
    J1903$-$7051$^*$  	& $54$ 	& $3.60$ 	& $19.6612(2)$ 	& $17.6(2)$ 	& $29(2)$ 	& $-7(2)$ 	& $8(2)$ 	& $0.60$ 	& $0.96(5)$ 	& $-1.8(2)$ 	& \citealt{ckr+15} \\ 
    J1904+0451    	& $7$ 	& $6.09$ 	& $182.76(2)$ 	&   --$^\dagger$  	& $4(3)$ 	& $0(20)$ 	& $<36$ 	& $14$ 	& $0.093(3)$ 	& $-2.25(12)$ 	& \citealt{skl+15} \\ 
    J1906+0055    	& $7$ 	& $2.79$ 	& $126.784(2)$ 	&   --$^\dagger$  	& $<5.1$ 	& $10(10)$ 	& $<29$ 	& $4.0$ 	& $0.122(4)$ 	& $-2.33(13)$ 	& \citealt{lbh+15} \\ 
    J1909$-$3744$^*$  	& $144$ 	& $2.95$ 	& $10.39077(4)$ 	& $-0.66(3)$ 	& $52(2)$ 	& $13(2)$ 	& $14(2)$ 	& $0.033$ 	& $1.80(9)$ 	& $-1.7(2)$ 	& \citealt{jbvk+03} \\ 
    J1910+1256    	& $6$ 	& $4.98$ 	& $38.0680(4)$ 	& $54.4(2)$ 	& $15(2)$ 	& $-0.5(38)$ 	& $13(4)$ 	& $0.82$ 	& $0.66(4)$ 	& $-1.0(2)$ 	& \citealt{fsk+04} \\ 
    J1911$-$1114$^*$  	& $6$ 	& $3.63$ 	& $30.9665(6)$ 	& $-28.0(5)$ 	& $21(2)$ 	& $-4(3)$ 	& $7(3)$ 	& $0.88$ 	& $1.00(7)$ 	& $-3.5(3)$ 	& \citealt{llb+96} \\ 
    J1914+0659    	& $7$ 	& $18.5$ 	& $225.34(3)$ 	&   --$^\dagger$  	& $<4.3$ 	& $7(7)$ 	& $<14.8$ 	& $75$ 	& $0.410(11)$ 	& $-1.31(11)$ 	& \citealt{lbh+15} \\ 
    J1918$-$0642$^*$  	& $39$ 	& $7.65$ 	& $26.5888(3)$ 	& $-58.72(11)$ 	& $19(2)$ 	& $-4(2)$ 	& $5(2)$ 	& $0.34$ 	& $1.70(5)$ 	& $-2.06(12)$ 	& \citealt{eb01b} \\ 
    J1921+0137    	& $6$ 	& $2.50$ 	& $104.9294(7)$ 	& $91(2)$ 	& $50(2)$ 	& $0(12)$ 	& $<23$ 	& $3.1$ 	& $0.097(3)$ 	& $-2.72(14)$ 	& \citealt{san16} \\ 
    J1921+1929    	& $7$ 	& $2.65$ 	& $64.7638(11)$ 	& $118.0(8)$ 	& $29(2)$ 	& $34(6)$ 	& $34(6)$ 	& $3.0$ 	& $0.198(10)$ 	& $-2.7(3)$ 	& \citealt{pkr+19} \\ 
    J1923+2515    	& $6$ 	& $3.79$ 	& $18.860(2)$ 	& $16(2)$ 	& $17(2)$ 	& $-3(5)$ 	& $<9.0$ 	& $1.3$ 	& $0.202(14)$ 	& $-3.5(4)$ 	& \citealt{lbr+13} \\ 
    J1930+2441    	& $6$ 	& $5.77$ 	& $69.582(2)$ 	& $90.7(8)$$^\dagger$  	& $12(2)$ 	& $-17(10)$ 	& $12(10)$ 	& $10$ 	& $0.103(2)$ 	& $-2.02(9)$ 	& \citealt{pkr+19} \\ 
    J1932+1756    	& $6$ 	& $41.8$ 	& $53.16(2)$ 	&   --$^\dagger$  	& $<5.5$ 	& $0(18)$ 	& $<36$ 	& $58$ 	& $0.035(2)$ 	& $-2.3(2)$ 	& \citealt{lbh+15} \\ 
    J1933$-$6211$^*$  	& $60$ 	& $3.54$ 	& $11.51476(12)$ 	& $9.7(3)$ 	& $13(2)$ 	& $-4(2)$ 	& $8(2)$ 	& $0.75$ 	& $0.95(7)$ 	& $-1.7(4)$ 	& \citealt{jbo+07} \\ 
    J1937+1658    	& $6$ 	& $3.96$ 	& $105.8442(7)$ 	& $-2.8(5)$ 	& $39(2)$ 	& $7(8)$ 	& $<16$ 	& $2.8$ 	& $0.169(9)$ 	& $-0.5(2)$ 	& \citealt{pkr+19} \\ 
    J1939+2134    	& $6$ 	& $1.56$ 	& $71.01515(2)$ 	& $7.8(2)$ 	& $33(2)$ 	& $0.9(20)$ 	& $<4.0$ 	& $0.0049$ 	& $13.9(5)$ 	& $-3.1(2)$ 	& \citealt{bkh+82} \\ 
    J1944+0907    	& $8$ 	& $5.19$ 	& $24.3505(7)$ 	& $-37.8(7)$ 	& $11(2)$ 	& $8(2)$ 	& $7(2)$ 	& $1.3$ 	& $2.1(2)$ 	& $-2.1(5)$ 	& \citealt{clm+05} \\ 
    J1944+2236    	& $6$ 	& $3.62$ 	& $185.450(7)$ 	&   --$^\dagger$  	& $<4.7$ 	& $0(10)$ 	& $<25$ 	& $15$ 	& $0.110(5)$ 	& $-1.9(2)$ 	& \citealt{csl+12} \\ 
    J1946$-$5403$^*$  	& $40$ 	& $2.71$ 	& $23.7314(2)$ 	& $13(2)$ 	& $5(2)$ 	& $-7(3)$ 	& $6(3)$ 	& $0.39$ 	& $0.35(3)$ 	& $-1.2(4)$ 	& \citealt{ckr+15} \\ 
    J1950+2414    	& $6$ 	& $4.30$ 	& $142.096(4)$ 	&   --$^\dagger$  	& $<4.5$ 	& $7(9)$ 	& $<18$ 	& $7.6$ 	& $0.163(3)$ 	& $-0.93(9)$ 	& \citealt{kls+15} \\ 
    J1955+2527    	& $6$ 	& $4.87$ 	& $209.961(2)$ 	&   --$^\dagger$  	& $<4.1$ 	& $-7(6)$ 	& $<11.4$ 	& $6.7$ 	& $0.229(3)$ 	& $-1.99(7)$ 	& \citealt{dfc+12} \\ 
    J1955+2908    	& $6$ 	& $6.13$ 	& $104.4906(9)$ 	& $15.3(5)$ 	& $22(2)$ 	& $-7(3)$ 	& $19(3)$ 	& $2.0$ 	& $0.98(3)$ 	& $-1.5(2)$ 	& \citealt{bbf83} \\ 
    J1959+2048    	& $6$ 	& $1.61$ 	& $29.1066(3)$ 	& $-68.4(5)$ 	& $14(2)$ 	& $13(5)$ 	& $8(5)$ 	& $0.41$ 	& $0.29(3)$ 	& $-3.0(4)$ 	& \citealt{fst88} \\ 
    J2007+2722    	& $6$ 	& $24.5$ 	& $126.782(12)$ 	& $-231.82(12)$ 	& $65(2)$ 	& $-4(2)$ 	& $<4.9$ 	& $9.0$ 	& $2.32(4)$ 	& $-0.62(10)$ 	& \citealt{kac+10} \\ 
    J2010$-$1323$^*$  	& $41$ 	& $5.22$ 	& $22.1619(2)$ 	& $-2.9(2)$ 	& $18(2)$ 	& $3(2)$ 	& $7(2)$ 	& $0.40$ 	& $0.70(2)$ 	& $-1.56(14)$ 	& \citealt{jbo+07} \\ 
    J2017+0603    	& $9$ 	& $2.90$ 	& $23.9232(13)$ 	& $-59(3)$ 	& $33(2)$ 	& $-1(5)$ 	& $<10.4$ 	& $0.71$ 	& $0.18(2)$ 	& $-3.7(4)$ 	& \citealt{cgj+11} \\ 
    J2019+2425    	& $6$ 	& $3.93$ 	& $17.201(2)$ 	& $-68(1)$ 	& $43(2)$ 	& $-0.4(45)$ 	& $<9.0$ 	& $3.8$ 	& $0.26(3)$ 	& $-3.6(5)$ 	& \citealt{ntf93} \\ 
    J2033+1734    	& $8$ 	& $5.95$ 	& $25.0864(12)$ 	& $-71.5(3)$ 	& $34(2)$ 	& $-3(4)$ 	& $10(4)$ 	& $1.9$ 	& $0.277(9)$ 	& $-2.7(2)$ 	& \citealt{rtj+96} \\ 
    J2039$-$3616$^*$  	& $40$ 	& $3.28$ 	& $23.9633(5)$ 	& $-14.5(4)$ 	& $34(2)$ 	& $1(2)$ 	& $4(2)$ 	& $0.95$ 	& $0.50(4)$ 	& $-2.0(4)$ 	& \citealt{mess+20} \\ 
    J2042+0246    	& $8$ 	& $4.53$ 	& $9.2670(4)$ 	& $-21(2)$$^\dagger$  	& $36(3)$ 	& $0(10)$ 	& $<25$ 	& $3.3$ 	& $0.059(7)$ 	& $-0.9(6)$ 	& \citealt{san16} \\ 
    J2043+1711    	& $8$ 	& $2.38$ 	& $20.712(7)$ 	& $-73.3(7)$ 	& $62(2)$ 	& $-2(6)$ 	& $<12.2$ 	& $0.43$ 	& $0.121(14)$ 	& $-4.4(5)$ 	& \citealt{hrm+11} \\ 
    J2124$-$3358$^*$  	& $39$ 	& $4.93$ 	& $4.5954(3)$ 	& $-0.5(2)$ 	& $29(2)$ 	& $-2(2)$ 	& $4(2)$ 	& $0.57$ 	& $4.7(2)$ 	& $-2.3(3)$ 	& \citealt{bjb+97} \\ 
    J2129$-$5721$^*$  	& $44$ 	& $3.73$ 	& $31.8480(4)$ 	& $22.7(2)$ 	& $52(2)$ 	& $-25(7)$ 	& $27(7)$ 	& $0.55$ 	& $0.97(7)$ 	& $-3.9(3)$ 	& \citealt{llb+96} \\ 
    J2144$-$5237  	& $6$ 	& $5.04$ 	& $19.5594(7)$ 	& $28.1(2)$ 	& $44(2)$ 	& $14(2)$ 	& $12(2)$ 	& $2.1$ 	& $1.4(2)$ 	& $0.1(6)$ 	& \citealt{bcm+16} \\ 
    J2145$-$0750$^*$  	& $40$ 	& $16.1$ 	& $9.0008(13)$ 	& $0.3(5)$$^\dagger$  	& $17(2)$ 	& $7(2)$ 	& $8(2)$ 	& $0.28$ 	& $5.5(3)$ 	& $-2.5(3)$ 	& \citealt{bhl+94} \\ 
    J2150$-$0326$^*$  	& $34$ 	& $3.51$ 	& $20.67332(4)$ 	& $7.7(5)$ 	& $16(2)$ 	& $-0.9(24)$ 	& $<4.7$ 	& $0.51$ 	& $0.48(3)$ 	& $-1.9(3)$ 	& \citealt{mess+20} \\ 
    J2222$-$0137$^*$  	& $38$ 	& $32.8$ 	& $3.2797(5)$ 	& $1.35(13)$ 	& $21(2)$ 	& $-1(2)$ 	& $5(2)$ 	& $0.57$ 	& $0.90(5)$ 	& $-2.1(3)$ 	& \citealt{blr+13} \\ 
    J2229+2643$^*$  	& $32$ 	& $2.98$ 	& $22.7282(4)$ 	& $-61.2(3)$ 	& $17(2)$ 	& $4(2)$ 	& $7(2)$ 	& $0.97$ 	& $0.80(7)$ 	& $-1.7(4)$ 	& \citealt{cam95a} \\ 
    J2234+0611    	& $7$ 	& $3.58$ 	& $10.7670(2)$ 	& $4(1)$ 	& $33(2)$ 	& $1(4)$ 	& $<7.3$ 	& $0.61$ 	& $0.46(10)$ 	& $0.5(9)$ 	& \citealt{dsm+13} \\ 
    J2234+0944$^*$  	& $33$ 	& $3.63$ 	& $17.8323(2)$ 	& $-11.5(5)$ 	& $23(2)$ 	& $6(2)$ 	& $6(2)$ 	& $0.53$ 	& $1.9(2)$ 	& $-2.5(5)$ 	& \citealt{rap+12} \\ 
    J2236$-$5527$^*$  	& $15$ 	& $6.91$ 	& $20.0871(3)$ 	& $18.9(4)$ 	& $22(2)$ 	& $0.2(30)$ 	& $9(3)$ 	& $2.8$ 	& $0.34(5)$ 	& $-2.0(6)$ 	& \citealt{bbb+13} \\ 
    J2241$-$5236$^*$  	& $47$ 	& $2.19$ 	& $11.41126(3)$ 	& $12.0(3)$ 	& $14(2)$ 	& $-3(2)$ 	& $6(2)$ 	& $0.038$ 	& $1.83(7)$ 	& $-3.2(2)$ 	& \citealt{kjr+11} \\ 
    J2317+1439$^*$  	& $32$ 	& $3.45$ 	& $21.8989(2)$ 	& $-9.8(4)$ 	& $30(2)$ 	& $7(2)$ 	& $6(2)$ 	& $0.62$ 	& $0.60(6)$ 	& $-2.5(4)$ 	& \citealt{cnt93} \\
    J2322+2057$^*$  	& $25$ 	& $4.81$ 	& $13.3833(4)$ 	& $-29(2)$ 	& $10(2)$ 	& $-1(2)$ 	& $5(2)$ 	& $2.0$ 	& $0.34(3)$ 	& $-2.1(4)$ 	& \citealt{ntf93} \\ 
    J2322$-$2650$^*$  	& $36$ 	& $3.46$ 	& $6.14906(13)$ 	& $8.8(3)$ 	& $23(2)$ 	& $-0.6(27)$ 	& $<5.4$ 	& $1.3$ 	& $0.212(9)$ 	& $-1.4(3)$ 	& \citealt{sbb+18} \\\hline
  \end{tabular}
  \end{footnotesize}
\end{table*}

\restoregeometry
\end{landscape}

Science observations for MeerTime commenced on 2019 February 12, and observations for this census were of a high priority, beginning with the lowest-declination sources that are inaccessible from telescopes in the Northern hemisphere. 
The main motivation for the census was to quickly establish which pulsars should be included in the \ac{pta} programme.
For each source, we planned a minimum of 6 observations in order to avoid random scintillation \res{extrema} biasing our results when deriving their timing potential and flux densities; due to delays and data quality issues, 4 sources have only 5 observations, and 1 was observed only 4 times. 
Sources that were found to have good timing precision ($<1\,\upmu$s uncertainty in $<2000$\,s)\footnote{In selecting pulsars for the regular timing programme, southern sources were prioritised over northern sources. Some northern sources with very high timing precision, such as PSR~J1713+0747, have been included.} were added to a list to be observed with a $\sim2$\,week cadence for the \ac{mpta}, with source observing times set to achieve sub-$\upmu$s timing precision on average \res{(with a minimum observing time of 256\,s to ensure a large fraction of telescope time is spent on-source)}. 
This list is updated regularly, to add new pulsars found to have high timing precision and to remove those found not to contribute significantly to the PTA.
There are currently 89 that are in the \ac{mpta}, and these, excluding one which was \res{included} in this analysis due to issues with the ephemeris, are denoted by an asterisk in Table~\ref{tab:mtc_basic1} below. 
In total, we analysed 3966 observations\footnote{\res{Due to various possible issues with the data, some observations were only usable for one part of this analysis or another. For example, an observation with failed polarization calibration might still be useful for the timing analysis, which only used Stokes \textit{I}. The number of such observations is $\sim200$. The numbers of observations listed in Table~\ref{tab:mtc_basic1} are the total unique observations used in any part of this analysis.}} of 189 MeerTime \acp{msp} from February 2019 to February 2021. 
This number includes observations performed for the MeerTime Relativistic Binaries sub-project \citep{ksvk+21}, with which there is considerable overlap of sources; data are shared between MeerTime sub-projects to maximise the science output of the entire Large Survey Project.

The processing of all observations for this project used \textsc{psrchive} \citep{hvm04,vsdo12}, in a pipeline developed by \citet{pbs+21} which includes \ac{rfi} excision with a modified version of the \textsc{CoastGuard} software \citep{lkg+20} that is optimised for the \ac{rfi} present at MeerKAT \citep[see, e.g.,][]{bja+20}. 
Observations taken prior to mid-April 2020 were polarization-calibrated with \res{the \textsc{psrchive} tool} \textsc{pac} using {\em post-facto} derived calibration solutions provided by the \ac{sarao}, whereas later observations are polarization-calibrated at the telescope prior to the data being received by the \ac{ptuse} computers \citep[see][\res{for a full descrption of both calibration methods}]{sjk+21}. 
All polarization values and derivatives follow standard pulsar conventions\footnote{The observed polarization position angles are measured from celestial North and increasing towards East, and the IEEE definition of circular polarization is used. Stokes \textit{V} is defined as left-circular minus right-circular.} as described in \citet{vsmjr10}.

\section{Polarization}
\label{sec:mtc_pol}

For each pulsar, we produce a high \ac{snr} profile by summing the brightest individual observations\footnote{To make the brightest profiles, we summed observations in descending order of $S/N$ until the next observation satisfied: ${S/N}_\textrm{obs} < {S/N}_\textrm{max}/3$ \textit{and} either ${S/N}_\textrm{obs} < 5$ or the number of observations summed was greater than 7.} \res{partially averaged in frequency} to 232 channels ($\approx$\,3.34\,MHz channel widths), and full time integration after summing. 
We show the polarization profiles in Figures~\ref{fig:mtc_profs_a}-\ref{fig:mtc_profs_u}, integrated to the centre frequency (after correcting for Faraday rotation as described below). 
All calculations of linear polarization, \textit{L}, have been de-biased \res{with a procedure modified from} \citet{ew01}, where, for every bin $i$ across the profile, 
\begin{equation}
    L_{\textrm{true},i} = 
    \begin{cases}
        \sqrt{L_{\textrm{meas},i}^2 - \sigma_I^2} & \quad \textrm{if}~ |L_{\textrm{meas},i}| \geq |\sigma_{I}| \\
        -\sqrt{|L_{\textrm{meas},i}^2 - \sigma_I^2|} & \quad \textrm{otherwise},
    \end{cases}
\end{equation}
where $L_\textrm{meas}$ is the measured linear polarization ($L^2 = Q^2 + U^2$) and $\sigma_I$ is the root-mean-square of the total intensity profile in the off-pulse region\footnote{\res{Note that \citet{ew01} apply a lower limit of 0 to each $L_\textrm{true,i}$, but we have chosen not to do so, just as the values of Stokes \textit{I} can be negative.}}. 
We determine the uncertainties on the polarization position angle (PA, $\psi$) following \citet{ew01}: for each bin $i$ across the profile, if $P_0=L_{\textrm{true},i}/\sigma_{I} > 10$, the uncertainty is $\sigma_{\psi,i} = \sigma_I/2L_i$; otherwise, we determine the uncertainty numerically from the probability distribution \citep{nkc93}
\begin{equation}
    G(\psi, P_0) = \frac{1}{\sqrt{\pi}}\left(\frac{1}{\sqrt{\pi}}+\eta \,e^{\eta^2}[1+\textrm{erf}(\eta)]\right)\,e^{-P_0^2/2},
\end{equation}
where $\eta=\frac{P_0}{\sqrt{2}}\cos2(\psi-\psi_\textrm{true})$ and ${\rm erf(\eta)}$ is the Gaussian error function. 
The uncertainty, $\sigma_\psi$, can therefore be determined by setting $\psi_\textrm{true}=0$ and adjusting the bounds of integration of the probability distribution to satisfy
\begin{equation}
    \int_{-1\sigma_\psi}^{1\sigma_\psi} G(\psi,P_0)\,d\psi=68.26\%.
\end{equation}

\res{When analysing our pulse profiles with \textsc{psrchive}, it is necessary to consider the noise baseline in each polarization channel, especially with complicated profiles with emission across the entire phase range. Accurate calibration will set the noise values appropriately, but, when calculating \snr or polarization fractions, \textsc{psrchive} will ``subtract the baseline'' by default.} 
For all but one pulsar in our sample, we use the ``minimum'' baseline estimator\footnote{These algorithms are described at \url{http://psrchive.sourceforge.net/manuals/psrstat/algorithms/}.}, which finds a continuous region with the minimum mean by smoothing the profile with a boxcar with custom widths varying from 1\% to $90$\% \res{(where the region width was determined manually for each pulsar)}. 
One pulsar for which the ``minimum'' baseline estimator was not optimal is PSR~J1421$-$4409 \res{\citep{sfj+20}}, which instead used the ``normal'' estimator\footnote{\res{We found that the ``normal'' baseline estimator performed worse than the ``minimum'' estimator with custom widths for most pulsars with $\gtrsim95$\% duty cycles.}}. 
Note that this pulsar still shows apparent over-polarization at pulse phases $\sim 0.6$-1.0 (shown in the central panels of Fig.~\ref{fig:mtc_profs_g}), as there is significant emission in both \textit{Q} and \textit{U} throughout that range. 
This pulsar may be similar to PSR~J0218+4232 with underlying unpulsed emission \citep{ndbf+95}, \res{which would contradict the general assumption that all polarization channels are consistent with noise over some phase range (the off-pulse baseline)}. 
Further study of this pulsar with MeerKAT could reveal the nature of this emission.

We measure significant fractional linear polarization ($L/I$; from fully time- and frequency-integrated profiles centred at 1284\,MHz, as shown in Fig.~\ref{fig:mtc_profs_a}-\ref{fig:mtc_profs_u}) for 180 pulsars in our sample ($>$\,1-$\sigma$ significance), and significant absolute fractional circular polarization ($|V|/I$) for 119. 
Our full results are listed in Table~\ref{tab:mtc_basic1}.

We calculate the \ac{rm} for each pulsar from the individual observations (fully integrated in time, with 29 frequency channels across the 775.75\,MHz band) or, for faint pulsars, from the summed observations (also fully time-integrated, with 29 frequency channels) and list all in Table~\ref{tab:mtc_basic1}. 
We use \textsc{rmfit} to determine \acp{rm} using a brute force linear polarization maximisation which is then refined using an iterative fit to the measured PA as a function of frequency. 
After calculating \ac{rm} from the individual observations, we then correct for the contributions of the ionosphere using \res{the \textsc{ionFR} software \citep{sbsh+13} with IGSG maps}\footnote{\res{IGSG maps for the ionospheric correction were downloaded from NASA's CDDIS database at \url{https://cddis.nasa.gov/archive/slr/data/npt_crd/lageos1} \citep{IGSG_noll10}.}}. 
\res{For each pulsar with $\geq10$ corrected RMs, we remove outliers using a basic drop-out algorithm comparing the standard deviation and median-absolute-deviation from the median. 
Outliers are removed manually for pulsars with fewer observations. 
We then report the mean of the corrected \ac{rm} values with the standard deviation as the 1-$\sigma$ uncertainty. }
For faint pulsars, because we measure the \ac{rm} from a summed profile, we report the \acp{rm} and uncertainties as given by \textsc{rmfit} after correcting by the mean of the ionospheric contributions per pulsar (the uncertainties on the ionospheric corrections are summed in quadrature with the uncertainties given by \textsc{rmfit}). 
No \acp{rm} are given for pulsars with linear polarization fractions less than $\approx0.05$ ($\sim$\,1.5-$\sigma$).

In total, we measure 171 non-zero \acp{rm} (at the 1-$\sigma$ level; 163 with 3-$\sigma$ measurements), increasing the number of \acp{msp} with known \acp{rm} by 88. 
To verify the reliability of our measurements, we compare the \acp{rm} in the \ac{atnf} pulsar catalogue (v1.64) for the 83 \acp{msp} with previously published values and calculate the significance of the difference of the values: $d = (\textrm{RM}_\textrm{psrcat} - \textrm{RM}_\textrm{census})/\sqrt{\sigma_\textrm{psrcat}^2 + \sigma_\textrm{census}^2}$. 
Nearly 50\% of the pulsars have \acp{rm} that are 1-$\sigma$ consistent with the previously published values, and 85\% of the pairs are consistent to 2.6-$\sigma$. 
Given that pulsar \acp{rm} are known to vary with time \res{due to inaccurate models of the ionospheric Faraday rotation} \citep[e.g.,][]{ymh+11}, \res{that not all values reported in the \ac{atnf} pulsar catalogue are corrected for the ionosphere}, and that measurements at different frequencies can result in statistically inconsistent \acp{rm} (see, e.g., the position angle fits in \citealt{dhm+15}), differences between our \acp{rm} and published values at this level are not unexpected. 
However, two pulsars have significantly different values in our census and the literature: PSR~J1502$-$6752 has a published value of $-$225(2)\,rad\,m$^{-2}$ from \citet{kjb+12}, while our measurement is $232.8(2)$\,rad\,m$^{-2}$; and PSR~J0154+1833 has a published value of 21.6(1)\,rad\,m$^{-2}$ \citep{mgf+19}, while our measurement is $-21.9(6)$\,rad\,m$^{-2}$. 
Given the consistency of our other \ac{rm} measurements, we are inclined to believe the \citet{kjb+12} and \citet{mgf+19} values suffer from a sign convention error.

\begin{figure*}
    \centering
    \includegraphics[width=\textwidth]{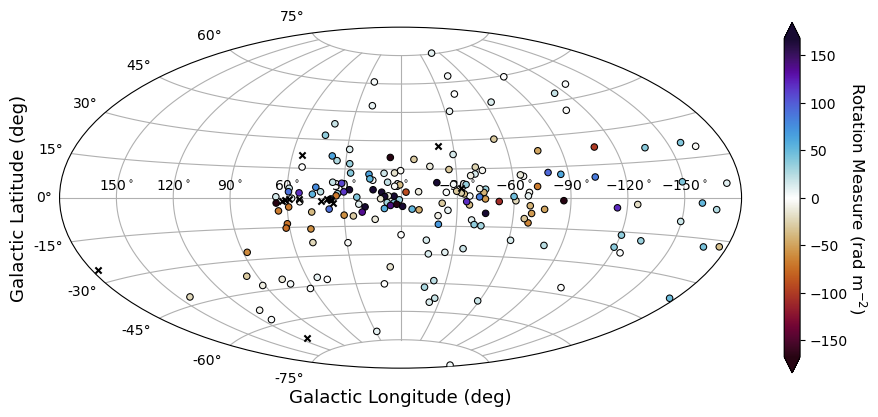}
    \caption[Galactic Distribution of MeerTime \acp{msp}]{The positions of the \acp{msp} in this sample on an Aitoff projection, with coloured circles indicating sources with measured \acp{rm} (including those formally consistent with zero) and black `x' markers indicating sources without measured \acp{rm}, as described in the text.} 
    \label{fig:mtc_skyrim}
\end{figure*}

We show in Figure~\ref{fig:mtc_skyrim} the positions of the \acp{msp} in our sample, coloured by their measured \ac{rm} (black marks indicate the positions of pulsars without measured \acp{rm}). 
Overall, the \acp{rm} for our sample do not indicate strong (average) magnetic fields along the lines of sight, with 90\% of values in the range $-$2.8 - 2.3\,$\upmu$G (full range $-$6.7 - 5.0\,$\upmu$G).

\section{Flux Densities and Spectral Indices}
\label{sec:mtc_flux}
Ideally, pulsar flux densities are obtained using an injected and pulsed broadband calibration signal that can be measured against a source of known flux density to derive a flux scale. 
\res{While attempting to derive a polarisation calibration solution using the MeerKAT L-Band receiver pulsed cal, we found that the degree of polarisation of the cal often exceeds 105\%. 
This is probably due to the strength of the cal being comparable to the system equivalent flux density and driving the system into a non-linear regime. 
This made this conventional method unsuitable for absolute flux calibration.}

An alternative calibration method involves using the radiometer equation and assuming that the rms of the pulsar baseline region is well-defined by the system equivalent flux density plus the galactic sky temperature divided by the antenna gain. 
In the presence of unrecognised sources of noise due to \ac{rfi}, this method can underestimate radio fluxes and \res{must} be approached with caution. 
Nevertheless, as shown in \citet{bja+20}, there are many parts of the MeerKAT L-band (856-1712\,MHz) that are almost always devoid of interference, and we found that many of our high \ac{dm} \res{(DM\,$\gtrsim100$\,pc\,cm$^{-3}$)} sources gave very reliable \ac{snr} \res{and hence flux} values from epoch to epoch across much of the band.
\res{To assess the reliability of this method, we computed the modulation indices of the derived flux densities for the five pulsars with the largest DMs in our sample and found them to be between 7-15\%, which means that the reliability of the flux densities from epoch to epoch is probably no worse than this figure.}

Absolute flux densities for each observation were thus calculated using the radiometer equation and assuming the \citet{hks+81} sky temperatures scale as the observing frequency to the power \res{$-2.55$ \citep{temp_lmop87}}. 
\res{To determine mean pulsar flux densities the integrated flux above a baseline is often computed by firstly converting the counts to Jy in a calibration procedure. 
The mean pulsed flux is then just the integral of the flux above a defined baseline divided by the number of bins.
Millisecond pulsars can have profiles that are very complex and this makes baseline estimation difficult if using a simple boxcar, especially if it extends into the wings of the pulse profile. 
For every MSP, the ideal boxcar width will differ, or even require multiple disjoint sections, so we used an interactive tool to define the baseline through visual inspection for each pulsar from high \ac{snr} observations to help determine the rms of the noise in the baseline.}
Software was written to cross-correlate the template with each profile, and the rms residual of the phase-aligned profiles in the baseline region was evaluated.

There are sections of the MeerKAT band that are rarely affected by \ac{rfi} (see \citealt{bja+20}), and these were used to determine the flux density in those bands. 
\res{We found that many of the high-DM MSPs that would be expected to have consistent epoch to epoch flux densities often exhibited dips in their mean spectra where RFI is regularly present in the fourth of the 8 sub-bands. 
As an example, for PSR J1017$-$7156, the off-pulse rms was roughly 5-21\% higher in this band with a mean 10\% higher after our attempts at RFI removal with \textsc{CoastGuard}. 
Scaling the flux densities of all channels by the ratio of their off-pulse rms counts to that of the median off-pulse rms counts led to mean spectra largely free of this dip in channel 4 and was performed on all the data.}

\begin{figure}
    \centering
    \includegraphics[width=\columnwidth]{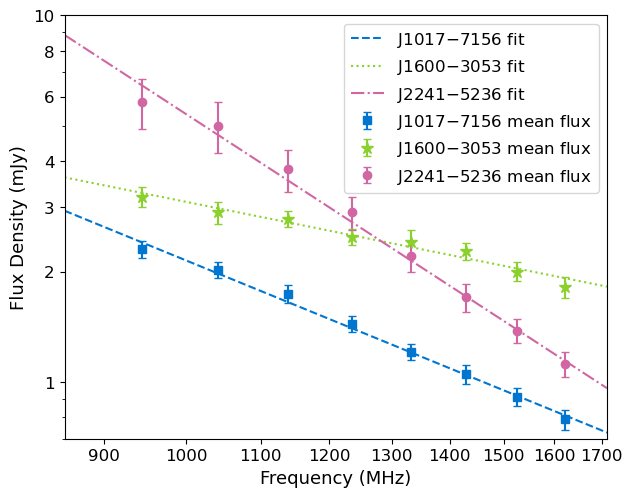}
    \caption[Example pulsar spectra with power law fits]{\res{Mean flux density measurements for 3 pulsars, PSRs~J1017$-$7156 (blue squares; $N_{\rm obs}=51$), J1600$-$3053 (green stars; $N_{\rm obs}=33$), and J2241$-$5236 (pink circles; $N_{\rm obs}=42$), fit with power law models (blue dashed, green dotted, and pink dash-dotted lines, respectively) to determine $S_{1400}$ and the spectral index, $\alpha$. The power law models for these pulsars are: $S_{1017} = 1.09\,{\rm mJy} \times(\nu/1400\,{\rm MHz})^{-2.01}$, $S_{1600} = 2.22\,{\rm mJy} \times(\nu/1400\,{\rm MHz})^{-0.99}$, and$S_{2241} = 1.83\,{\rm mJy} \times(\nu/1400\,{\rm MHz})^{-3.20}$. The error bars indicate the formal error on the means. }
    }
    \label{fig:mtc_exspec}
\end{figure}

After integrating all of the observations \res{for each pulsar}, the vast majority of our pulsars were shown to have a smooth power law variation of mean flux density with frequency as is often seen in the population between 1-1.7\,GHz, especially those at high dispersion ($\gtrsim 100$\,pc\,cm$^{-3}$) where strong scintillation events across $\sim$100-MHz bands are not seen \res{(\citealt{dhm+15}; see also Fig.~\ref{fig:mtc_exspec}, showing PSRs~J1017$-$7156, J1600$-$3053, and J2241$-$5236, with DMs of 94.2, 52.3, and 11.4\,pc\,cm$^{-3}$, respectively)}.
To derive spectral indices and normalise our flux density measurements to a standard frequency (1400\,MHz), we fitted a power-law model, $S = S_{1400}\,(\nu/1400\,\textrm{MHz})^\alpha$, to the mean flux density in each of 8 sub-bands across the 775.75\,MHz band\footnote{We performed the power-law fitting in log space, weighting each channel mean by the standard deviation of the observed values.}, and our results are listed in Table~\ref{tab:mtc_basic1}. 
In Figure~\ref{fig:mtc_exspec}, we show an example of the flux density measurements and best-fitting models for 3 pulsars \res{with varying \ac{dm} values}, selected for comparison with the study by \citet{dhm+15} on pulsars in the \res{\ac{ppta}} \citep{krh+20}. 

\res{We refer the reader to Gitika et al. (in prep.) for a thorough analysis of the flux density measurements from these and more recent MeerTime MSP observations, and here we provide a simple comparison of our results with overlapping samples from \citet{dhm+15} and \citet{aab+21} to verify the accuracy of our method. }
Note that the \citet{dhm+15} spectral indices were derived from flux density measurements in 3 bands centred at 732, 1369, and 3100\,MHz, and therefore they may be sensitive to, e.g., spectral turn-over, which is possibly seen in their analysis of PSR~J1600$-$3053. 
With this caveat, we find good agreement between our $S_{1400}$ values and spectral indices and those of \citet{dhm+15}, with PSR~J1744$-$1134 showing the greatest disparities in both $S_{1400}$ and spectral index (although the values are consistent at the 2-$\sigma$ level and this pulsar has a very small \ac{dm} which makes it prone to large flux density variations).
\res{The flux density of this pulsar in the 97-Mhz sub-band centred at 1429 MHz varied from 0.14 to 14\,mJy with a mean of 3.7\,mJy and a standard deviation of 4.0\,mJy. 
Thus its difference in mean flux density is not that worrying.}

Recently, the \res{\ac{nanograv}} collaboration presented a comprehensive study of flux densities for its sample with which we have significant overlap \citep{aab+21}. 
Of the 15 non-PPTA pulsars for which we have common data, 12 had flux densities within $\sim$30\% of our values and those with the largest deviations (\res{differing by roughly a factor of 2 from our values}) have low \acp{dm}.
\citet{aab+21} pointed out that the median flux they obtained for PSR~J2317+1439 ($0.45^{+0.35}_{-0.19}$\,mJy) differed from the catalogue flux density of 4(1)\,mJy significantly\footnote{\res{The ATNF catalogue flux density for PSR~J2317+1439 was from \citet{kxl+98} and represents the mean of their measured distribution. The median quoted in that work was 2\,mJy, hinting at significant scintillation.}}. 
Our mean value 0.60(6)\,mJy (\res{median of 0.37\,mJy}) is much closer to the \citet{aab+21} result, and we note that this pulsar scintillates strongly. 
The pulsar with the most observations (134 with reliable fluxes) was PSR~J1909$-$3744. 
\citet{aab+21} measured a median flux density \res{at 1400\,MHz} of 1.03\,mJy, \res{matching our median value precisely}, which gives us great confidence in our relative flux density scales. 
This pulsar experiences extreme scintillation events, and our measured mean flux density was 1.80(9)\,mJy.
\citet{dhm+15} recorded a mean flux density for PSR~J1909$-$3744 of 2.5(2)\,mJy at 1400\,MHz. 
\res{Note also that different observing strategies can lead to biases in observed flux density distributions. 
MeerKAT observations use a queue system with minimal intervention (and NANOGrav observations are run with minimal intervention), while observations for the PPTA are done manually, and observers are encouraged to switch sources when the source \snr is low and to repeat observations when the \snr is high. 
This leads to better uncertainties on the \acp{toa} but biases the flux density distributions to higher values, which is consistent with our results.}

\begin{figure}
    \centering
    \includegraphics[width=\columnwidth]{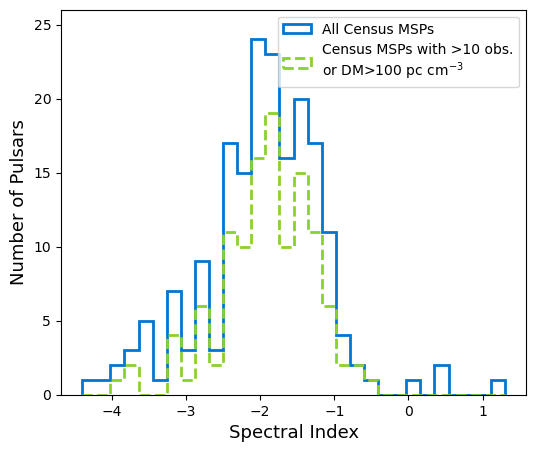}
    \caption{Histogram of spectral indices. In green, we show the spectral indices for pulsars with DM\,$>\,100$\,pc\,cm$^{-3}$ (and therefore low scintillation) and those with $>\,10$ observations to reduce the impact of scintillation on the measurements. We show the remaining spectral indices in our sample in the blue histogram.}
    \label{fig:mtc_spind}
\end{figure}

Overall, we find a mean spectral index of $-1.92(6)$, with a standard deviation of 0.6, for our sample of 189 pulsars, and we show the full distribution of spectral indices in Figure~\ref{fig:mtc_spind}. 
This mean is consistent at the 2-$\sigma$ level with that found by \citet{dhm+15}, who found a value of $-1.81(1)$, as well as previous results for \ac{msp} studies \citep{tbms98,kll+99} and studies of normal pulsars \citep{jvsk+18}. 
The typical uncertainties on the spectral indices for pulsars with $>10$ observations are $\sim0.1$, with 95\% of the uncertainties $<0.4$. 
\res{These results could be improved in future using the MeerKAT UHF and S-band receivers to increase the frequency coverage.}

\section{Timing Results}
\label{sec:mtc_time}

\begin{figure}
    \centering
    \includegraphics[width=\columnwidth]{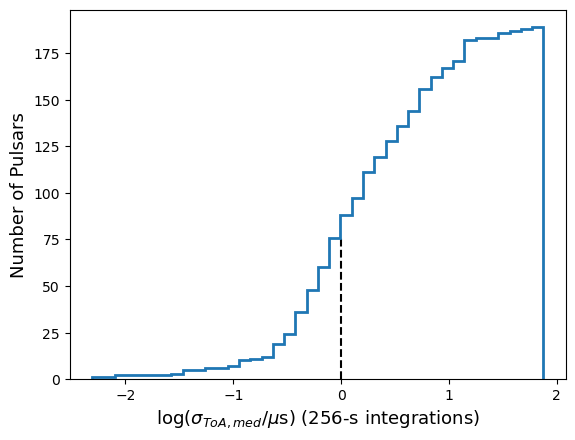}
    \caption[Cumulative Histogram of \ac{toa} Uncertainties from MeerTime]{Cumulative histogram of median \ac{toa} uncertainties from 97-MHz, 256-second sub-integrations, scaled to the full 775.75\,MHz, on a logarithmic scale. From these data, 77 pulsars have sub-microsecond (median) timing precision in less than 256\,seconds.}
    \label{fig:mtc_toahist}
\end{figure}

For all pulsars in this study, we list basic parameters in Table~\ref{tab:mtc_basic1}, and provide our measured median \ac{toa} uncertainties.
The templates used for our standard timing procedure are produced from summed, temporally-integrated observations with 232 frequency channels across the 775.75-MHz band. 
From these high-\ac{snr} observations, we then derive pulse ``portraits'' using the \textsc{PulsePortraiture} software \citep{pennucci19} and produce noise-free templates with 8 sub-bands for timing\footnote{True wide-band timing with the \citet{pennucci19} method is left to future work.}. 
We use the ``Fourier Domain with Markov chain Monte Carlo'' (FDM) fitting algorithm from \textsc{pat} \res{in \textsc{psrchive}} to produce \acp{toa} from archives partially-integrated to 256-second sub-integrations and 8 sub-bands across the 775.75-MHz band \res{($\sim97$-MHz sub-bands)}.  
We compute ``typical'' \ac{toa} uncertainties per pulsar \res{by first normalising the measured uncertainties to the nominal minimum observing time, 256\,s, then calculating the error on the weighted mean ToA for each sub-integration as 
\begin{equation}
    \frac{1}{\sigma_\textrm{sub}^2} = \sum_i \frac{1}{\sigma_{\textrm{sub,}i}^2} ,
\end{equation}
where \textit{i} represents the frequency channels, and finally returning the median of the resulting $\sigma_\textrm{sub}$ values as the typical ToA uncertainty. We note that this method approximates the ToA uncertainty achievable by proper wide-band timing using a 2-dimensional template (i.e., returning a single ToA per sub-integration). } 
We summarise the resulting median \ac{toa} uncertainties in Figure~\ref{fig:mtc_toahist}, showing the cumulative histogram of the median uncertainties. 
Note that the median \ac{toa} uncertainties should not be confused with the ultimate fitted rms residuals, as they do not factor in pulse jitter nor the deleterious effects of red noise.

\subsection{Timing Forecast}
In assembling a large dataset of \ac{msp} data across a minimum timespan of six months, we are able to determine the timing potential for each source and inform future studies. 
The potential for MeerTime to contribute to global efforts to detect \acp{gw} with the \ac{ipta} can be estimated from these data. 
As mentioned in section~\ref{sec:mtc_method}, we continue to regularly observe pulsars that have low \ac{toa} uncertainties, $\sigma_\tau<1\,\upmu$s, calculated using frequency-scrunched templates\footnote{Note that frequency-dependent profile evolution and scintillation can result in poor \ac{toa} fits, and therefore higher uncertainties, using frequency-scrunched templates than when using frequency-resolved templates. This effect, and the differing datasets used in this Census and the \ac{mpta} programme, leads to different numbers of pulsars with sub-microsecond timing precision.} and observations with integration times, $\res{256} < T_\textrm{obs}<2000$\,s.

To compare the sensitivity of \ac{pta} observations with MeerKAT to other current programmes, and its \res{potential} impact on global efforts, we consider the detection of a \ac{gw} background (assuming a strain spectrum of $h_c(f) = 1.9\times 10^{-15} (f/1\,{\rm yr})^{-2/3}$; e.g., \citealt{sejr13}) through correlations between pulsars in the array \citep{hd83}. 
The background is chosen to be the amplitude of the apparent common red noise found in analysis of the \ac{nanograv} 12.5-year data release \citep{aab+21wb}.
We can approximate the sensitivity of a given \ac{pta} to both the auto-correlated signal (the ``pulsar term'') and the cross-correlated signal (the ``Earth term''), following \citet{sejr13}\footnote{\res{\citet{src+13} used the same approach to place the bound on a gravitational wave background.}}. 
\res{They construct matched detection statistics in the Fourier domain (see their Equation 17) and calculate how the signal to noise ratio of the detection statistic depends on the noise properties of the array.}

\res{The expected signal-to-noise ratio in the autocorrelations is
\begin{align}
  \rho_{ACF}^2 = \sum_{p=1}^{N_p} \sum_{k=1}^{N_f}  \frac{P_g^2(f_k)}{P_{p}^2(f_k)},  
\end{align}
where $P_g(f_k)$ is the power spectral density of the GWB in frequency channel $f_k$, where $k=1,N_f$ is a Fourier series starting $1/T_{eff}$ at the reciprocal of the common effective observing $T_{\rm eff}$. 
$P_{p}(f_k)=P_g(f_k) + P_{p,r}(f_k) + P_{p,w}$ is the total power (the sum of the GWB and the red noise ($P_{p,r}$) and the white noise ($P_{w,r}$) in the same channel for pulsar $p$.}

\res{The expected signal-to-noise ratio in the cross correlations is found as the S/N of a matched filter with the Hellings-Downs correlation function
\begin{align}
 \rho^2 = \sum_{i=1}^{N_p-1} \sum_{j=i+1}^{N_p} \chi^2_{ij} \sum_{k=1}^{N_f} \frac{P^2_g(f_k)}{P_{p_i}(f_k) P_{p_j}(f_k)},
\end{align}
where $\chi_{ij}$  are the angular correlation coefficients  \cite[][]{hd83}
\begin{equation}
\begin{split}
\chi_{ij} = \frac{3}{4} \left(1 - \cos(\theta_{ij})\right) &\log \left(1 - \cos(\theta_{ij})\right) \\
&-\frac{1}{8} \left( 1 - \cos(\theta_{ij}) \right) + \frac{1}{2}
\end{split}
\end{equation}}

\begin{figure}
    \centering
    \includegraphics[width=\columnwidth]{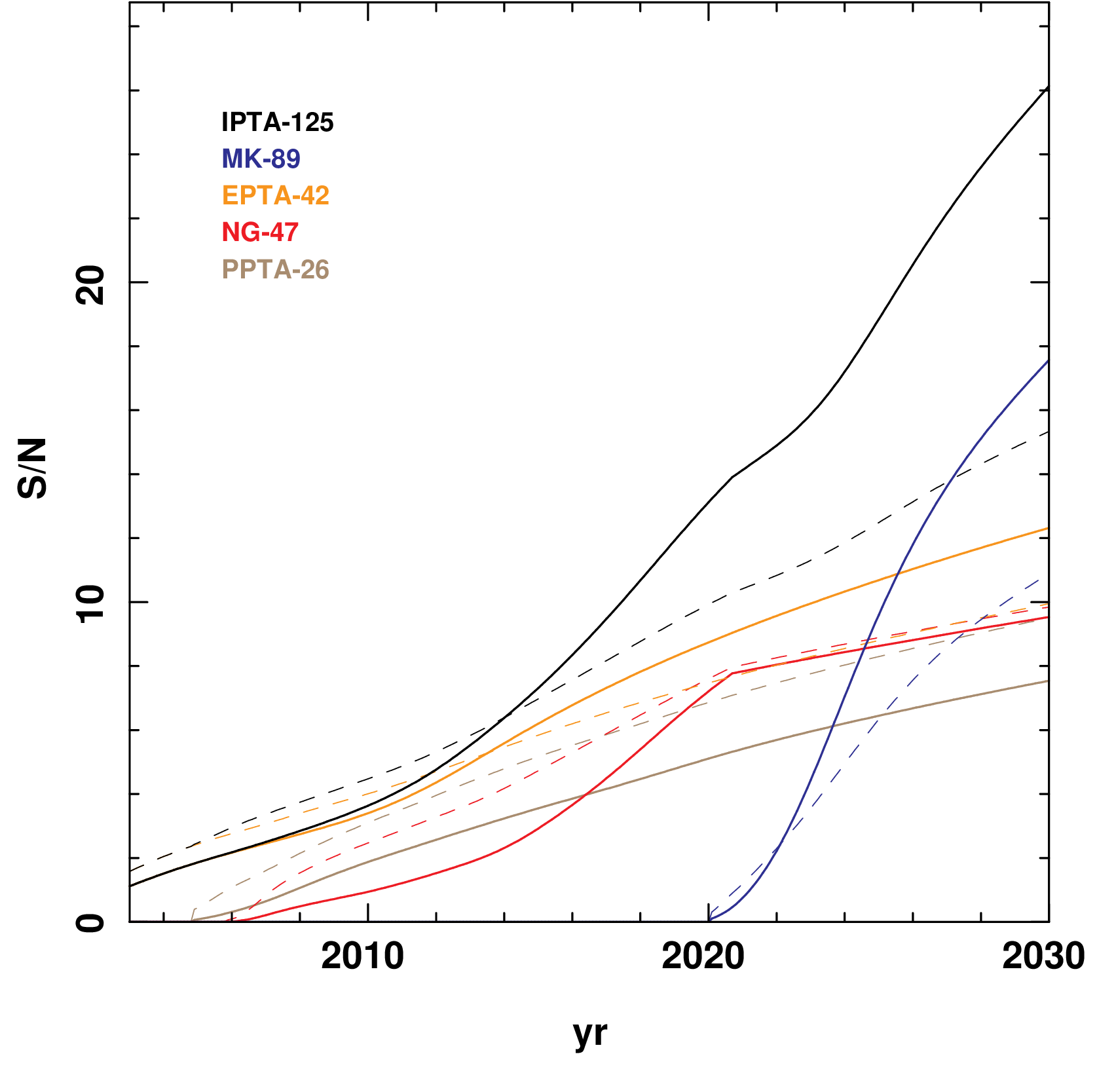}
    \caption[Simulated MeerTime contribution to the \ac{ipta}]{Comparison of PTA sensitivity to the cross-correlated component of the \ac{gw} background (solid lines) and the uncorrelated signal (dashed lines), using the current MeerTime timing programme with 89 pulsars (blue lines), the \ac{epta} programme with 42 pulsars (orange lines), the \ac{nanograv} programme with 47 sources (red lines), the \ac{ppta} with 26 (\res{brown} lines), or the \ac{ipta} as the union of the four (black lines).}
    \label{fig:mtc_pta}
\end{figure}

We use the published source lists, median pulsar timing precisions, timespans, cadences, and observational parameters from the European Pulsar Timing Array \citep[EPTA;][]{dcl+16}, \ac{nanograv} \citep{aab+21wb}, and the \ac{ppta} \citep{krh+20} to form a representative sample of the \ac{ipta}. \acused{epta}
For sources in multiple \acp{pta}, the data set with the longest time span is used. 
We show how the \ac{snr} for this background increases with time in Figure~\ref{fig:mtc_pta}.

For the \ac{nanograv} sensitivity curves, we have assumed that the Arecibo timing programme ceased in August 2020 and that the pulsars are not observed elsewhere, which demonstrates the importance of Arecibo to international pulsar timing efforts. 
The modelling includes red noise in the sensitivity estimates, for pulsars where red noise has \res{previously been detected \citep{abb+20ng,gsr+21}. 
Undetected red noise is likely to be present in some of the pulsars timed by MeerKAT, which will affect our sensitivity estimates.}

These calculations imply that the \ac{mpta} will be comparable in sensitivity to the \ac{ppta} by 2023, \ac{nanograv} by 2024, and the EPTA by $\approx2025$. 
In less than four years, the \ac{mpta} will be contributing to the \ac{ipta} effort. 
As the MeerKAT timing array programme has observed a large number of pulsars from its start, the sensitivity to the cross correlated signal (dashed lines) is always comparable to the auto-correlated signal, \res{similar to the EPTA with its 42 pulsars}. 
In contrast, the \ac{ppta} project includes only 26 pulsars, and \ac{nanograv} started out with a smaller number of pulsars before gradually increasing in size starting in 2009, and, in the most recent data release (12.5-yr), they listed 47 pulsars \citep{aab+21wb}. 
\res{The MPTA could be expanded in the future with timing of additional pulsars in other bands or timing of MSPs discovered by MeerKAT and other facilities. 
Future work by Middleton et al. (2022, in prep.) will examine the sensitivity of the MPTA and potential improvements. }

\section{Summary}
\label{sec:mtc_conc}

In this work, we demonstrated the outstanding potential of MeerKAT for precision pulsar timing of \acp{msp} with this initial census, analysing polarization properties, flux densities and spectral indices, and timing precision. 
We presented a dataset of nearly 4000 observations of 189 pulsars, spanning 24 months, including polarization-calibrated profiles and high-precision \acp{toa}. 
Our timing simulations offer simple predictions of future results, and we predict that the MPTA will be a major contributor to global PTA efforts within the next 5 years. 
\res{More work is forthcoming, including the first MPTA data release with full timing and noise analyses (Miles et al., 2022, in prep.), an analysis of flux density measurements for MeerTime MSPs (Gitika et al., in prep.), and a study of the sensitivity of the MPTA and potential for improvements (Middleton et al., 2022, in prep.). }

\section*{Data Availability}\label{sec:mtc_data}
We have made available all 189 polarization profiles, as shown in Figures~\ref{fig:mtc_profs_a}--\ref{fig:mtc_profs_u}, in PSRFITS format with full phase resolution (1024 bins), 4 polarization channels, and 8 frequency channels across the 775.75-MHz band, as well as the table containing the analysis results. 
The table, in csv format with 189 rows, contains mean flux densities (and standard deviations) in 8 frequency bands per pulsar, spectral indices and fit $S_{1400}$ values with uncertainties, median timing uncertainties as described in section~\ref{sec:mtc_time}, polarization fractions and rotation measures, dispersion measures and reference epochs, and basic pulsar properties (rotation properties and positions) for easy reference. 
The dataset will be made available prior to publication of this manuscript, under the DOI 10.5281/zenodo.5347875.

\section*{Acknowledgements}
\res{The authors thank the expert reviewers for providing thorough and thoughtful comments on this manuscript.} 
The authors \res{additionally} thank R.~Eatough, D.~Lorimer, and J.~Swiggum for providing phase-connected pulsar ephemerides for unpublished pulsars. 
The MeerKAT telescope is operated by \ac{sarao}, which is a facility of the National Research Foundation, an agency of the Department of Science and Innovation. 
PTUSE was developed with support from the Australian SKA Office and Swinburne University of Technology, with financial contributions from the MeerTime collaboration members.
This research was funded partially by the Australian Government through the Australian Research Council (ARC), grants CE170100004 (OzGrav) and FL150100148. 
RMS acknowledges support through ARC Future Fellowship FT190100155. 
This work used the OzSTAR national facility at Swinburne University of Technology. 
OzSTAR is funded by Swinburne University of Technology and the National Collaborative Research Infrastructure Strategy (NCRIS). 
This research made use of \textsc{astropy}, a community-developed core Python package for Astronomy \citep{astropy18}, the \textsc{matplotlib} package \citep[v1.5.1;][]{hunter07}, and \textsc{psrqpy}, a Python package for searching the ATNF Pulsar Catalogue \citep{psrqpy}.



\bibliographystyle{pasa-mnras}
\bibliography{MT_MSP_census}



\appendix

\section{Polarization Profiles}


\begin{figure*}
    \centering
    \includegraphics[width=0.91\textwidth]{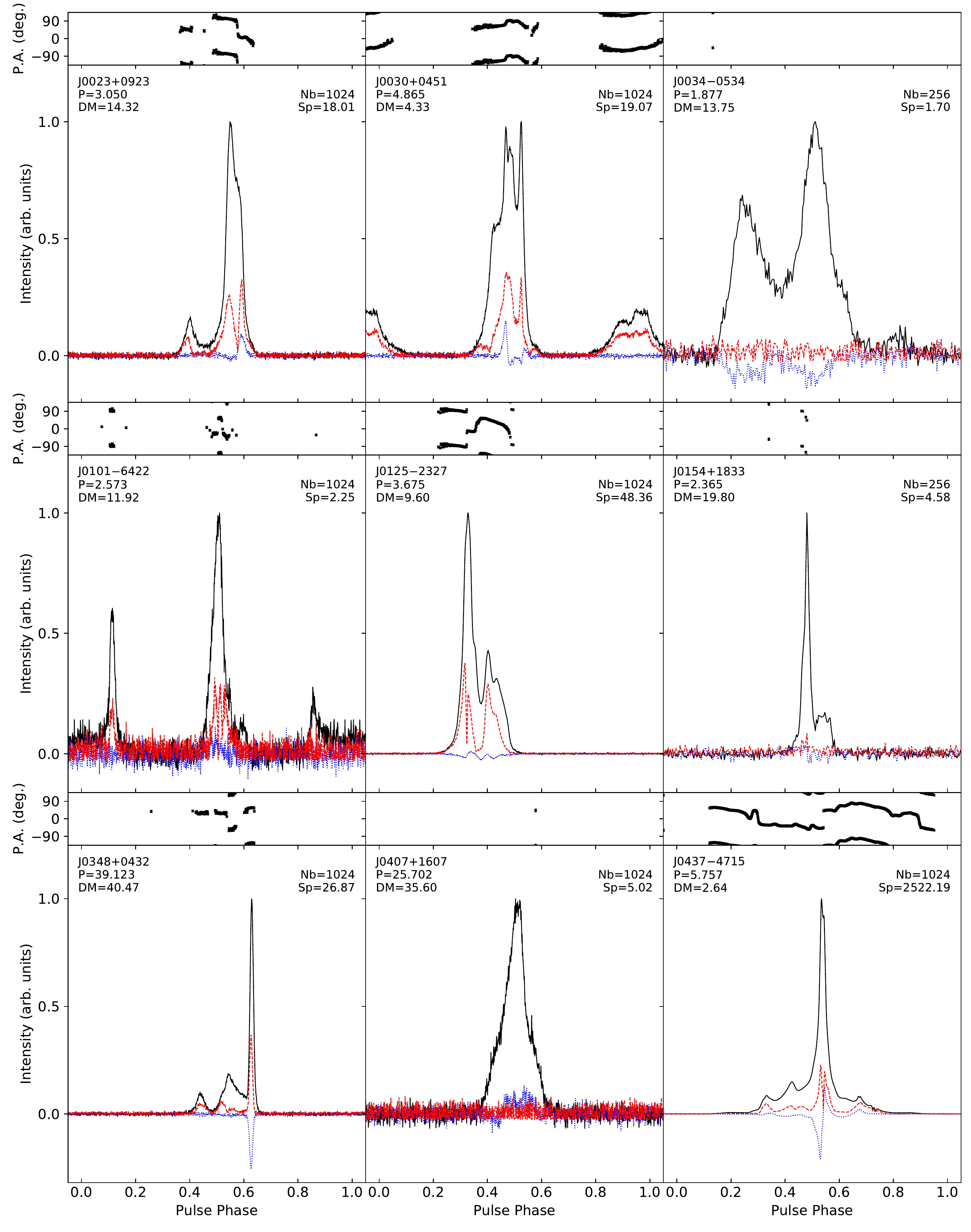}
    \caption[MeerTime Census Polarization Profiles - 1]{Polarization profiles for \acp{msp} from the MeerTime Census project, fully integrated from the 775.75-MHz band centred at 1284\,MHz, with polarization position angles in the upper panels (not corrected to infinite frequency). The black solid line indicates the total intensity, the red dashed line shows the linearly polarized emission, and the blue dotted line shows the circularly polarized emission. All profiles show the phase range ($-$0.05, 1.05), and P.A.s are plotted over the range ($-$135, 135)\,degrees. The period and \ac{dm} of each source is listed in the upper left corner, in units of ms and pc\,cm$^{-3}$, respectively. In the upper right corner, we indicate the number of phase bins used (256 for $P<15$\,ms and low \snr) and the normalisation factor (peak flux density at 1284\,MHz in mJy). }
    \label{fig:mtc_profs_a}
\end{figure*}

\begin{figure*}
    \centering
    \includegraphics[width=0.94\textwidth]{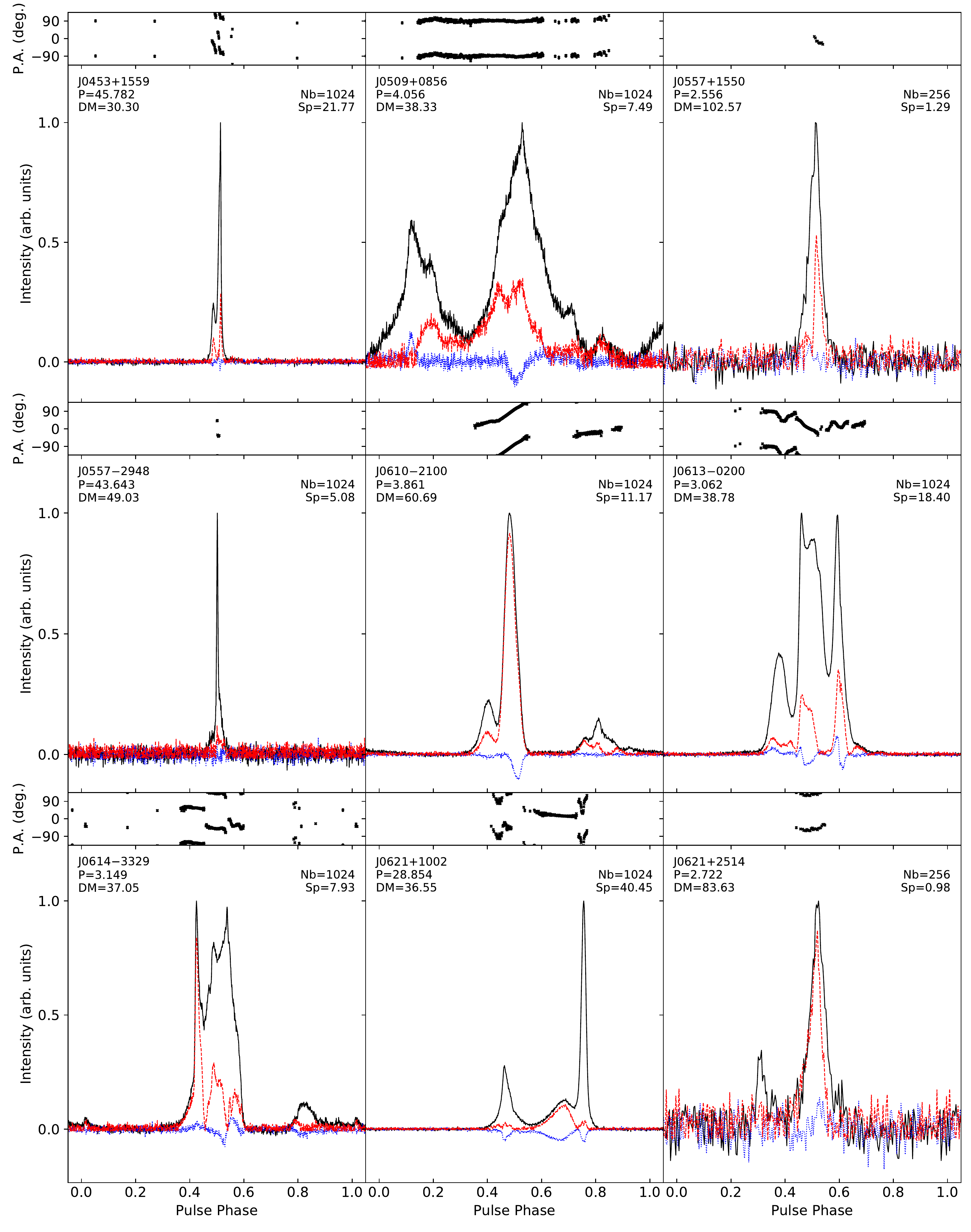}
    \caption[MeerTime Census Polarization Profiles - 2]{Polarization profiles for \acp{msp} from the MeerTime Census project, as Fig.~\ref{fig:mtc_profs_a}. }
    \label{fig:mtc_profs_b}
\end{figure*}

\begin{figure*}
    \centering
    \includegraphics[width=0.94\textwidth]{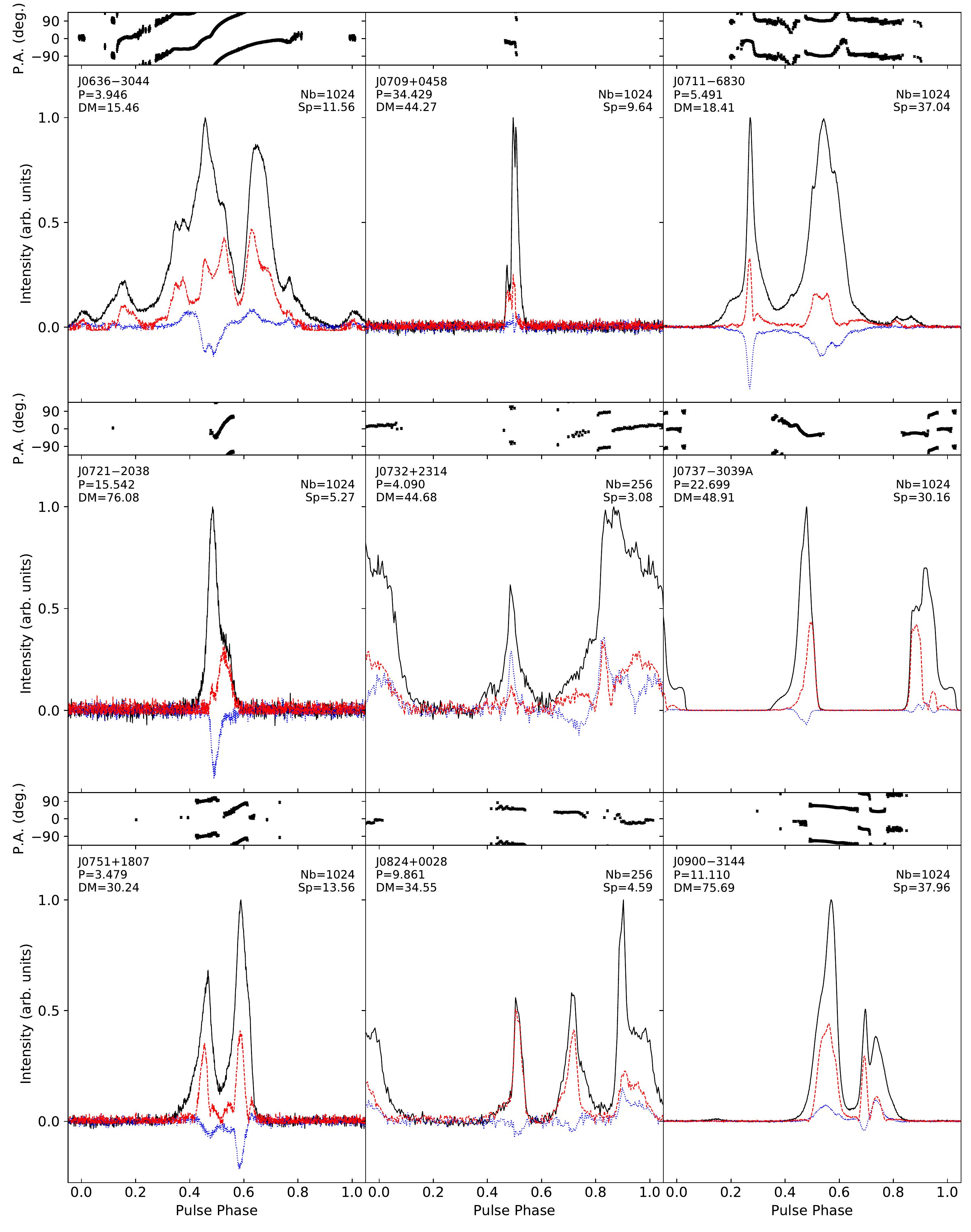}
    \caption[MeerTime Census Polarization Profiles - 3]{Polarization profiles for \acp{msp} from the MeerTime Census project, as Fig.~\ref{fig:mtc_profs_a}.}
    \label{fig:mtc_profs_c}
\end{figure*}

\begin{figure*}
    \centering
    \includegraphics[width=0.94\textwidth]{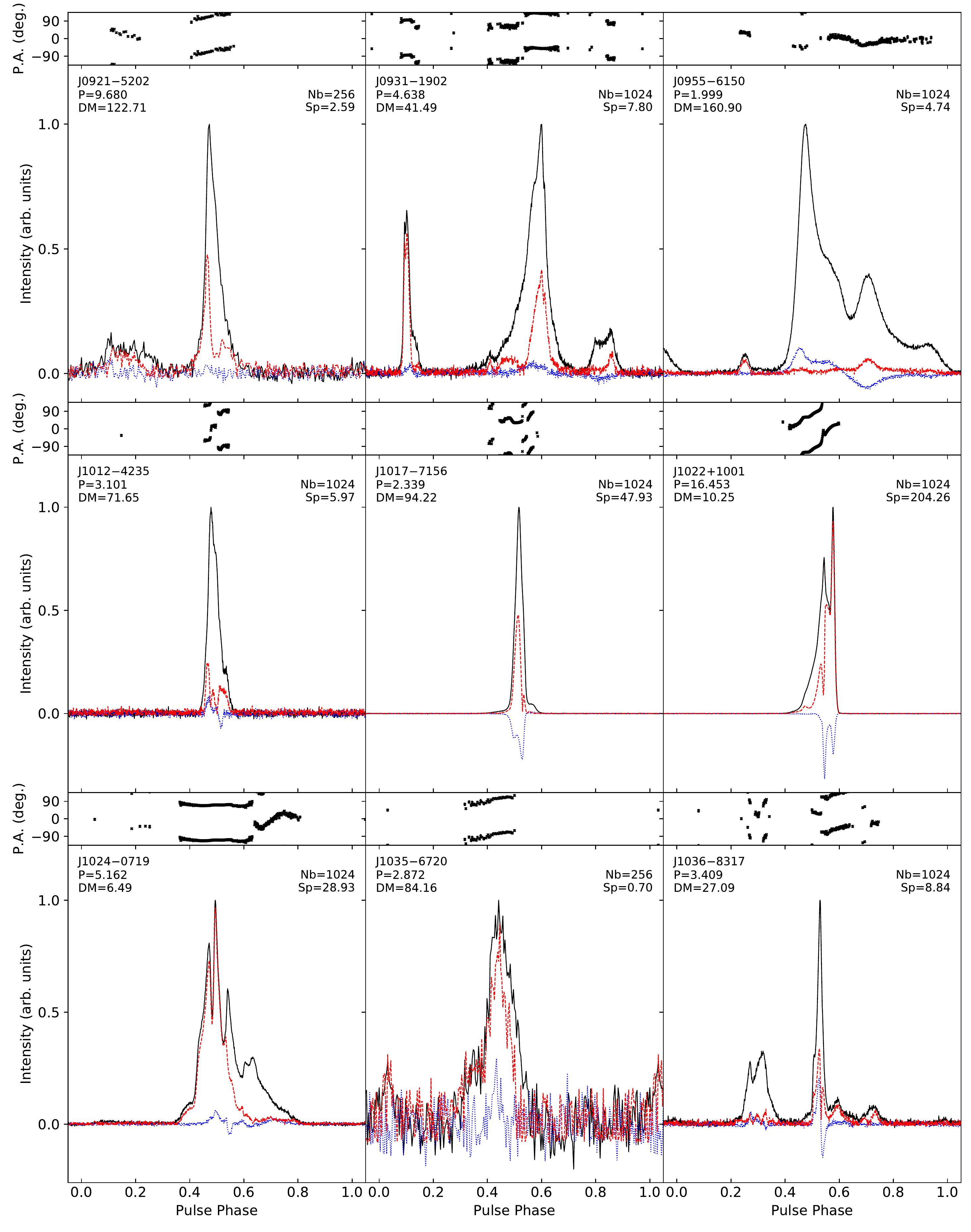}
    \caption[MeerTime Census Polarization Profiles - 4]{Polarization profiles for \acp{msp} from the MeerTime Census project, as Fig.~\ref{fig:mtc_profs_a}.}
    \label{fig:mtc_profs_d}
\end{figure*}

\begin{figure*}
    \centering
    \includegraphics[width=0.94\textwidth]{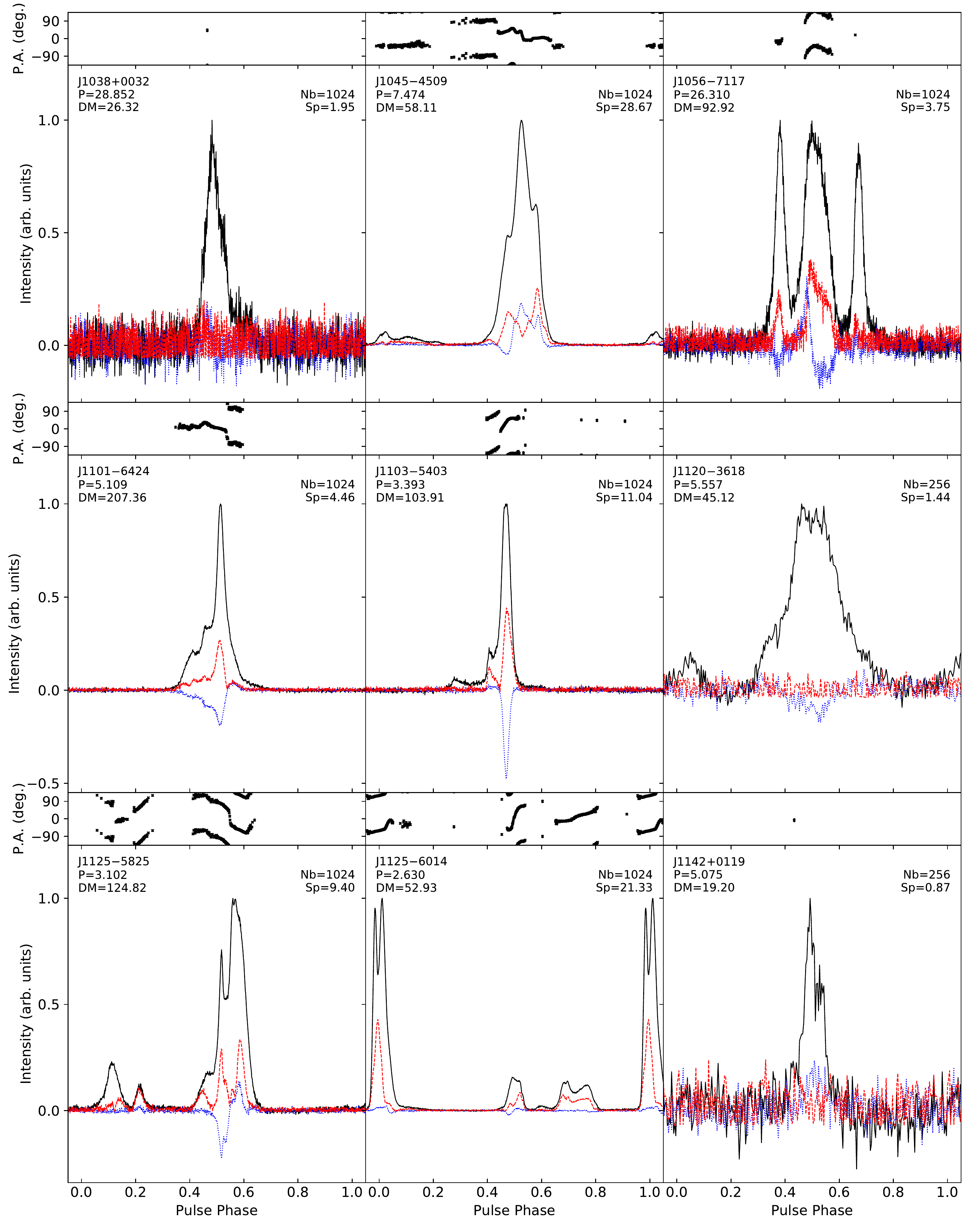}
    \caption[MeerTime Census Polarization Profiles - 5]{Polarization profiles for \acp{msp} from the MeerTime Census project, as Fig.~\ref{fig:mtc_profs_a}.}
    \label{fig:mtc_profs_e}
\end{figure*}

\begin{figure*}
    \centering
    \includegraphics[width=0.94\textwidth]{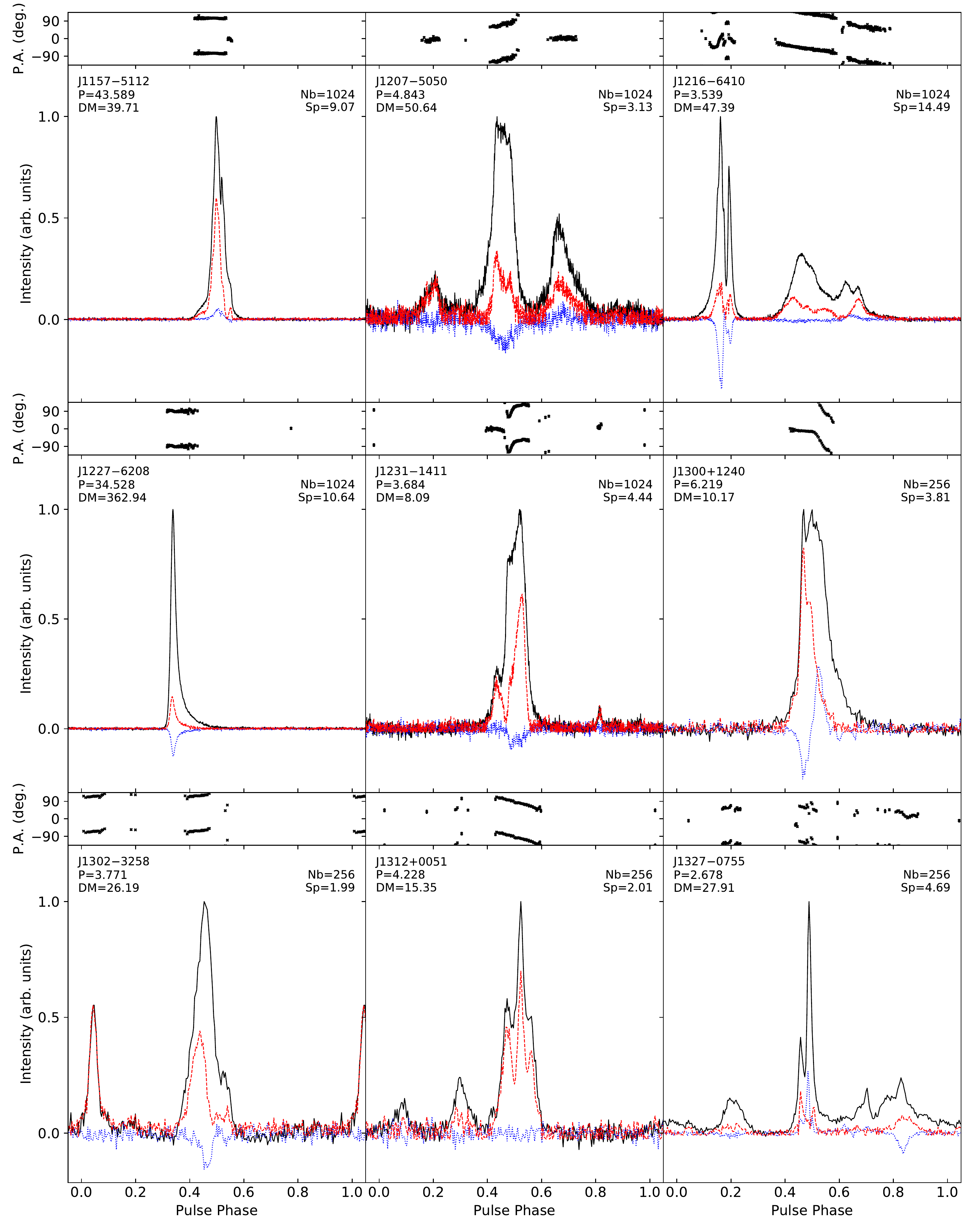}
    \caption[MeerTime Census Polarization Profiles - 6]{Polarization profiles for \acp{msp} from the MeerTime Census project, as Fig.~\ref{fig:mtc_profs_a}.}
    \label{fig:mtc_profs_f}
\end{figure*}

\begin{figure*}
    \centering
    \includegraphics[width=0.94\textwidth]{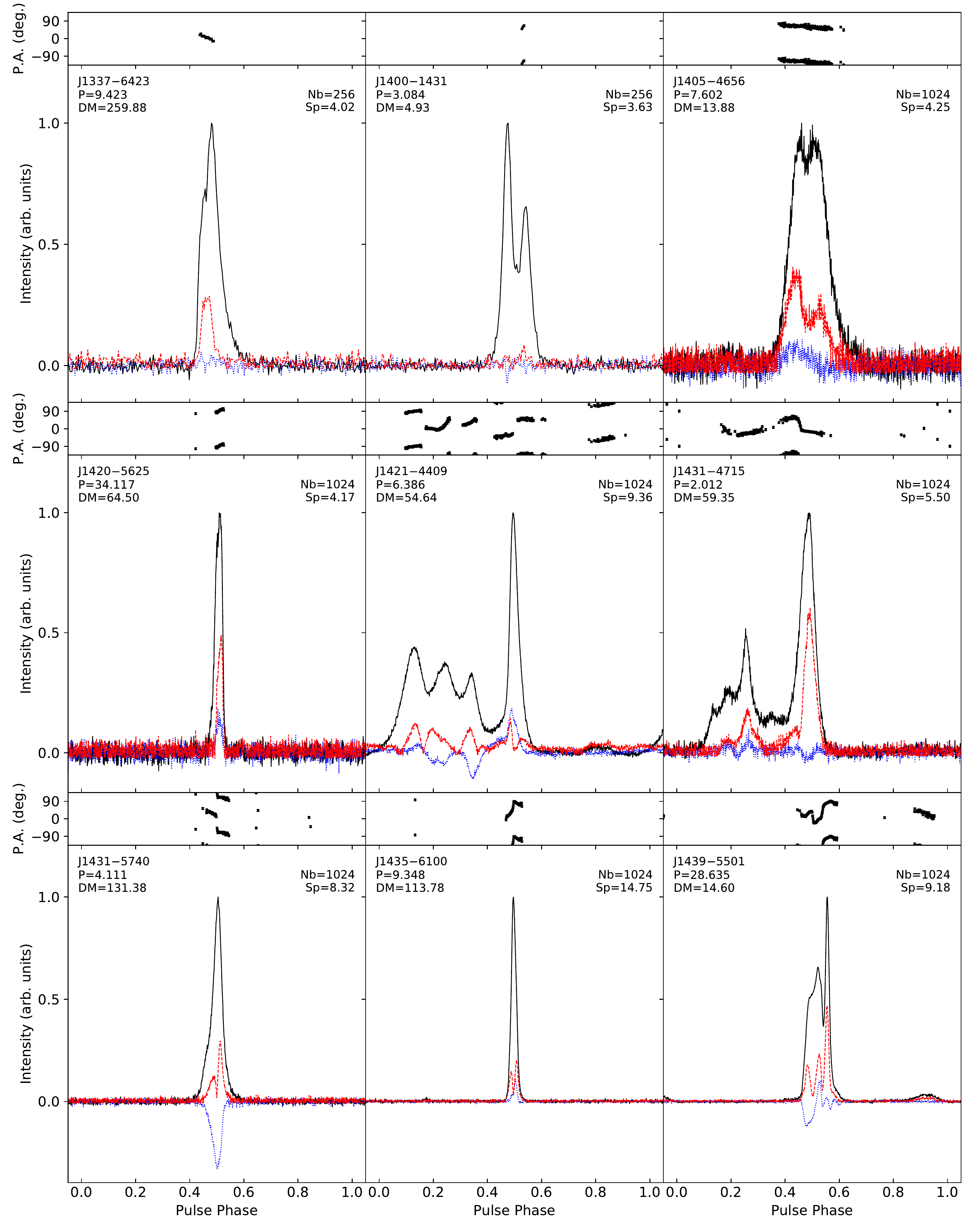}
    \caption[MeerTime Census Polarization Profiles - 7]{Polarization profiles for \acp{msp} from the MeerTime Census project, as Fig.~\ref{fig:mtc_profs_a}.}
    \label{fig:mtc_profs_g}
\end{figure*}

\begin{figure*}
    \centering
    \includegraphics[width=0.94\textwidth]{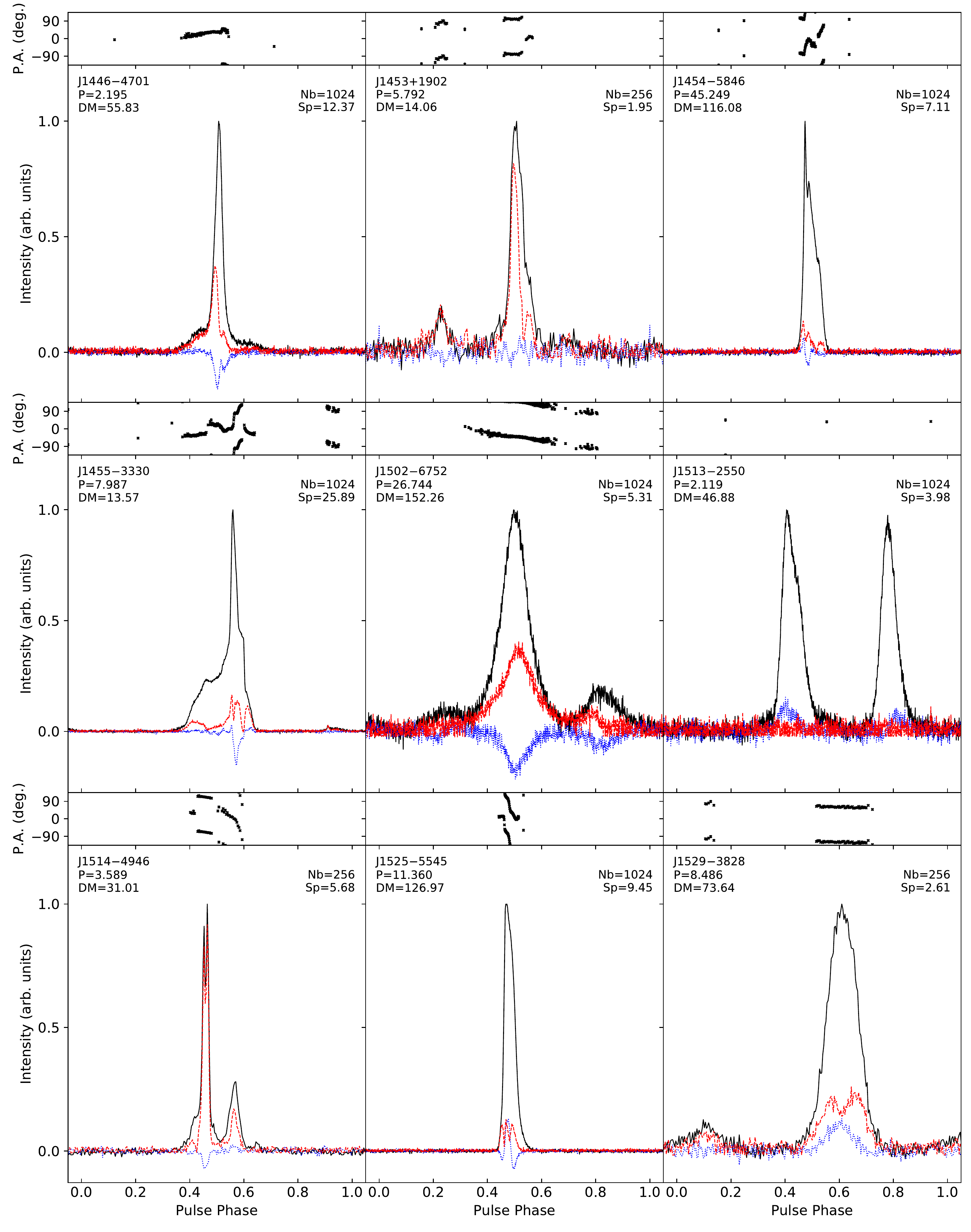}
    \caption[MeerTime Census Polarization Profiles - 6]{Polarization profiles for \acp{msp} from the MeerTime Census project, as Fig.~\ref{fig:mtc_profs_a}.}
    \label{fig:mtc_profs_h}
\end{figure*}

\begin{figure*}
    \centering
    \includegraphics[width=0.94\textwidth]{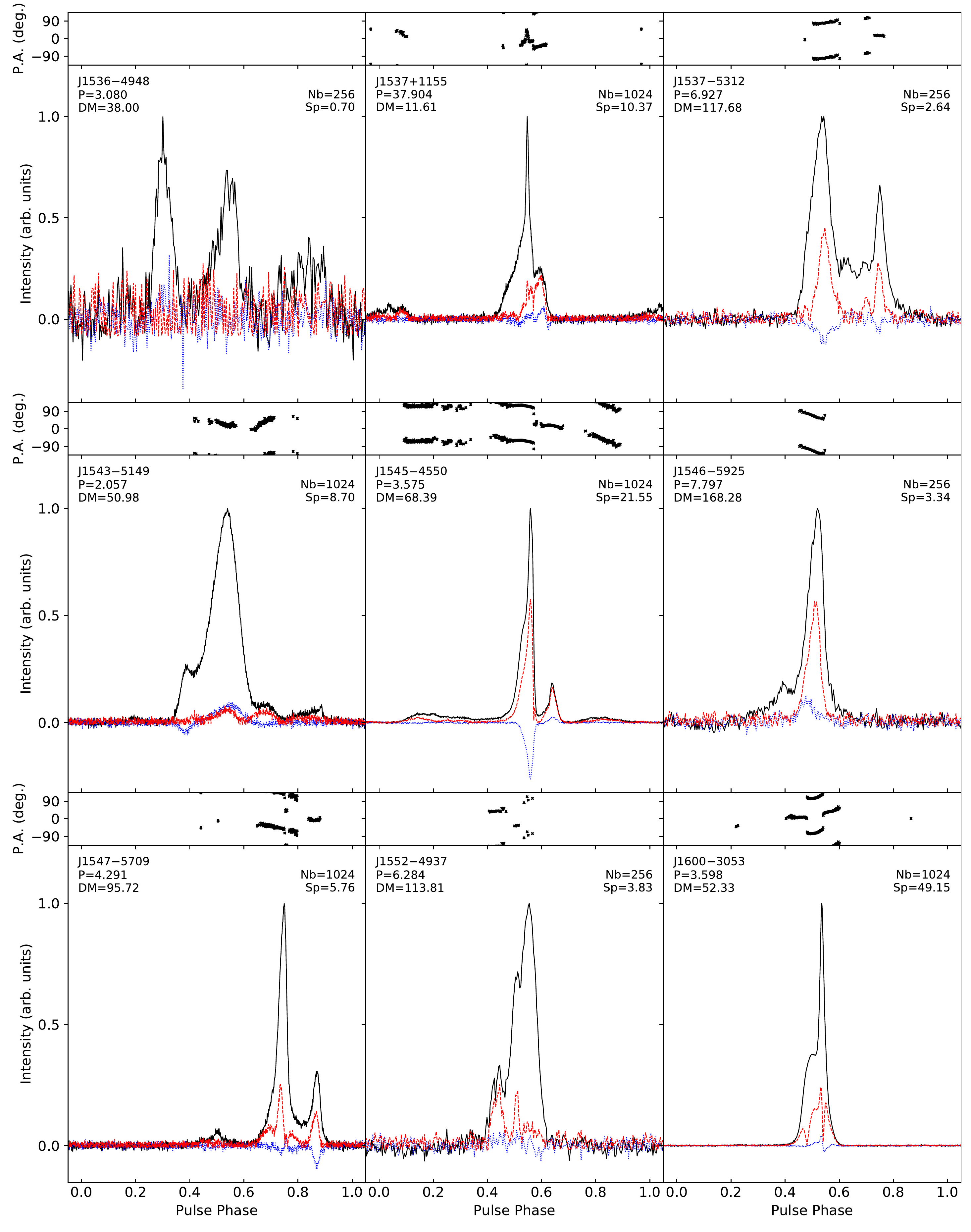}
    \caption[MeerTime Census Polarization Profiles - 9]{Polarization profiles for \acp{msp} from the MeerTime Census project, as Fig.~\ref{fig:mtc_profs_a}.}
    \label{fig:mtc_profs_i}
\end{figure*}

\begin{figure*}
    \centering
    \includegraphics[width=0.94\textwidth]{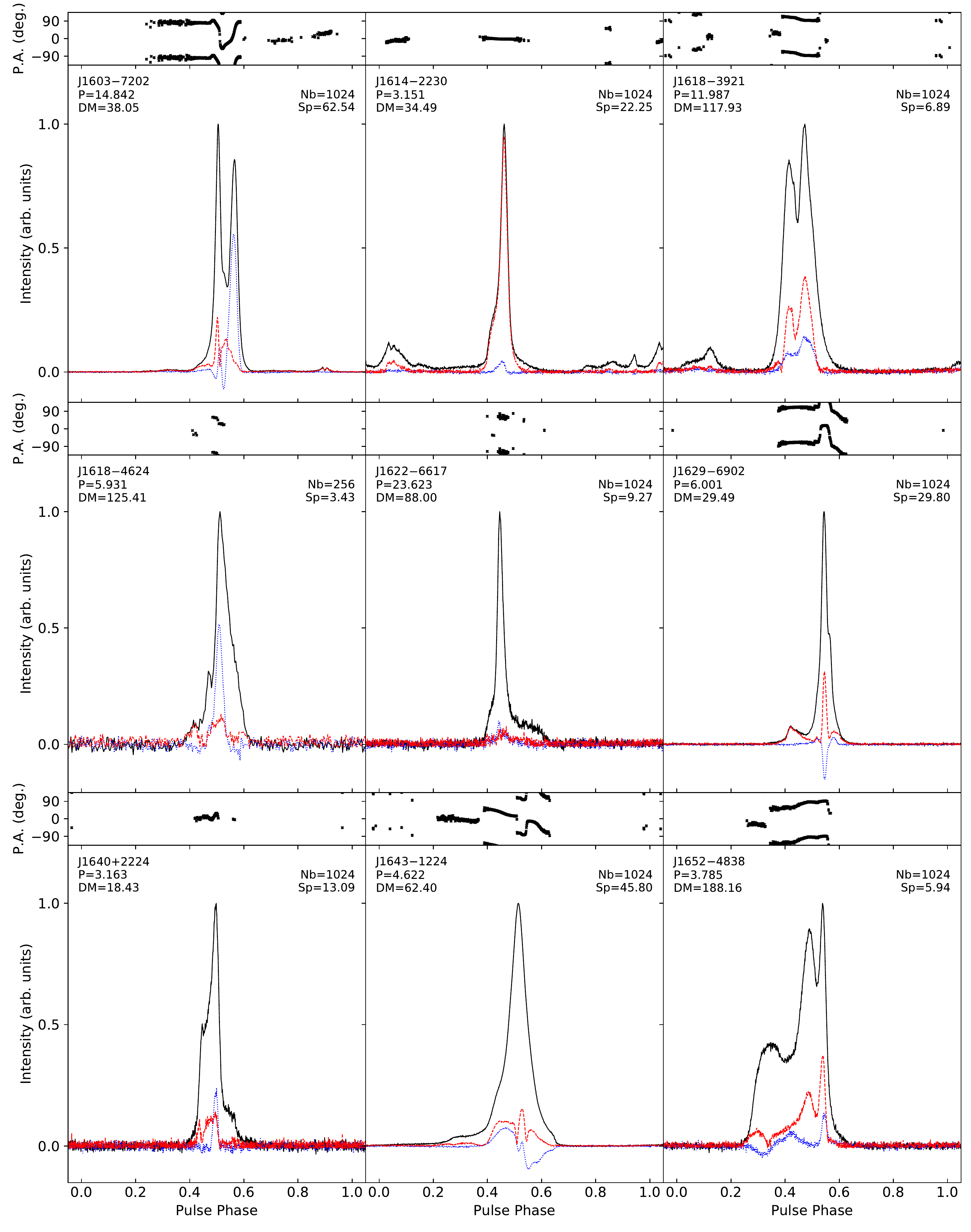}
    \caption[MeerTime Census Polarization Profiles - 10]{Polarization profiles for \acp{msp} from the MeerTime Census project, as Fig.~\ref{fig:mtc_profs_a}.}
    \label{fig:mtc_profs_j}
\end{figure*}

\begin{figure*}
    \centering
    \includegraphics[width=0.94\textwidth]{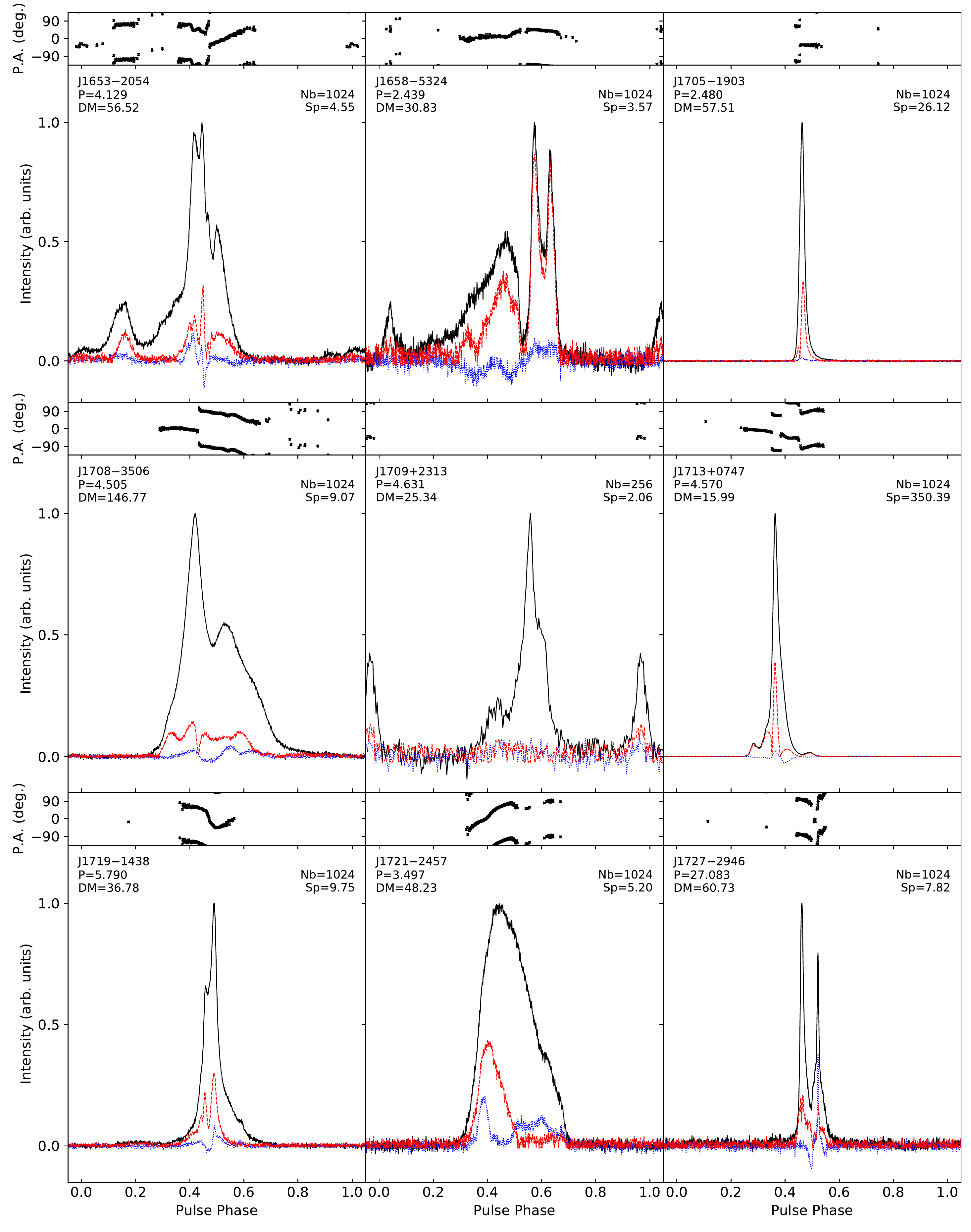}
    \caption[MeerTime Census Polarization Profiles - 11]{Polarization profiles for \acp{msp} from the MeerTime Census project, as Fig.~\ref{fig:mtc_profs_a}.}
    \label{fig:mtc_profs_k}
\end{figure*}

\begin{figure*}
    \centering
    \includegraphics[width=0.94\textwidth]{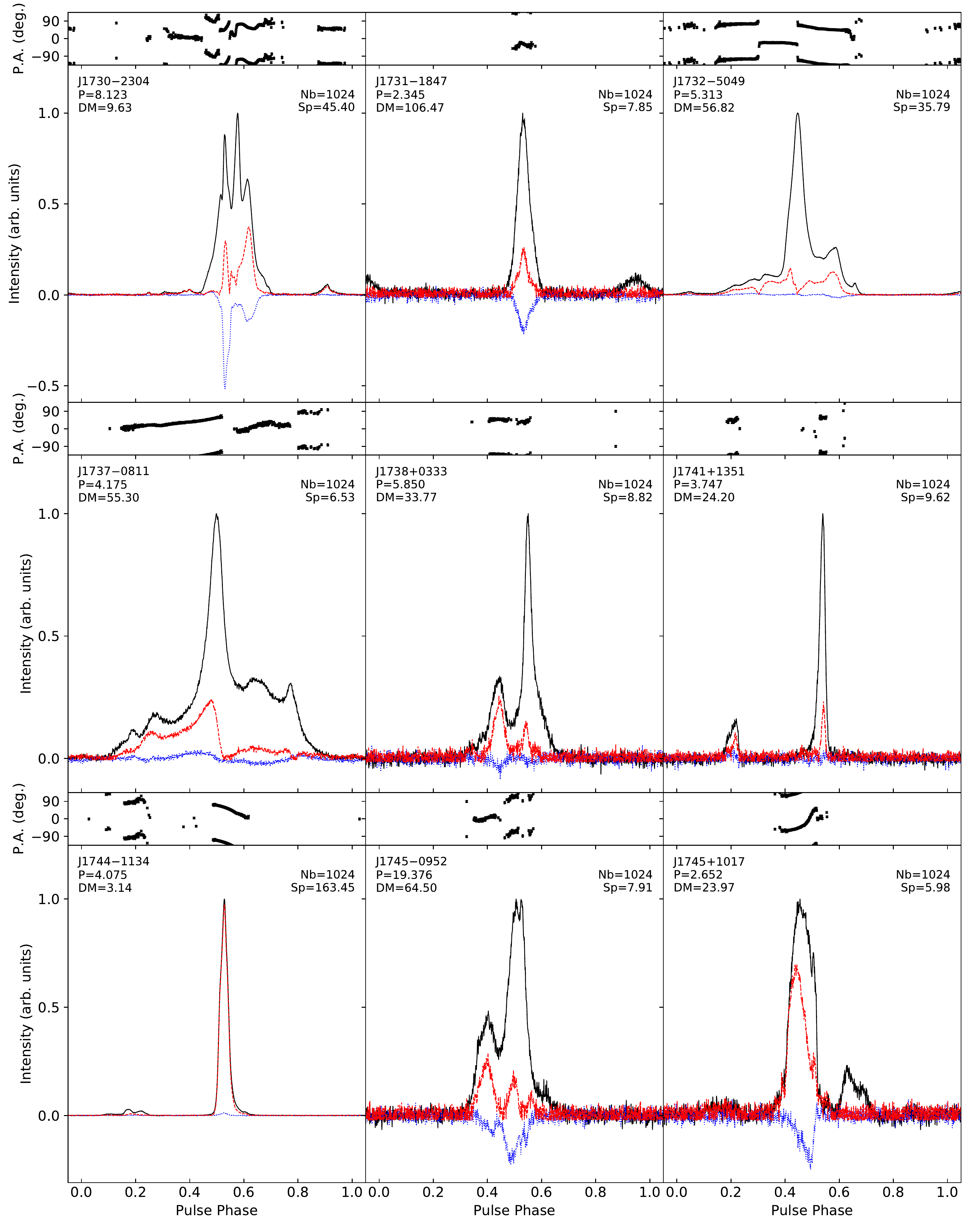}
    \caption[MeerTime Census Polarization Profiles - 12]{Polarization profiles for \acp{msp} from the MeerTime Census project, as Fig.~\ref{fig:mtc_profs_a}.}
    \label{fig:mtc_profs_l}
\end{figure*}

\begin{figure*}
    \centering
    \includegraphics[width=0.94\textwidth]{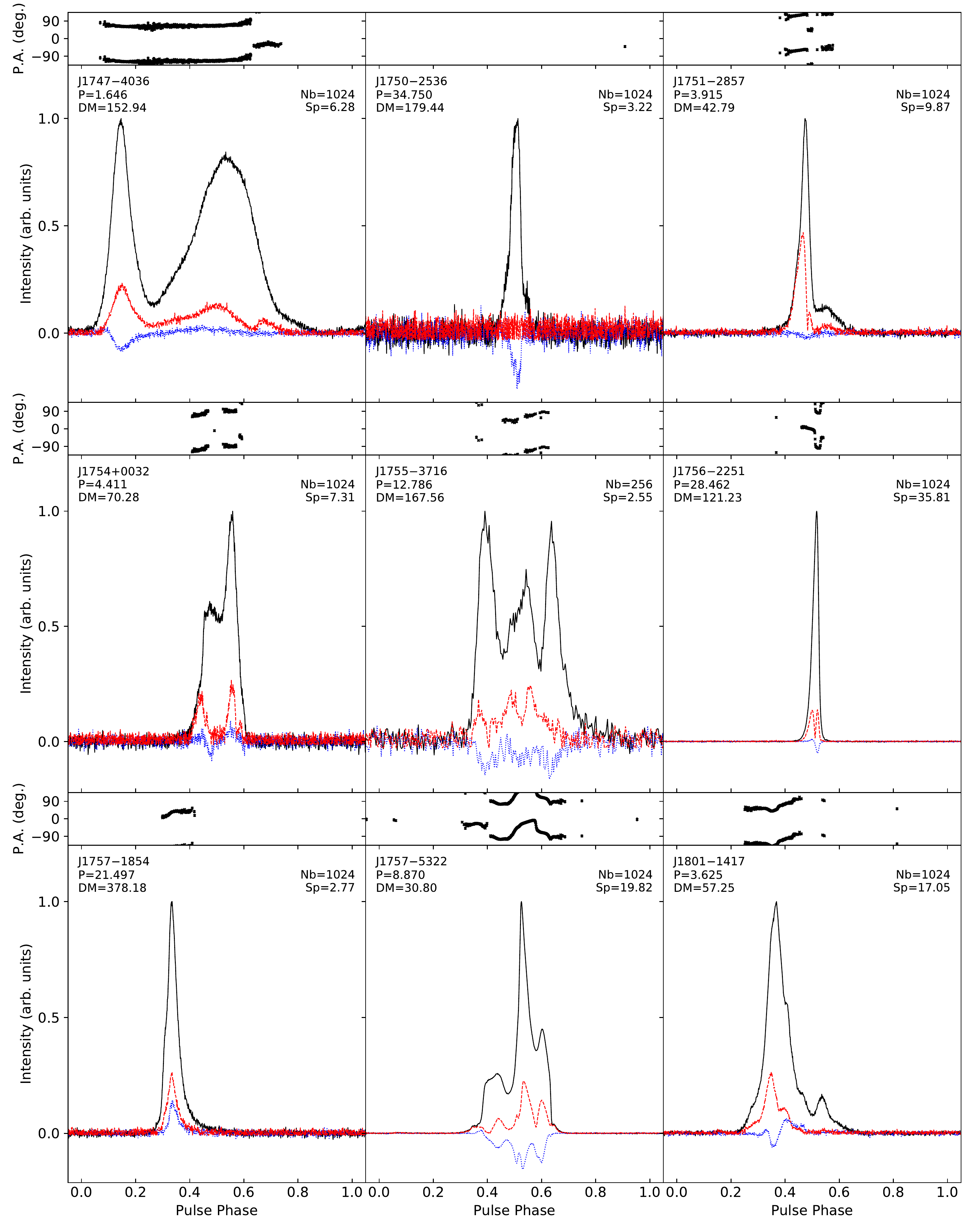}
    \caption[MeerTime Census Polarization Profiles - 13]{Polarization profiles for \acp{msp} from the MeerTime Census project, as Fig.~\ref{fig:mtc_profs_a}.}
    \label{fig:mtc_profs_m}
\end{figure*}

\begin{figure*}
    \centering
    \includegraphics[width=0.94\textwidth]{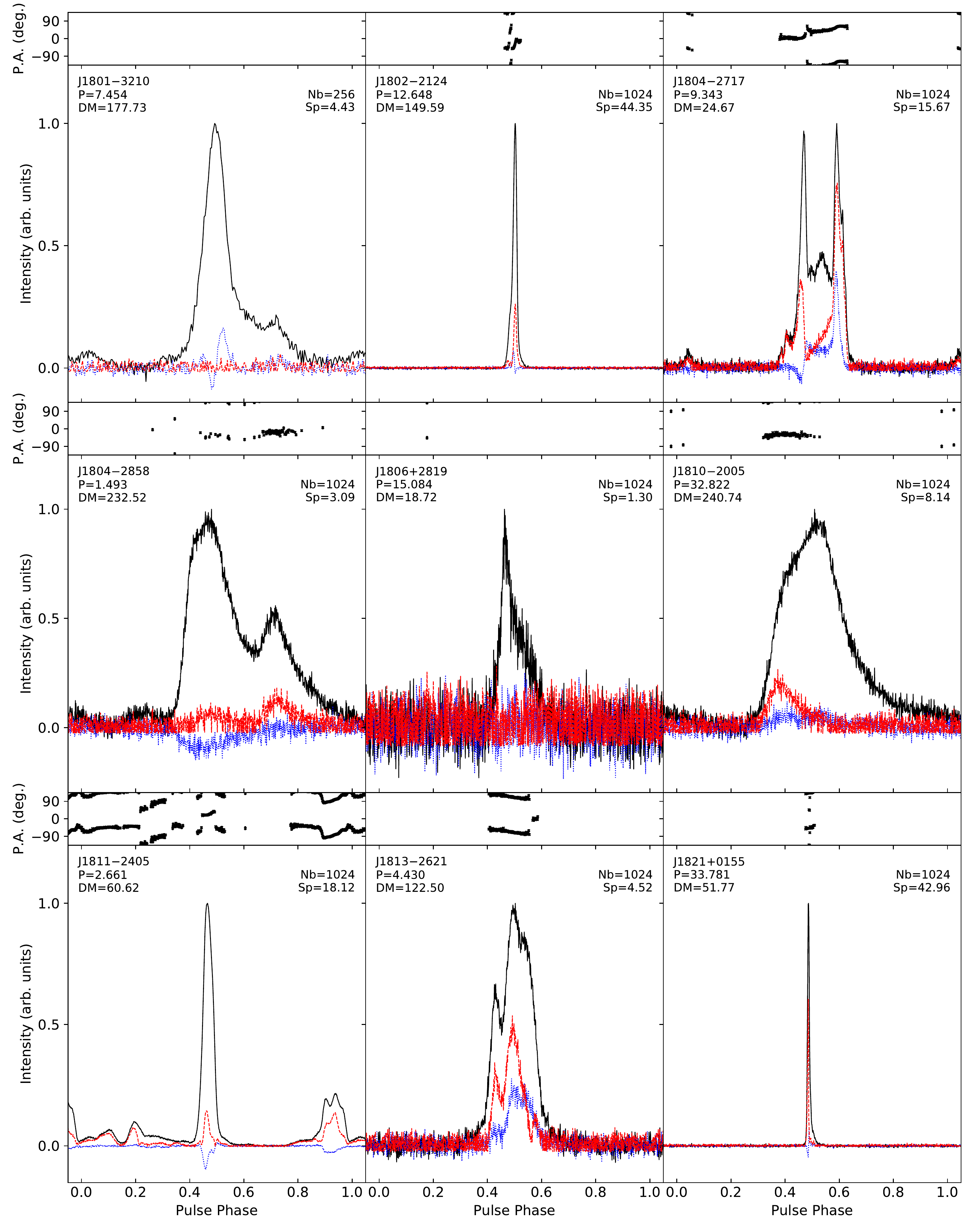}
    \caption[MeerTime Census Polarization Profiles - 14]{Polarization profiles for \acp{msp} from the MeerTime Census project, as Fig.~\ref{fig:mtc_profs_a}.}
    \label{fig:mtc_profs_n}
\end{figure*}

\begin{figure*}
    \centering
    \includegraphics[width=0.94\textwidth]{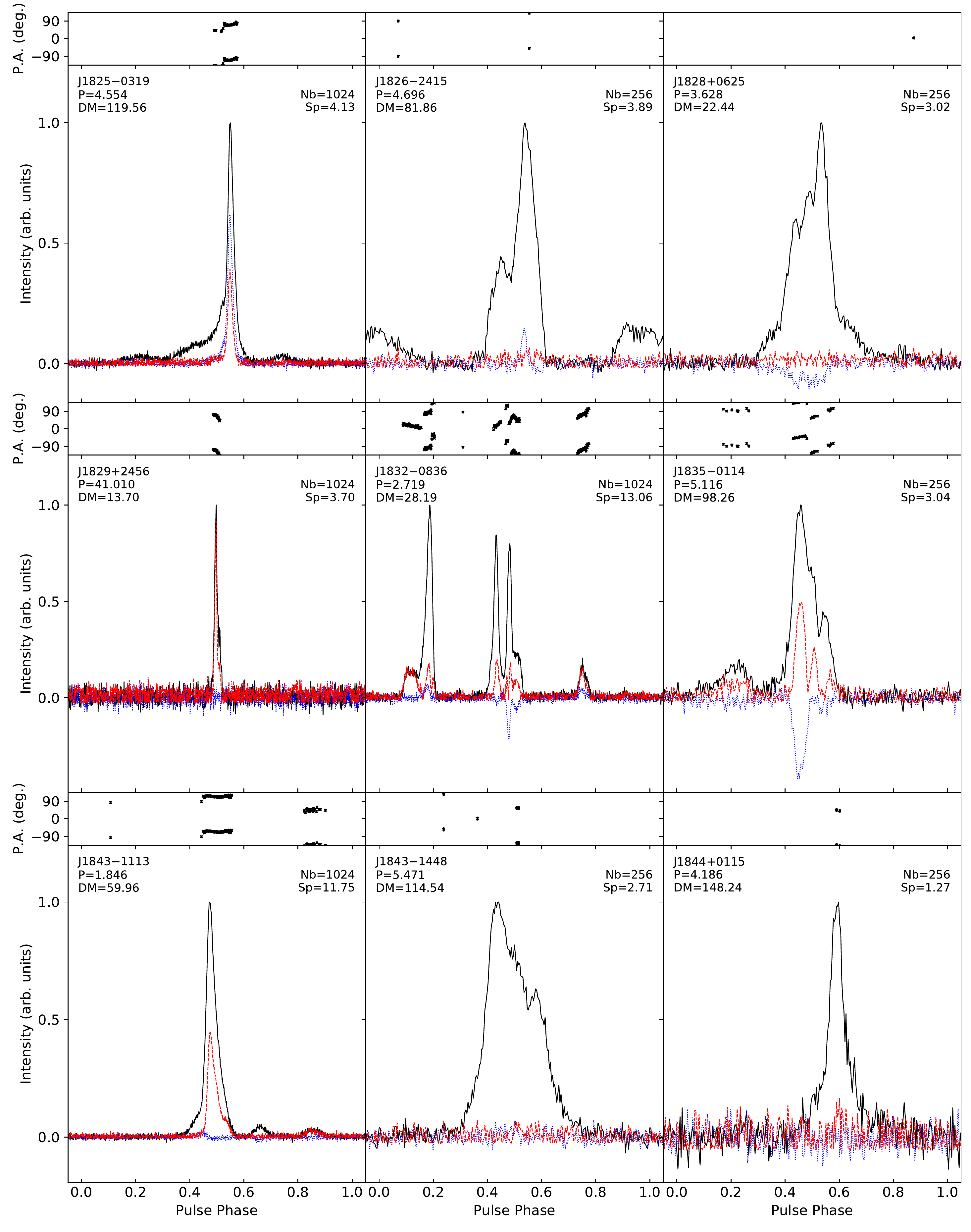}
    \caption[MeerTime Census Polarization Profiles - 15]{Polarization profiles for \acp{msp} from the MeerTime Census project, as Fig.~\ref{fig:mtc_profs_a}.}
    \label{fig:mtc_profs_o}
\end{figure*}

\begin{figure*}
    \centering
    \includegraphics[width=0.94\textwidth]{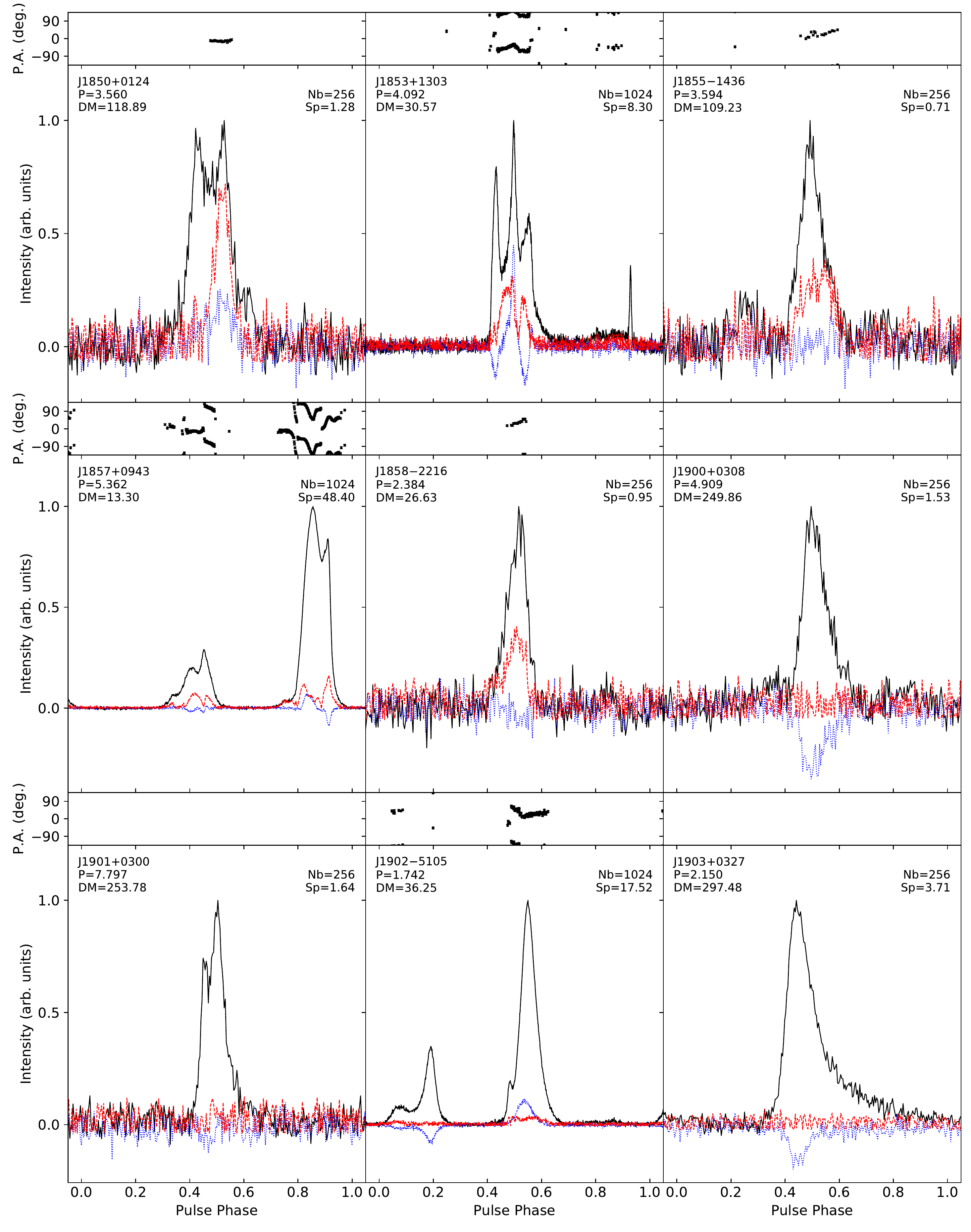}
    \caption[MeerTime Census Polarization Profiles - 16]{Polarization profiles for \acp{msp} from the MeerTime Census project, as Fig.~\ref{fig:mtc_profs_a}.}
    \label{fig:mtc_profs_p}
\end{figure*}

\begin{figure*}
    \centering
    \includegraphics[width=0.94\textwidth]{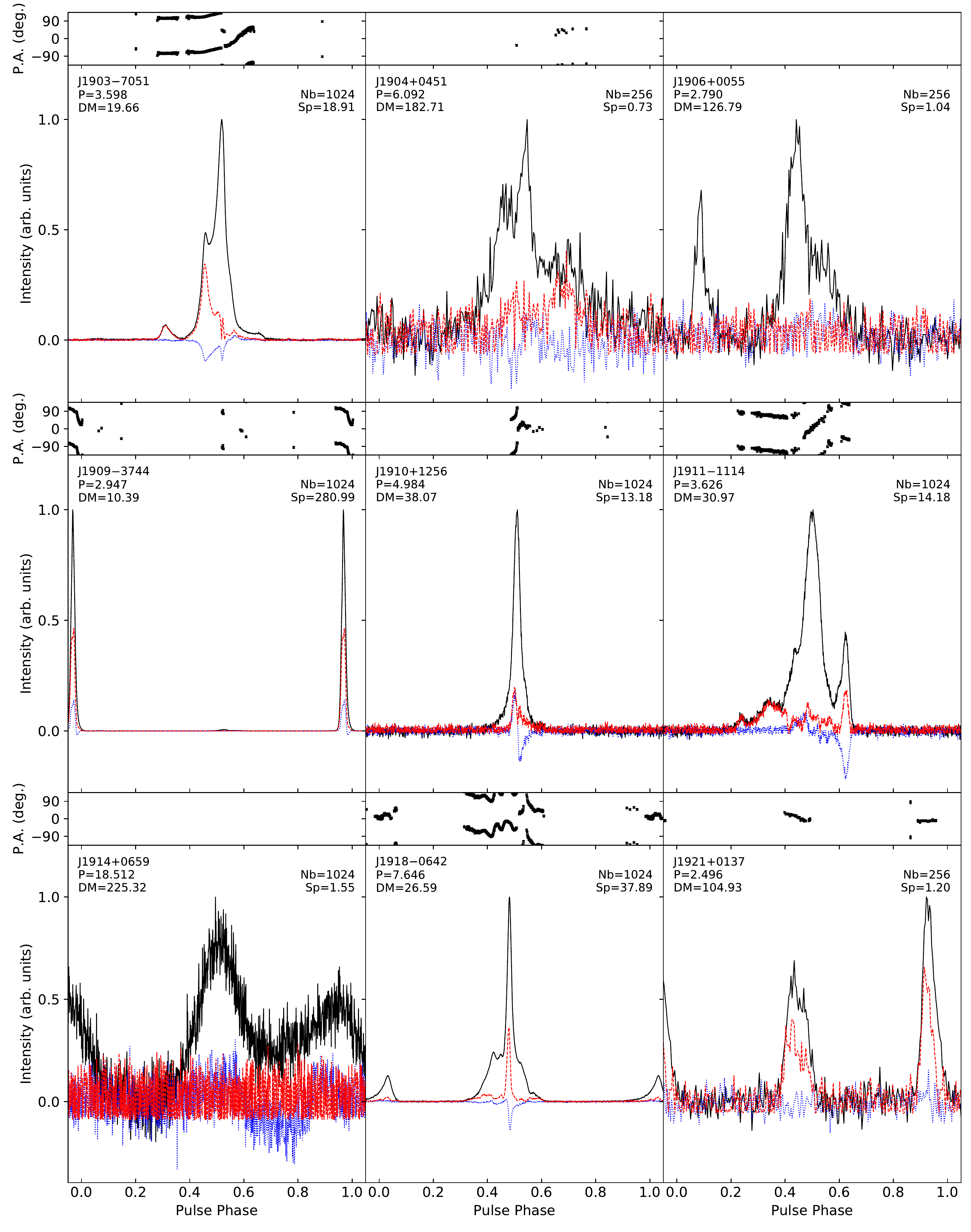}
    \caption[MeerTime Census Polarization Profiles - 17]{Polarization profiles for \acp{msp} from the MeerTime Census project, as Fig.~\ref{fig:mtc_profs_a}.}
    \label{fig:mtc_profs_q}
\end{figure*}

\begin{figure*}
    \centering
    \includegraphics[width=0.94\textwidth]{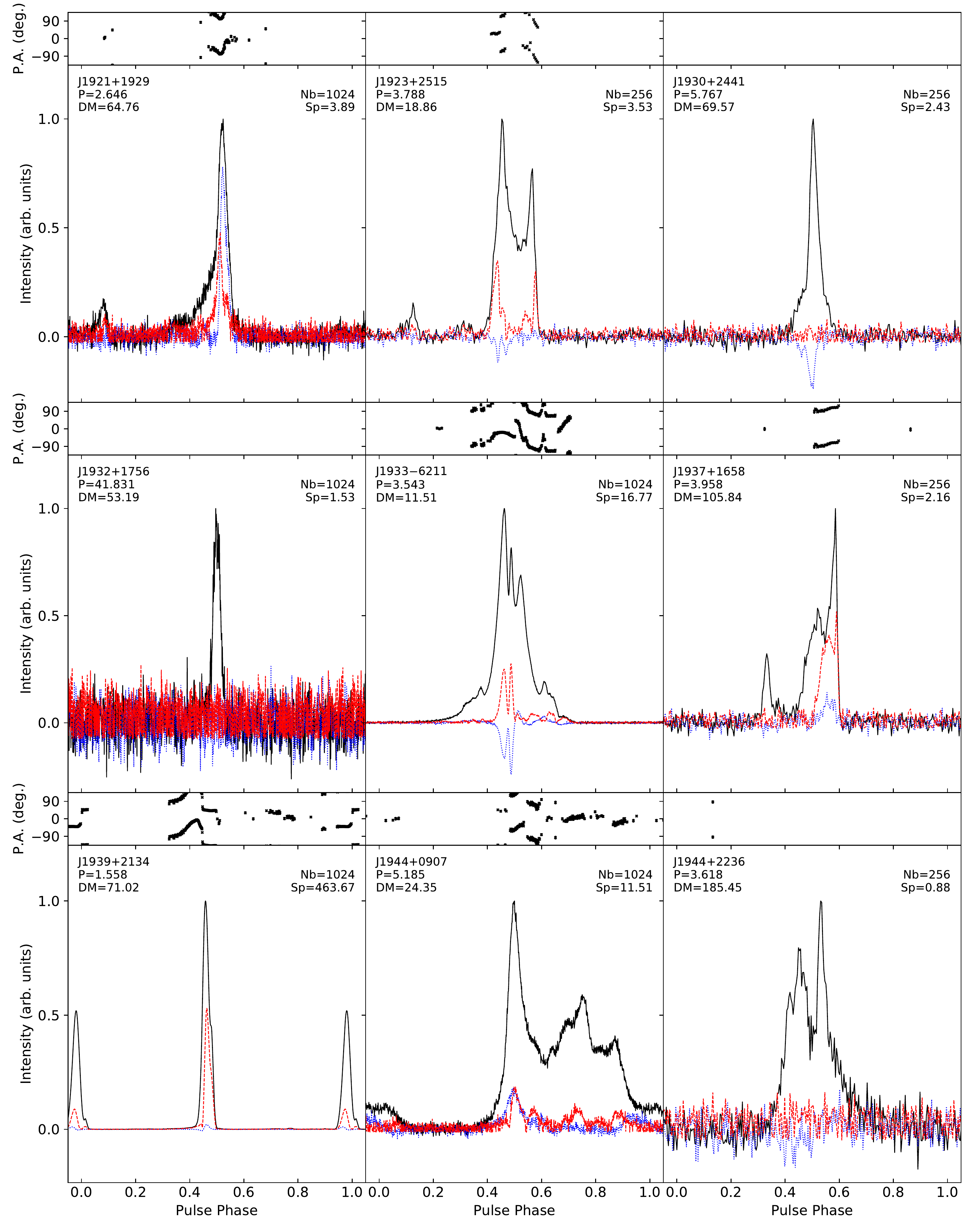}
    \caption[MeerTime Census Polarization Profiles - 18]{Polarization profiles for \acp{msp} from the MeerTime Census project, as Fig.~\ref{fig:mtc_profs_a}.}
    \label{fig:mtc_profs_r}
\end{figure*}

\begin{figure*}
    \centering
    \includegraphics[width=0.94\textwidth]{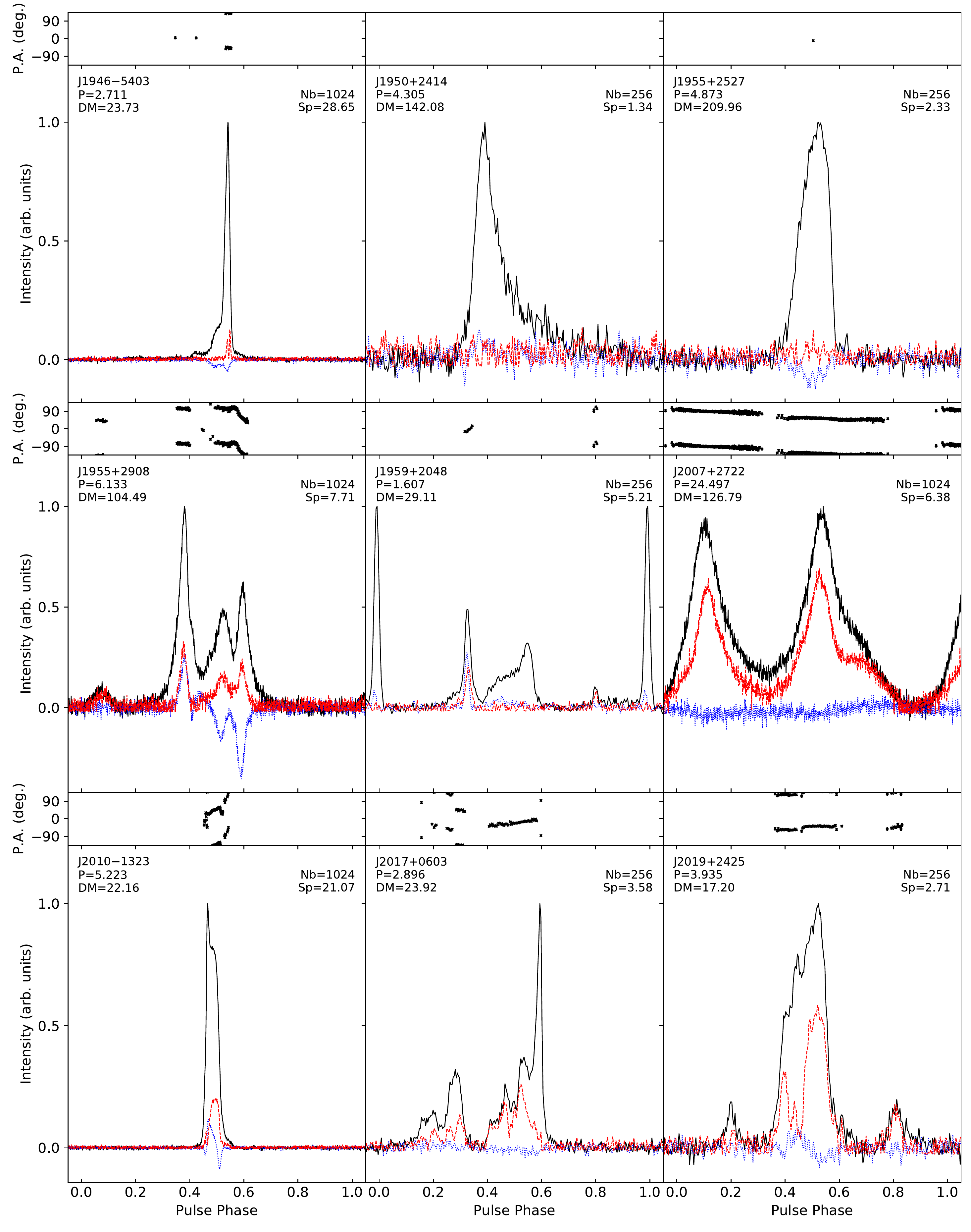}
    \caption[MeerTime Census Polarization Profiles - 19]{Polarization profiles for \acp{msp} from the MeerTime Census project, as Fig.~\ref{fig:mtc_profs_a}.}
    \label{fig:mtc_profs_s}
\end{figure*}

\begin{figure*}
    \centering
    \includegraphics[width=0.94\textwidth]{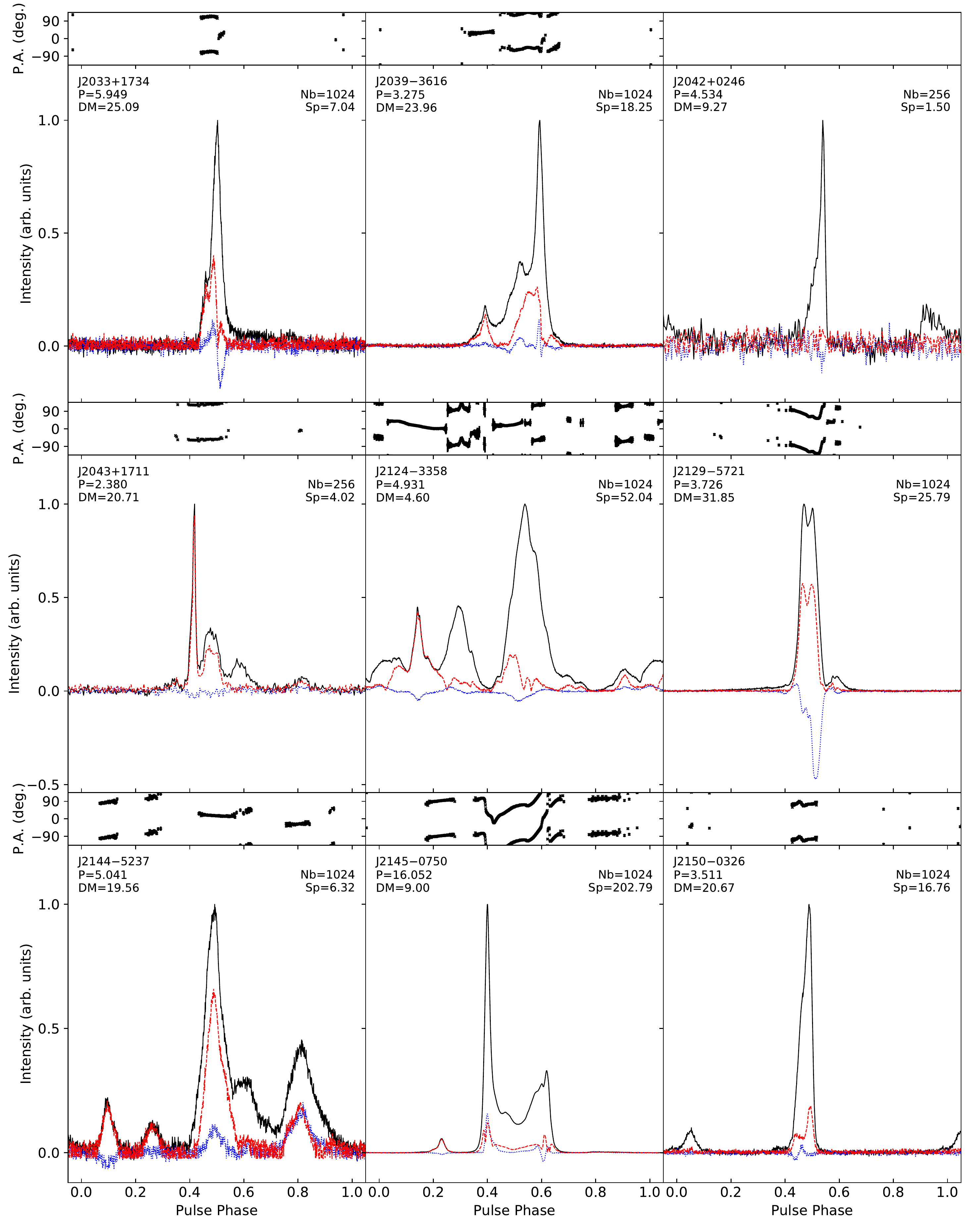}
    \caption[MeerTime Census Polarization Profiles - 20]{Polarization profiles for \acp{msp} from the MeerTime Census project, as Fig.~\ref{fig:mtc_profs_a}.}
    \label{fig:mtc_profs_t}
\end{figure*}

\begin{figure*}
    \centering
    \includegraphics[width=0.94\textwidth]{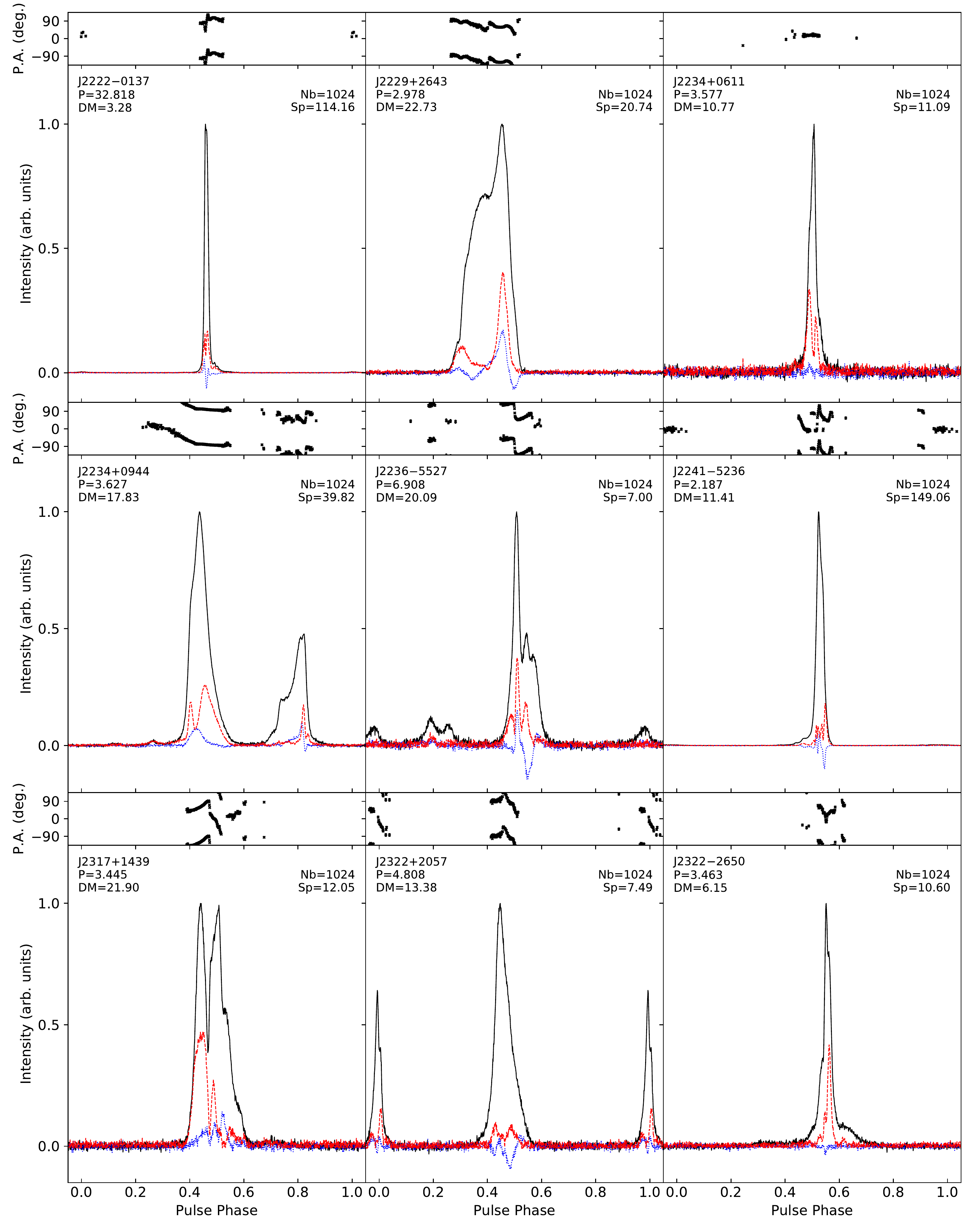}
    \caption[MeerTime Census Polarization Profiles - 21]{Polarization profiles for \acp{msp} from the MeerTime Census project, as Fig.~\ref{fig:mtc_profs_a}.}
    \label{fig:mtc_profs_u}
\end{figure*}


\end{document}